\documentclass[9pt]{article}
\usepackage{lscape}
\usepackage[utf8]{inputenc}
\usepackage{authblk}
\usepackage{cite}
\usepackage{array}
\usepackage{amsmath}
\usepackage{stackengine}
\usepackage{multicol}
\usepackage[table]{xcolor}
\usepackage{amsfonts}
\usepackage{amssymb}
\usepackage{graphicx}
\usepackage{multirow}
\usepackage{tabularx}
\newcolumntype{C}[1]{>{\centering\arraybackslash}p{#1}}
\usepackage[bottom]{footmisc}
\usepackage{hyperref}
\usepackage{secdot}
\usepackage{caption}
\usepackage{lscape}
\usepackage{rotating}
\usepackage{epstopdf}
\usepackage{cleveref}
\usepackage[font=sf]{caption}
\usepackage{titlesec}
\usepackage{wrapfig,lipsum,booktabs}
\usepackage[utf8]{inputenc}
\usepackage{microtype}
\usepackage[left=0.6in, right=0.6in, top=0.5in, bottom=0.6in]{geometry}
\usepackage{float}
\usepackage{url}
\usepackage{hyperref}

\title{}
\date{}
\begin{document}

%\maketitle

\begin{center}
\begin{Large}
    \textbf{Burstiness and interpersonal foraging between human infants and caregivers in the vocal domain}\\
\end{Large}

\begin{footnotesize}
    VPS Ritwika\textsuperscript{a,1}, Sara Schneider\textsuperscript{b}, Lukas D. Lopez\textsuperscript{c}, Jeffrey Mai\textsuperscript{a}, Ajay Gopinathan\textsuperscript{d}, Christopher T. Kello\textsuperscript{e}, Anne S. Warlaumont\textsuperscript{a,f,1}
\end{footnotesize}
\end{center}

\begin{scriptsize}
\noindent{}\textsuperscript{a}Department of Communication, University of California, Los Angeles; \textsuperscript{b}Nicotine \& Cannabis Policy Center, Health Sciences Research Institute, University of California, Merced; \textsuperscript{c}Department of Family and Consumer Studies, University of Utah; \textsuperscript{d}Department of Physics, University of California, Merced; \textsuperscript{e}Department of Cognitive and Information Sciences, University of California, Merced; \textsuperscript{f}Department of Psychology, University of California, Los Angeles

\noindent{}\textsuperscript{1}Corresponding author(s); e-mail: ritwika@ucmerced.edu OR warlaumont@ucla.edu

\noindent{} \textit{Preprint elements}: Figures~1-3, Appendices~1-9 (including Figures~1-36 and Tables~1-16) as supplemental PDF.

\noindent{} \textit{Keywords}: infant-caregiver interactions $|$ vocal communication $|$ burstiness $|$ behavioral clustering $|$ foraging  $|$ contingent responses 
\end{scriptsize}

\bigskip
\bigskip
\begin{center}
\begin{large}
    \textbf{Abstract}
\end{large}
\end{center}
Adult caregiver vocal responses are believed to promote more frequent and more advanced infant vocalizations. However, the inherent hierarchical temporal clustering of infant and caregiver vocal interactions over the course of a day creates a confound for reliably measuring immediate turn-by-turn effects of social vocal input within real-world vocal behavior. We first quantify temporal clustering in infant and caregiver vocalization events in daylong naturalistic infant-centered (ages 3, 6, 9, and 18 months) audio recordings using positive correlations between successive inter-event intervals (IEIs). Informed by flight time analyses in foraging studies, we develop an analytical framework to assess the effect of social input on the timing of subsequent infant and caregiver vocalizations after controlling for endogenous temporal dynamics as quantified by positive IEI-to-IEI correlations. For both infants and adults, receiving a social response predicts a shorter subsequent inter-event interval (IEI) than would be expected in the absence of a response. These results support a conceptualization of infant-caregiver vocal interactions as an ‘interpersonal foraging’ process with inherent multiscale dynamics wherein social responses serve as foraged-for resources. The analytic approaches introduced here offer a broadly applicable framework to study interactive communication in other modalities, contexts, and species by decoupling endogenous and exogenous dynamics.

\bigskip
\begin{center}
\begin{large}
    \textbf{Significance statement}
\end{large}
\end{center}
Many behaviors\textemdash including cognitive processes\textemdash can be understood as foraging processes, with individuals balancing resource gathering with broader exploratory search. 
We propose that infant-caregiver vocal interactions can be understood as `interpersonal foraging', with infants and caregivers seeking each others' vocal engagement and producing more vocal activity once engagement is obtained.
Over the course of the day, both human infant and adult caregiver vocalizations occur in hierarchically clustered bursts and lulls. 
Taking this clustering into account, we find evidence that adult responses to infants' speech-related vocalizations promote subsequent infant vocalization and vice versa, supporting the interpersonal vocal foraging view. 
This provides an ecological perspective on infant-caregiver vocal interaction, connecting this communication domain to a broad array of other search behaviors.

\bigskip
\section*{Introduction}

Human infants produce large numbers of vocalizations that are not yet words but are also not reflexive responses to specific emotional states (e.g., cries and laughs). 
These vocalizations, known as protophones or vocal babbling, are speech precursors \cite{locke1989babbling, oller2000emergence}. 
Infants babble when they are not engaged in social interactions \textit{and} during proto-conversations with adult caregivers \cite{long2020social}, where infant and adult vocalizations closely follow each other \cite{goldstein2008social, goldstein2003social, albert2018social, lipkind2020development, smith2008infant, pretzer2019infant, long2020social, nguyen2022systematic}. 
These proto-conversations and the contingent adult responses  within them are  important for speech and language learning \cite{kuhl2004early, casillas2016turn-taking,warlaumont2020infant,rohlfing2020multimodal, elmlinger2025immature, luchkina2024does}. 
Longitudinal studies have  shown that caregiver responses and infant-caregiver turn-taking predict later language outcomes, suggesting an enduring influence of infant-caregiver vocal interactions on language development \cite{tamis-lemonda2001maternal,rollins2003caregivers,ramirez2019parent,nguyen2023your,zhang2024developmental}.\\

It has been proposed that this connection between infant-caregiver vocal interactions and infant speech-language development is due to a social feedback loop wherein contingent adult vocalizations serve as rewards  guiding infants' exploratory vocal practice. 
The social feedback loop theory asserts that cycles of mutually contingent actions guide infant and adult vocal behavior, reinforcing infant vocalization in general and stimulating some vocalization types more than others (e.g., speech-related sounds versus cries and fusses) \cite{bloom1987turn, goldstein2003social, leezenbaum2013maternal, gros-louis2014maternal, warlaumont2014social, warlaumont2020infant, elmlinger2020learning}.
Interactive vocal input from caregivers may be reinforcing to infants in part because it provides rich learning opportunities and infants appear to learn the value of their own vocal productions as elicitors of this social input within the first six months after birth \cite{elmlinger2020learning}.\\

A related way to conceptualize infant-caregiver vocal interactions is as an `interpersonal foraging' process where infants and caregivers act as interacting foragers in the vocal domain, with social vocal engagement serving as a possible foraged-for resource \cite{ritwika2020exploratory}. 
This perspective is similar to the social feedback loop theory in that social input is hypothesized to modify infant and adult vocal behavior.
The foraging framing is supported by the hierarchical temporal clustering observed in  both infant and caregiver vocalizations over the course of a day.
Both infant and caregiver vocalizations tend to occur in bursts\textemdash multiple vocalizations in quick succession\textemdash followed by periods of silence.
These bursts and lulls  are hierarchically clustered in time, spanning multiple timescales, with coordinating clustering patterns between infants and adults \cite{abney2017multiple, warlaumont2022daylong}.
Such hierarchical clustering is common in  foraging processes (e.g., black-browed and winged albatrosses, spider monkeys, and marine predators such as sharks) in terms of distances between stops during resource gathering (`steps’ or `flights'), the time taken to traverse these steps (`flight times’), or both \cite{viswanasthan1996levy, viswanathan2011physics, ramos2004levy, humphries2012foraging, sims2008scaling}.
Adult humans’ semantic memory search\textemdash in the context of a memory retrieval task that required listing as many different types of animals as possible\textemdash also showed hierarchical clustering in steps in semantic space and corresponding inter-event intervals (i.e., time between successive retrievals) \cite{montez2015role, hills2012optimal}. 
Further, humans' spatial foraging patterns have been found to be correlated with participants' tendency toward divergent versus convergent cognitive styles \cite{malaie2024divergent}. \\

Foraging across these domains has been modeled as area-restricted search (ARS): a search strategy in which an agent adaptively switches between intensive local search (short spatial steps) and more exploratory global search (longer steps).
Intensive search is utilized in response to resource encounters or expectations thereof, to  exploit a resource-rich region while longer, exploratory steps are taken when resource encounters decline, to facilitate the discovery of resource-rich regions. 
Short steps correspond to short flight times\textemdash or analogously, inter-event intervals (IEIs), in the context of discrete, event-based processes\textemdash and longer steps correspond to longer flight times/IEIs \cite{hills2012optimal,dorfman2022guide}.
ARS has been shown to be  effective  when agents are able to respond to changing environmental cues and retain memory of past events \cite{bond1980optimal, visser1988host, hills2004dopamine}.
While human infant behavior has been explicitly characterized as foraging primarily in  visual foraging studies via eye movements \cite{robertson2004dynamics,robertson2012attentional}, the idea that exploration plays a dominant role in infant behavior and learning is widely held (e.g., \citenum{smith2005cognition,robertson2012attentional}).\\

Adapting this view to infant-caregiver vocal interactions, IEIs for vocalization events and corresponding inter-vocalization steps (in appropriate spatial representations of the vocal domain) are expected to be shorter during  interpersonal vocal engagement, reflecting transitions from exploratory to exploitative behavior following resource acquisition \cite{ritwika2020exploratory}. 
Accordingly, infants' and caregivers'  back-and-forth turn-taking  \cite{tamis-lemonda2001maternal,bornstein2015mother,hilbrink2015early,gratier2015early,dominguez2016roots,nguyen2023your,zhang2024developmental} can be viewed as periods when infants and adults are capitalizing on finding themselves in high social engagement contexts (i.e., resource-rich contexts).
Therefore, within an interpersonal foraging framework, shorter IEIs following social vocal interactions would reflect exploitative behavior, while  longer IEIs in the absence of vocal interactions would represent more exploratory behavior.\\

This explore-exploit trade-off modulated by an agent's  (social) reward reception is one of several similarities shared by foraging and reinforcement learning (RL) \cite{kolling2017reinforcement}. 
RL is in turn closely related to the social feedback loop theory\textemdash both are inspired by operant conditioning \cite{dorfman2022guide, warlaumont2014social}. 
Prior foraging research  has often focused on distributional properties of forager movements and  resource distributions, and models of foraging behavior typically assume simpler underlying processes \cite{viswanathan2011physics,kolling2017reinforcement}.
In contrast, RL models often assume more complex cognitive processes underlying agents' decision-making \cite{kolling2017reinforcement}.
The framing of social engagement as a temporally dynamic resource that shapes interactive dynamics through an underlying foraging process provides a complementary perspective, and differences between these perspectives lie in the behavioral patterns emphasized and the analytical tools employed.\\

The idea that caregiver responses serve as motivating rewards for infant vocal babbling is supported by computational models wherein reinforcement by a human or simulated listener results in emergent speech-like vocal characteristics \cite{howard2007computational, warlaumont2016learning}. 
Experimental studies that manipulated caregiver behavior  during short ($<$ 30 minutes) sessions have found that stimulating infants through contingent adult vocalizations increased the rate of infant vocalizations following manipulation \cite{rheingold1959social,weisberg1963social,nathani1996conditioning,goldstein2008social}.
Moreover, selective reinforcement of infant utterances containing only vowels or containing both vowels and consonants has been found to increase the reinforced infant vocal type \cite{routh1969conditioning}.
However, experimental approaches have limitations in understanding the posited bi-directional nature of infant-caregiver vocal interactions\textemdash it is impossible to give infants real-time instructions on when to produce vocal responses to an adult, to study the effects of infant responses on adult vocal behavior.
In addition, these experiments are limited in terms of event-level insights\textemdash information on the effect of a single instance of caregiver input on the infant’s next vocal action\textemdash and only provide session-level (e.g., over a period of several minutes) evidence that contingent responses increase subsequent vocalization frequency.\\ 

\begin{figure*}[h!]
\centering
    \includegraphics[width=12.85cm]{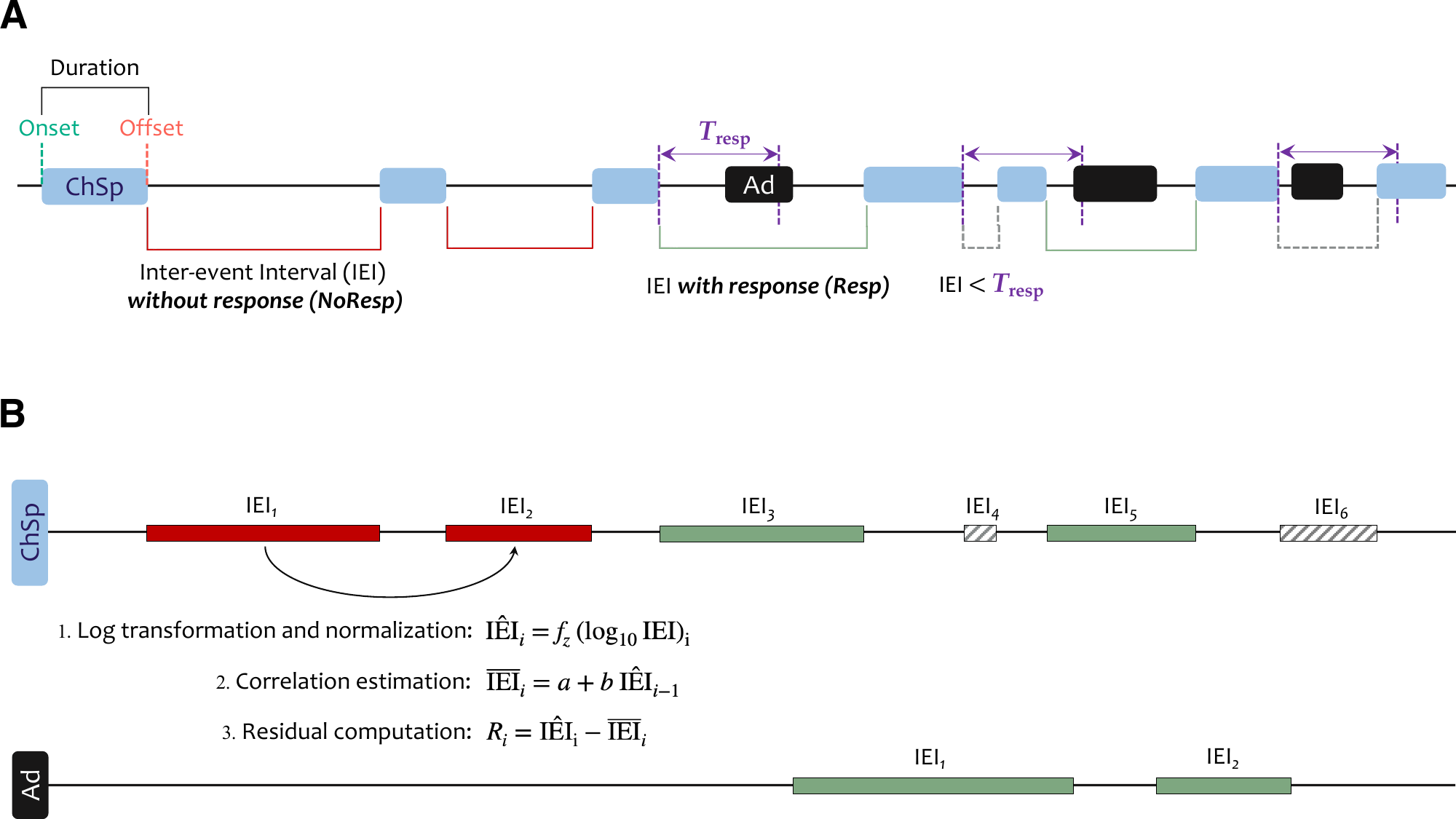}
    \caption{\textbf{Analysis procedure. (A)} shows a hypothetical sample time series of infant speech-related (ChSp; blue) and adult (Ad; black) vocalizations.
    ChSp inter-vocalization event intervals (IEIs) are computed as the offset-to-onset time difference between successive ChSp vocalizations; adult IEIs are computed similarly.
    The response window, $T_{\rm resp}$ (0.5 s to 10 s, with 1 s increments between 1 s and 10 s), is  in purple and determines whether an IEI is associated with a response. 
    Child speech-related (ChSp) IEIs are excluded from response analyses (dashed gray brackets) if less than or equal to $T_{\rm resp}$ or if the offset of the first ChSp in the sequence is followed by a child non-speech related vocalization (ChNsp; not shown) onset within $T_{\rm resp}$.
    Otherwise, a ChSp IEI is labeled as associated with a response (`Resp' IEIs; green brackets) if the offset of the first ChSp vocalization  is followed by an adult vocalization onset within time $T_{\rm resp}$.
    A ChSp IEI is labeled as not associated with a response (`NoResp' IEIs; red brackets) if there is no adult vocalization onset within time $T_{\rm resp}$ from the offset of the first ChSp vocalization.
    Adult (Ad) IEIs (not labeled in panel A) and their associations with ChSp responses are computed similarly.
    \textbf{(B)} shows the time series of ChSp and Ad IEIs corresponding to the vocalization series in A.
    IEIs associated with a response are in green, IEIs not associated with a response are in red, and IEIs excluded from response analyses are in gray stripes.
    The arrow from ChSp IEI\textsubscript{1} to IEI\textsubscript{2} represents the correlation between successive IEIs.
    Response receipt\textemdash and therefore, the choice of $T_{\rm resp}$\textemdash is not relevant to this analysis step;  \textit{all}  IEIs go into the correlation estimation.
    The equations summarize the analysis steps that go into correlation estimation and subsequent residual computation. 
    Visualizations of these steps are provided in Supplementary Figs. \ref{fig:ExtendedSchematics_Pt1} and \ref{fig:ExtendedSchematics_Pt2}.
    For more details, see Methods as well as Supplementary Sections \ref{subsec:RespVsNoRespIeis}, \ref{subsec:StatTests}, and \ref{subsec:ResponseAnalysis}.
    Colors used in this figure and subsequent figures in the main text have been tested for color accessibility.} 
    \label{fig:Schematic}
\end{figure*}

At the event level, prior research has indeed found higher-than-chance infant vocalization rates  following an adult vocal response and vice versa \cite{bornstein2015mother,harbison2018new,pretzer2019infant,ritwika2020exploratory}.
However, these findings are confounded by  infant and adult vocalizations'  coordinated temporal clustering, i.e., series of vocalization events separated by relatively short IEIs.
Specifically, the lack of statistical controls for such clustering means that despite infant and adult vocalizations' temporal proximity, one or both  agents may not actually be responsive to the other.
For example, adult responses  during an  infant vocalization burst can result in the misattribution of shortened infant IEIs to adult social input, even with no actual modification of infant vocal behavior due to adult vocal engagement.  
Indeed, a previous study found that infant cries tend to follow adult vocalizations\textemdash a finding that could be due to a tendency for cries to occur in bouts and for adults to vocalize during those cry bouts \cite{pretzer2019infant}.
Confounds  from inherent behavioral clustering pose unique challenges for studying the effects of resource acquisition on interpersonal foraging as conceptualized here: foraged-for resources are time-dependent and agent-generated (through their foraging behaviors),  making resources more likely to occur during activity bursts.\\

Here, we propose a method to address this confound and more reliably test for the effect of adult vocal input on subsequent infant vocal behavior and vice versa, by controlling for the temporal clustering of infant and adult vocalizations as evidenced by positive correlations between consecutive IEIs (Fig.~\ref{fig:Schematic}; see also Supplementary Section \ref{sec:burstiness} for a summary of  clustering in our data using a burstiness measure \cite{goh2008burstiness}).
IEI residuals from this correlation analysis provide a time series without the (first-order) event-to-event  clustering  in infant and caregiver vocalizations and therefore, can be utilized to assess the effect of social vocal interactions on infant and caregiver vocal behavior \textit{after} controlling for inherent temporal structure. 
Vocal responses  predicting shorter IEIs after controlling for IEI-to-IEI correlations  would be consistent with adult vocal input following an infant vocalization serving as a reward, causing the infant to vocalize sooner than they otherwise would, and vice versa for the effect of infant vocal input on adult vocalizations.
Analyses were performed at the daylong and five-minute levels using the LENA\texttrademark  \cite{gilkerson2020guide} system's automated sound segmentation and labeling and at the five-minute level using human annotations.
The large-scale, automatically labeled daylong dataset allows us to test for the effects of social vocal responses on infant and caregiver vocal behavior in naturalistic settings over the course of a typical day, across typical auditory and linguistic contexts experienced by infants, while the 5-minute human-labeled dataset along with the corresponding 5-minute automatically labeled dataset allows for validation of both the automatic labeling and observed trends from the daylong dataset. 
This study, therefore, expands on previous work \cite{ritwika2020exploratory} exploring a similar vocal foraging perspective by \textbf{(1)} providing an analytical framework to explicitly control for  inherent temporal structure that may otherwise bias assessments of the bi-directional excitatory effects of social vocal engagement; \textbf{(2)} demonstrating the utility of this framework on  a larger-scale longitudinal dataset  of naturalistic daylong audio recordings; and \textbf{(3)} validating analyses results using a larger human listener-labeled dataset.

\section*{Methods}

Extended methods are in Supplementary Section \ref{sec:SuppMethods}, with a summary provided below.
\textbf{Data.} 
We used data from 200 daylong (10+ hour) home audio recordings collected longitudinally (infant ages: 3, 6, 9, and 18 months) from infant-worn recorders (the LENA\textsuperscript{TM} Pro system), totaling over 2,400 hours of audio.
Participating families were recruited from the San Joaquin Valley region of California,
USA, and infants were primarily exposed to English and/or Spanish, although other languages were also spoken in some of the infants' homes (see Supplementary Sections \ref{subsec:DataCollec} and \ref{sec:DataOverview_FileNumAndAge} for details).
\textbf{Data processing.}
Recorded audio was automatically segmented and labeled by the LENA Pro software.
This study focuses on \textbf{Ch}ild \textbf{sp}eech-related (ChSp) and \textbf{Ad}ult (Ad) vocalizations as identified by  LENA, in addition to using \textbf{Ch}ild \textbf{N}on-\textbf{sp}eech-related (ChNsp) vocalizations to assess the receipt of a social response (see Fig. \ref{fig:Schematic}; see also Supplementary Section \ref{subsec:VocIding} for details).
\textbf{Data analysis.}
We first calculate infant and adult caregiver vocalization inter-event intervals (IEIs), distinguishing between IEIs based on whether the first vocalization event received a vocal response or not (see Fig.~\ref{fig:Schematic} and Supplementary Section \ref{subsec:RespVsNoRespIeis} for details).
We then assess the extent of temporal clustering in infants' and adults' spontaneous vocal productions using the full set of IEIs (without considering response receipt; see Fig. \ref{fig:Schematic} and Supplementary Section \ref{subsec:StatTests} for details).
We group these IEI residuals based on response receipt to test whether adult responses predict shorter infant IEIs after controlling for infant vocal clustering (Supplementary Fig. \ref{fig:ExtendedSchematics_Pt2}) and whether infant responses predict shorter adult IEIs after controlling for adult vocal clustering.
We also test for age effects for all assessments mentioned above (see Supplementary Section \ref{subsec:StatTests} for more details on statistical tests performed, and Supplementary Sections \ref{sec:DataOverview_FileNumAndAge} and \ref{sec:SegmentNumAndDuration} for details of data collection and sample sizes).
All data preparation and analyses were carried out in MATLAB (R2024a) and R (4.2.0).
\textbf{Validation using human listener-labeled 5-minute excerpts.}
To validate results based on LENA-labeled daylong data, we used a subset of the daylong data from 45 infants.
The validation dataset consists of 420 5-minute sections from 148 daylong recordings (up to three 5-minute sections per recording) independently segmented and labeled by trained human listeners (see Supplementary Section \ref{subsec:ValDataMethods} for details).

\begin{figure*}[t!]
    \centering
    \includegraphics[width = 12 cm]{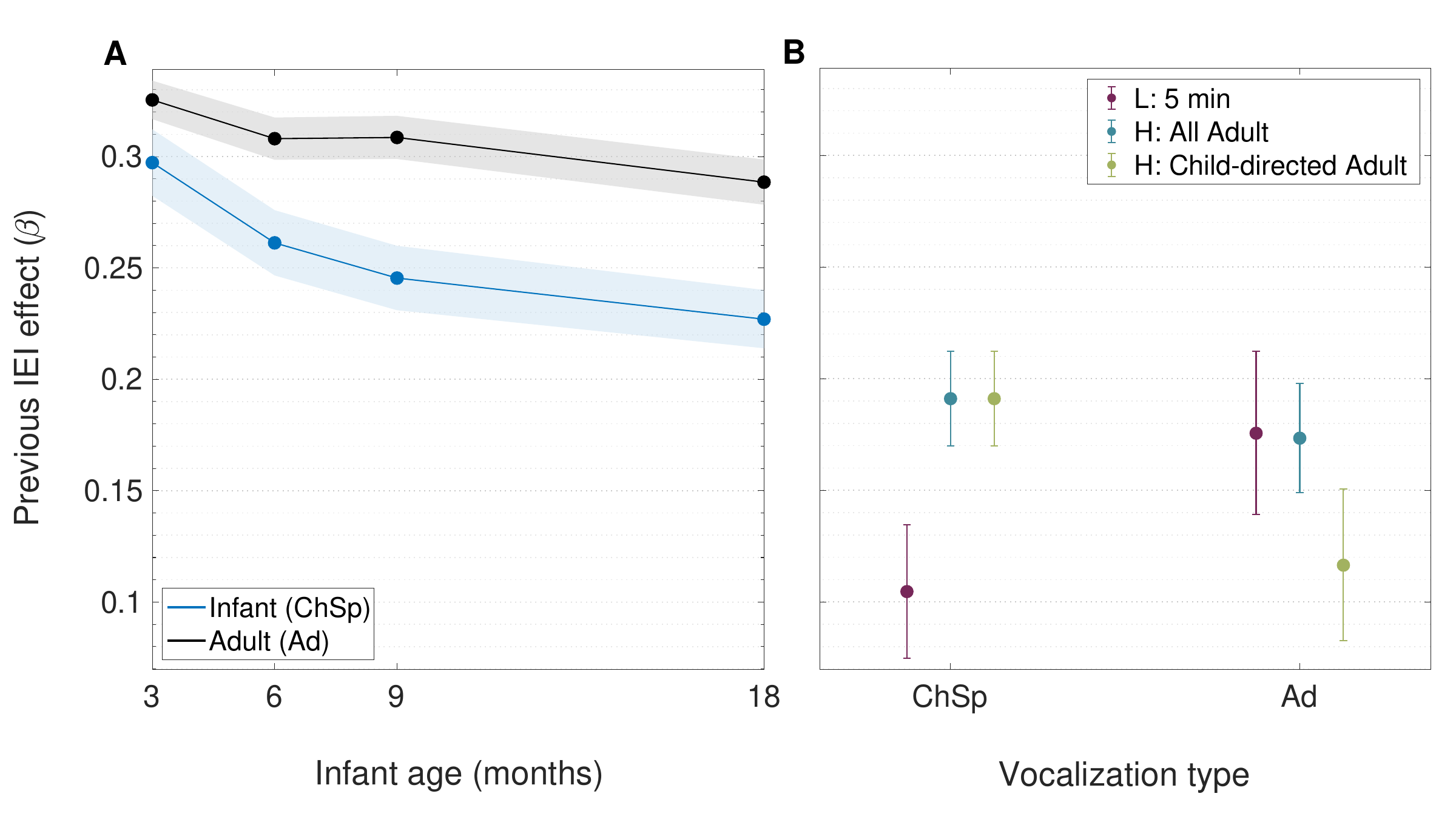}
    %all points at 0.001 sig level, 99.9 percent CI 
    \caption{\textbf{Positive correlations between previous IEI and current IEI. } 
    \textbf{(A)} Standardized regression coefficients ($\beta$; y-axis) from linear mixed effects analyses testing IEI-to-IEI correlations  for infant speech-related (ChSp; blue) and adult  (Ad; black) vocalizers are shown as a function of infant age in months (x-axis), for daylong vocalization data automatically labeled by LENA.
    All computed IEIs go into this analysis step, regardless of response reciept.
    For A and B, 99.9\% confidence intervals are shown as shaded areas and significant $\beta$ values (at $p < 0.001$) are indicated as solid circles.
    Analyses were performed separately for each age group (3, 6, 9, and 18 months).
    Infant ID was treated as a random effect  (model structure: Current IEI $\sim$ Previous IEI + (1\textbar Infant ID)).
    \textbf{(B)} $\beta$s (y-axis) from linear mixed effects analyses testing IEI-to-IEI correlations  for ChSp and Ad vocalizations are shown for validation datasets: human listener-labeled 5 minute sections with all adult vocalizations included (H: All Adult; dark cyan), human listener-labeled 5 minute sections with only infant-directed adult vocalizations included (H: Child-directed Adult; green), and corresponding 5 minute sections as labeled by LENA (L: 5 min; maroon). 
    Due to limited sample size, validation data were not separated into age-blocks; infant age (in months), infant ID, and infant ID-infant age interaction were treated as random effects (model structure: Current IEI $\sim$ Previous IEI + (1\textbar Infant ID) + (1\textbar Infant Age) + (1\textbar Infant ID:Infant Age)). 
     $\beta$ values for ChSp vocalizations from H: Child-directed Adult and H: All Adult datasets are the same, since the child vocalizations included are the same for both datasets.
    }
    \label{fig:PrevStSiBetas}
\end{figure*}

\section*{Results}\label{sec2}

\subsection*{Successive IEIs are positively correlated} 
 
For daylong LENA-segmented data, previous IEIs and current IEIs were positively correlated for both infant speech-related (ChSp) and adult (Ad) vocalizations, for all infant ages (Fig. \ref{fig:PrevStSiBetas}A): standardized regression coefficients ($\beta$s) ranged from 0.23 to 0.30 with a mean of 0.26 for infants, and from 0.29 to 0.33 with a mean of 0.31 for adults, with $p <$ 0.001 in all cases (see Supplementary Tables \ref{tab:Lday_PrevStepBeta_ChSp}, \ref{tab:Lday_PrevStepBeta_Ad}). 
Mixed effects regressions predicting recording-level $\beta$s\textemdash recording-level correlations between current  and previous IEIs\textemdash as a function of infant age found a significant negative effect of infant age for infants (unstandardized regression coefficient, $b =$ -0.01, $p <$ 0.001; Supplementary Table \ref{tab: PrevStepBeta_AgeEffs}) but not for adults ($b =$ -0.003, $p =$ 0.31; Supplementary Table \ref{tab: PrevStepBeta_AgeEffs}). 
Infant age$^2$ effects were non-significant at the $p < 0.001$ level for both infants and adults. \\

Fig. \ref{fig:PrevStSiBetas}B shows  correlations between previous and current IEIs for the 5-minute excerpts segmented and labeled by human listeners, as well as for the corresponding LENA-labeled data. 
For  human listener-labeled data, there are statistically significant positive correlations for infants ($\beta =$ 0.19, $p <$ 0.001; Supplementary Table \ref{tab:Valdata_PrevStepBeta_Chsp}) and for adults, whether all adult vocalizations ($\beta =$ 0.17, $p <$ 0.001; Supplementary Table \ref{tab:Valdata_PrevStepBeta_Ad}) or only infant-directed adult vocalizations ($\beta =$ 0.12, $p < $ 0.001) are considered.
These results indicate that the positive correlation between successive IEIs is not an artifact of the idiosyncrasies of the LENA labeling system  and is robust to labeling method.\\

On the other hand, validation analyses  suggest that the stronger current IEI $\sim$ previous IEI correlation for adults vs. infants for the daylong LENA data (Fig. \ref{fig:PrevStSiBetas}A) is not robust to labeling method. 
The correlation between current  and previous IEIs is stronger for adults than infants for the LENA-labeled 5-minute excerpts while the human listener-labeled 5-minute excerpts show either no significant difference between infants and adults (when all adult vocalizations are included) or the opposite pattern (when only infant-directed adult vocalizations are included; Fig. \ref{fig:PrevStSiBetas}B).
While these differences could be due to labeling errors or potential differences in how infant- and non-infant-directed adult vocalizations overlap with other sounds, it is also possible that the temporal structure of infant- versus non-infant-directed adult vocalizations is not different.
Investigating this could be a fruitful direction for future research, which may nevertheless be limited by the availability of labeling systems that can classify child-directedness at scale.\\

Age trends for infant and adult current IEI $\sim$ previous IEI correlations were not significant for the 5-minute human-labeled data (both when all adult vocalizations are considered and when only infant-directed adult vocalizations are considered) or the LENA 5-minute data (Supplementary Table \ref{tab: PrevStepBeta_AgeEffs}), underscoring the lack of statistical power to resolve age trends in the validation datasets  due to limited sample size (see Supplementary Section \ref{sec:SegmentNumAndDuration} for sample size details).

\subsection*{Social inputs are associated with shorter IEIs, controlling for previous IEI}

\begin{figure*}[ht!]
    \centering
    \includegraphics[width=17cm]{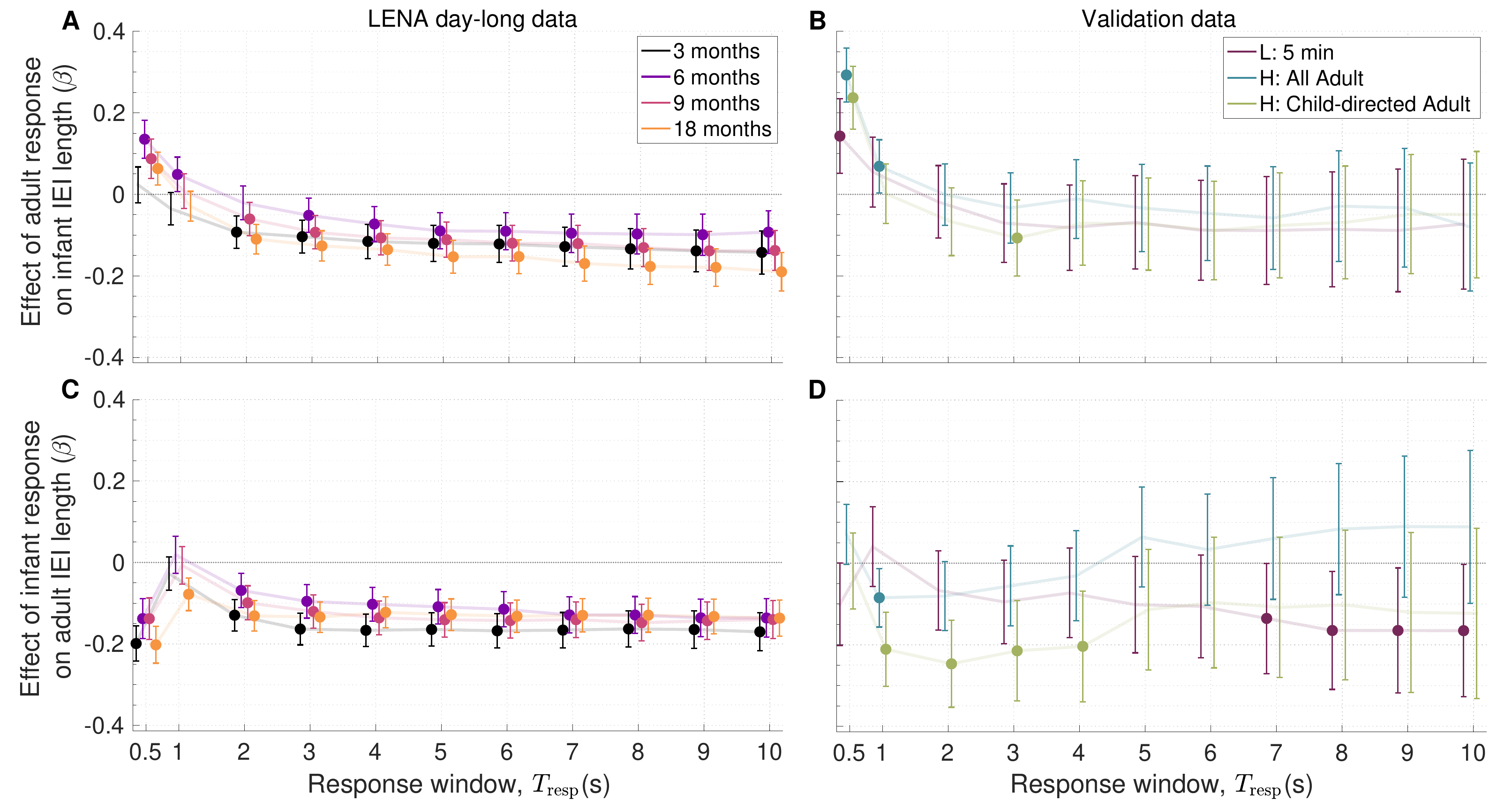}
    \caption{\textbf{Response effects on IEIs after controlling for previous IEI} 
    \textbf{(A)} Standardized regression coefficients ($\beta$; y-axis) from linear models testing the effect of  adult (Ad) response receipt on the length of infant speech-related (ChSp) IEIs\textemdash after controlling for IEI-to-IEI positive correlations (see Fig. \ref{fig:PrevStSiBetas}A)\textemdash are shown as a function of the response window, $T_{\rm resp}$ (seconds; x-axis), for LENA daylong data (model structure: Residual IEI $\sim$ Response Receipt).
    For all panels, 99.9\% confidence intervals are shown as error bars, significant $\beta$ values (at $p < 0.001$) are indicated as solid circles, and $\beta$ values are staggered around the relevant $T_{\rm resp}$ value for easier visualization.
    Statistical analyses were performed separately for each infant age group: 3, 6, 9, and 18 months (indicated by line and circle color; see legend). 
    \textbf{(B)} $\beta$s (y-axis scale shared with panel A) from linear models analogous to A (see Methods and Figs. \ref{fig:Schematic}, \ref{fig:PrevStSiBetas} for details on the current IEI $\sim$ previous IEI control) are shown as a function of the response window, $T_{\rm resp}$ (seconds; x-axis) for the three different 5-minute validation datasets (see legend): human listener-labeled 5 minute sections with all adult vocalizations included (H: All Adult; dark cyan), human listener-labeled 5 minute sections with only infant-directed adult vocalizations included (H: Child-directed Adult; green), and corresponding 5 minute sections as labeled by LENA (L: 5 min; maroon). 
    Validation data were not separated into age blocks due to limited sample size. 
    \textbf{(C)} $\beta$ values (y-axis) from a similar model as in A, testing the effect of infant response on adult IEI length are shown for LENA daylong data. 
    \textbf{(D)} $\beta$ values (y-axis shared with panel C) from a similar model as in B testing the effect of infant response on adult IEI length are shown for validation datasets.  
} \label{fig:RespBetas}
\end{figure*}

Measures of how adult caregiver responses affect subsequent infant IEIs and how infant responses affect subsequent caregiver IEIs can be expected to be confounded by the robust positive correlation between successive IEIs (Fig. \ref{fig:PrevStSiBetas}). 
We thus ran analyses testing whether response receipt predicts the following IEI being shorter than expected, after controlling for this positive correlation.
Linear models testing for differences in IEIs associated with the receipt of a response (Resp IEIs) versus absence of a response (NoResp IEIs) were run on the residuals after predicting the current IEI based on the previous IEI, for response window ($T_{\rm{resp}}$) values ranging from 0.5 to 10 s (Figure \ref{fig:RespBetas}).  
$T_{\rm{resp}}$ sets the window of time to determine response receipt as well as the threshold for the minimum IEI included in response analysis (see Methods and Fig. \ref{fig:Schematic} for details).\\
 
We also ran mixed-effects regressions testing whether recording-level response effects were predicted by infant age and age\textsuperscript{2}, controlling for infant ID as a random effect.
No age effects predicting response $\beta$s were significant at $p < 0.001$ (see the \href{https://osf.io/5xp7z/files/8sjvm}{relevant table}\footnote{\url{https://osf.io/5xp7z/files/8sjvm}} in the associated \href{https://osf.io/5xp7z/overview}
{OSF repository}\footnote{\url{https://osf.io/5xp7z/overview}}).
However, there may be age trends that require larger sample sizes and/or more sensitive measures, such as being able to differentiate adult vocalizations that are infant-directed versus not at the full-day level.
Finally, we ran supplementary analyses to test the effect of incorporating more temporal history into the previous IEI control step on the assessment of response effects. 
These results (see Supplementary Section \ref{subsec: Prev1vsPrev2Iei}) show that  correlations between the immediately preceding IEI and the current IEI is sufficient to control for the temporal structure in vocalization time series when assessing interactive effects.

\subsubsection*{Effect of adult responses on infant IEIs}

For $T_{\rm{resp}} \ge 2$ s, infant IEIs were significantly shorter ($p < 0.001$) after receiving an adult response for LENA daylong data, as evidenced by the negative response $\beta$s in Fig. \ref{fig:RespBetas}A.
This pattern of negative $\beta$s holds for all age groups for the full day LENA-segmented data, except for 6 month-old infants at $T_{\rm resp}=2$ s, in which case the trend was consistent but did not reach significance at $p < 0.001$ (see Supplementary Fig. \ref{fig:RespEffFigs_wAndWoCtrl_Lday_SI_99_9CI} for significant $\beta$ values at less stringent significance thresholds). 
For the representative case of $T_{\rm resp}=5$ s (Supplementary Fig. \ref{fig:InfIEI_WrWorResidsByAge}), $\beta$s ranged from -0.15 to -0.09 across all age groups (see Supplementary Table \ref{tab:Daylong_ChSpIEI_5s_RespBeta_Lday} as well as the \href{https://osf.io/5xp7z/files/qzurx}{relevant OSF table}\footnote{\url{https://osf.io/5xp7z/files/qzurx}}).\\

For validation datasets  (Fig. \ref{fig:RespBetas}B), the negative $\beta$ was significant only for the human-labeled dataset with only infant-directed adult vocalizations included at  $T_{\rm resp}=3$ s.
All other $\beta$s trended in a direction consistent with the negative $\beta$s found for the LENA daylong  data for $T_{\rm{resp}} \ge 2$ s but were not statistically significant at the $p < 0.001$ threshold (see Supplementary Fig. \ref{fig:RespEffFigs_wandwoCtrl_Valdata_99_9CI} for significant $\beta$ values at less stringent significance thresholds and Supplementary Table \ref{tab:Valdata_ChSpIEI_5s_RespBeta} for $\beta$ values for the representative case of $T_{\rm resp} = 5$ s). 
The corresponding LENA-labeled 5-minute portions also had non-significant (at $p < 0.001$) negative response $\beta$s for $T_{\rm{resp}} \ge 2$~s, suggesting that the non-significance may primarily be an issue of insufficient statistical power with these much smaller datasets.\\

For $T_{\rm resp} < 2$ s, infant IEIs were longer after an adult response for LENA daylong data, as evidenced by positive $\beta$ values for all age groups.
This trend was significant at $p < 0.001$ at $T_{\rm resp} = 0.5$ s for infants at 6, 9, and 18 months, and at $T_{\rm resp} = 1$ s for 6 month old infants.
The tendency for adult vocalizations to be longer than most other sound types (see Supplementary Section \ref{sec:SegmentNumAndDuration}) may contribute to positive $\beta$ values at $T_{\rm resp} < 2$ s, which correspond to cases where the shortest IEIs in the data are included in response analyses (see Supplementary Section \ref{subsec:RespEffectResults} for a detailed discussion, and  Supplementary Fig. \ref{fig:NumResponses_Ldaylong} for a summary of the proportion of IEIs excluded from response analyses as a function of $T_{\rm resp}$). %Note: Re: S29, for larger response windows, about 50-60 percent IEIs are excluded and we still get robust results matching the negative beta trend and still statistically significant  
This pattern of positive $\beta$s for $T_{\rm resp} < 2$ s also holds for the validation datasets, with significant $\beta$s at $T_{\rm resp} = 0.5$ s for all three validation datasets and at $T_{\rm resp} = 1$ s for the human-labeled 5-minute excerpts with all adult vocalizations included.
The conserved patterns of response effects at short response windows for the daylong and validation datasets suggest that these patterns are likely not an artifact of minimum vocalization durations as set by LENA.\\

Thus, as long as the response window parameter, $T_{\rm{resp}}$, is sufficiently long, results are consistent with a tendency for adult vocal responses to promote infant vocalization, causing infants to vocalize again sooner than would be expected based on the infant's previous IEI.

\subsubsection*{Effect of infant response on adult IEIs}

Adult IEIs were shorter following an infant response for $T_{\rm{resp}} \ge 2$ s. 
This effect is significant for all infant age groups for the LENA daylong data (Fig. \ref{fig:RespBetas}C). 
$\beta$s ranged from -0.16 to -0.11 across all infant age groups for the representative case of $T_{\rm resp}=5$ s (see Supplementary Table \ref{tab:Daylong_AdIEI_5s_RespBeta_Lday}).\\

For the human-labeled 5-minute datasets (Fig. \ref{fig:RespBetas}D), the response effect is significant for 1 s $\le T_{\rm{resp}} \le 4$ s when only infant-directed adult vocalizations are included in the analysis, with the significant effect sizes being stronger than the effect sizes for the human-labeled 5-minute datasets with all adult vocalizations included, indicating that infant responses affect infant-directed adult vocalizations more than non-infant-directed adult vocalizations.
For human-labeled 5-minute portions with all adult vocalizations included and for LENA labels for those same 5-minute portions, response effects are significant for a more limited range of $T_{\rm{resp}}$ values.\\

For very short $T_{\rm resp}$ values ($\leq 2$ s), response effects for the LENA daylong  data are non-monotonic as a function of $T_{\rm resp}$: $\beta$s for $T_{\rm resp} = 0.5$ s are negative and statistically significant at $p < 0.001$ for all age groups while $\beta$s for $T_{\rm resp} = 1$ s are greater than those at $T_{\rm resp} = 0.5$ s but only significant at 18 months (Fig. \ref{fig:RespBetas}C). 
This non-monotonicity is not mirrored in the human-labeled validation data. 
Therefore, it is possible that the pattern of increase followed by decrease observed for LENA daylong response $\beta$s at very short response windows is an artifact resulting from a combination of LENA’s minimum duration thresholds for different sound types and differences in typical vocalization durations (see Supplementary Section \ref{subsec:RespEffectResults} for a detailed discussion).\\

%Note: we only report the modal duration for CHNSP, but I just confirmed that this is also true for CHNNSP
For the research questions that are the focus of this study, the results are most stable and interpretable for $T_{\rm resp} \ge 2$ s. 
When this is the case, for adult IEIs, similar to infant IEIs, results are consistent with a tendency for infant vocal responses to promote adult vocalization, causing adults to vocalize again sooner than they otherwise would have.

\section*{Discussion}

The modification of foraging trajectories in response to resource acquisition is a key feature of adaptive foraging processes such as area-restricted search (ARS).
The adaptive hallmark of ARS is a switch from exploratory search (longer steps and IEIs) to exploitative search (shorter steps and IEIs) in response to 
resource acquisition. 
This trade-off between exploitation and exploration is observed in adaptive foraging processes in physical space and in cognitive domains where search processes aren't anchored to an explicit physical space \cite{pirolli1999information,viswanathan2011physics, rhodes2007human, dorfman2022guide, hills2015exploration}. 
In the context of framing infant-caregiver vocal interactions as interpersonal foraging, we posit that resources being sought for include social vocal engagement \cite{ritwika2020exploratory}.
However, prior research  has not resolved whether adult social responses causally increase the likelihood of subsequent infant vocalization or whether adult responses are merely embedded within  vocalization clusters  in infants' spontaneous vocal production, without infants actively modifying their behavior due to receiving adult input \cite{pretzer2019infant}.\\

To address this confound, we analyzed infant and adult IEIs, informed by the analysis of step sizes and flight times in foraging studies.
We first established that successive infant and adult IEIs are indeed positively correlated, for the full-day automatically labeled dataset and the validation datasets.
We also found that successive infant IEIs were less correlated as infant age increased for the LENA daylong data, indicating increased baseline exploratory vocal practice by infants as they get older (see also Supplementary Fig. \ref{fig:IEIDist_ChnspLday}B).
%Note: the infant IEI distributions for different ages also seem to provide support for this, with longer and heavier tails ccdf tails as infants get older (no stats done)
This trend, however, was not statistically significant for the human-labeled validation data and would require more validation data for verification.\\

Correlation between successive IEIs as demonstrated here serves as a metric for assessing clustering that complements other measures of temporal clustering in infant vocalizations \cite{abney2017multiple,yoo2024infant} as well as other behavioral domains \cite{goh2008burstiness} such as language production (e.g., \citenum{altmann2009beyond}), 
infant play behavior (e.g., \citenum{karmazyn-raz2022sampling}),
and animals' foraging in space (e.g., \citenum{viswanathan2011physics}).
This underscores the broad potential utility of correlations between consecutive IEIs as a measure of event-level clustering in assessing the temporal structure of event time series.
This successive IEI correlation metric is advantageous in studying interactive behavioral time series (such as infant-caregiver vocal interactions) since it provides a method to control for temporal clustering when looking for immediate effects of social engagement.\\

Analyses testing for the effects of social vocal input on the focal agent's vocalization behavior after controlling for the successive IEI correlation revealed that infants vocalized sooner following an adult response across all age groups in the daylong automatically labeled data for response window values ($ T_{\rm{resp}}) \geq 3$ s.
Similarly, adult IEIs were significantly shorter following an infant response across infant ages for the LENA daylong data for $T_{\rm{resp}} \geq 2$ s. 
Results from  the validation datasets supported these findings, although not all validation results reached statistical significance. 
These results align with the observation that foraging agents switch from exploratory search (longer steps and IEIs) to exploitative search (shorter steps and IEIs) following resource acquisition in adaptive foraging contexts \cite{dorfman2022guide, viswanathan2011physics}. 
Thus, within the interpersonal foraging framework, shorter IEIs  associated with response receipt suggests that both infants and adults are switching from exploratory to exploitative behaviors following social vocal interactions that follow their own vocal productions. 
This may reflect infants’ and adults’ tendency to engage more frequently and exploit a ``patch'' of interactivity\textemdash much like animals foraging for food exploit a recently discovered patch of food\textemdash when the other vocal agent is likely to be in a vocally interactive state.\\

This pattern of shorter IEIs following social input breaks down when the $T_{\rm{resp}}$ parameter is very small ($\leq$ 1 s). 
This may be due to the fact that response analysis for very short response windows allows for the inclusion of very short IEIs, the length of which may be dictated primarily by the duration of the intervening sound type.
Using larger $T_{\rm{resp}}$ values also provides increased flexibility in capturing responses that are delayed due to competing demands on caregivers' attention, or that are effortful and take time to activate and execute, especially for infants at early stages of speech production \cite{hilbrink2015early, casillas2016turn-taking}. 
Further, since our method does not distinguish between effects specific to contingent responses versus more general stimulation due to social input, larger $T_{\rm{resp}}$ values include more of the latter type of events, resulting in increased sample size for IEIs associated with a response and consequently, statistical power. 
On the other hand, the longer the response window, the more IEIs are excluded from analysis (see Supplementary Section \ref{subsec:ResponseAnalysis}). 
Performing a sweep across a range of $T_{\rm{resp}}$ values as we did in the present study allows for empirically choosing suitable values of $T_{\rm{resp}}$ to achieve balance across these considerations.
Specifically, the inclusion of $T_{\rm{resp}}$ values from 1 s to 5 s is motivated by the use of these values in prior literature on infant-caregiver vocal interactions (e.g., \citenum{pretzer2019infant}), while the full sweep (0.5 s to 10 s) is motivated by the expectation that this would provide comprehensive information about the sensitivity of the response analyses to different values of the $T_{\rm{resp}}$ parameter.\\

There are several analytical approaches (e.g., simultaneous regression approaches, traditional autoregressive models) that may be considered as alternatives to our two-step analysis framework. 
A discussion on how these alternative approaches compare to the current analysis framework is provided in Supplementary Section \ref{sec: AltAnalysis}.
Another methodological dimension for future exploration is the automatic labeling. 
Some of the LENA algorithm's properties\textemdash such as minimum duration thresholds for various vocalization types and how overlapping sounds are treated\textemdash affect IEI lengths as well as the assessment of response receipt in this study, and may affect the results. 
Inaccuracies in the labels are also present (see Supplementary Section \ref{sec:Reliability} for reliability metrics). 
While results from the human listener-labeled dataset validate our main findings when the minimum duration requirement is relaxed and for a labeling system that can be expected to be more accurate, reanalyzing the daylong data based on sound segmentation and labeling using different labeling systems (e.g., {BabyHuBERT-VTC} \cite{charlot2025babyhubert}) would be informative.\\

Taken together, our findings support the framing of infant-caregiver vocal communication as an interpersonal foraging process, indicating that social engagement from the other agent promotes subsequent vocal activity by both infants and adults in real-world contexts experienced over the course of a day.
Within the interpersonal foraging framework, our results are consistent with area-restricted search, showing that the time elapsed between foraging stops tends to be shorter when the foraging agent is retrieving resources at a high rate and longer when resource obtainment slows \cite{hills2012optimal,dorfman2022guide}.
Further, these results complement the social feedback loop framework of studying infant-caregiver vocal interactions, thereby establishing a theoretical link between social reinforcement learning and foraging-inspired perspectives of studying this behavioral domain.
Thus, these perspectives are not antagonistic to each other, but rather, provide an expanded methodological and theoretical framework to study infant vocal development as the emergent outcome of a complex, adaptive, interactive behavioral system \cite{slone2023temporal}.
In particular, the two-step analysis protocol we propose is directly inspired by the tradition of analyzing step sizes and/or inter-event intervals in foraging studies \cite{viswanathan2011physics}.
The demonstration that infant and caregiver vocal behavior can be analyzed from a foraging perspective and that vocal IEIs are shorter during periods of social engagement is consistent with the framing of social vocal engagement as a reward, and provides a foundation for future work to  test predictions  for this system in line with other findings in the foraging literature.
For instance, IEI distributions may differ as a function of resource sparsity, and IEI distributions for infants whose social environments provide fewer rewards overall may have heavier tails \cite{viswanathan2011physics}.
Similarly, in line with marginal value theorem, the vocalization-promoting effects of social engagement may be expected to be stronger after longer periods of disengagement from the social partner \cite{kolling2017reinforcement}.\\

We also note that burstiness is evident in diverse behavioral systems and that the discussion of the causes and functions of burstiness is broad and not restricted to foraging.
For example, bursty sequences of nonlinguistic actions \cite{karmazyn2023discourse} and referential  words \cite{slone2023temporal} during dyadic, object-directed play may support infants learning object names by constraining and organizing domain-general attentional and memory processes.
Reciprocally, it has been suggested that these attentional and memory processes may contribute to the temporal clustering observed in behavioral event series within the broader complex, multimodal interactive system that is the infant-caregiver dyad \cite{slone2023temporal}. 
It has also been shown that coordinated temporal clustering of infant and caregiver vocalizations is dynamically coupled to infants' and adults' physiological arousal \cite{smith2023vocalization}.
Across these various aspects of behavior, methodological approaches, and conceptual frameworks, there is a common theme of viewing the infant-caregiver dyad as a complex interacting system, where complexity is a driver and consequence of bursty behavioral patterns.
These and related works provide insights regarding how clustered behavior may be generated by complex adaptive processes that include physiological arousal as a contributing factor. 
The interpersonal foraging framework is compatible with this complex dynamical systems perspective in that it suggests a mechanism by which social interaction may additionally contribute (for example, by activating reward-response networks), and relates the pattern of lulls and bursts in behavior to shifts between exploratory and exploitative modes. 
We see these different theories as complementary, and propose that future work elaborating on their relationships, perhaps from an explanatory pluralism perspective \cite{dale2009explanatory}, could be fruitful.\\

The pattern of responses being associated with shorter IEIs could reflect a process of learning to vocalise to elicit responses or a general tendency for social vocal input to stimulate infant (and adult) vocal activity. 
This has frequently been noted as a limitation of experimental studies manipulating adult responses to infant vocalizations \cite{rheingold1959social,weisberg1963social,bloom1984distinguishing,goldstein2008social}.
Indeed, at least one experiment has found that a general vocal elicitation effect of adult input influenced infant volubility more than temporal contingency.
This ambiguity also applies to infant response effects on adult IEIs. 
Applying our analysis approach to experimental or simulation data could help assess and refine the approach with respect to this distinction.
In a similar vein, while our results provide strong support for the bi-directional excitatory effects of social vocal responses on infant and adult vocalization streams, we do not investigate higher-order effects arising from potential cross-dependency of infant and adult vocal time series. 
Future research focused on the coupled dynamics of the infant-caregiver dyad could provide insights into such higher-order dependencies.\\

Even as we demonstrate that our choice of controlling for the previous IEI is robust in terms of the effects of historical temporal structure beyond the immediately preceding IEI, the fact remains that the hierarchical clustering of infant and caregiver vocalizations is likely to create rich temporal structure with long-range correlations.
While the effect(s) of long-term history on the current IEI may be small (see Supplementary Fig. \ref{fig:Prev1vs2IeiFig}), determining the precise range and structure of resultant correlations may enhance the prediction of current IEI based on prior behavior by facilitating the incorporation of longer-term temporal effects, which may in turn amplify the true signal when testing for the effects of social input.
The current approach also does not address the possibility of vocal extinction bursts\textemdash a phenomenon observed in the still face experimental paradigm where infants exhibit a sequence of increased then decreased vocal activity following cessation of engagement by an adult \cite{goldstein2009value,franklin2014effects,elmlinger2020learning}. 
How extinction manifests in infants' real-world experiences and how to identify instances of extinction bursts in real-world interactions remain open questions.
We also note that while the focus of this study is on the event-to-event temporal dynamics\textemdash and therefore, the quantity\textemdash of vocalizations, infant-caregiver vocal interactions can be described along multiple dimensions.
Previous work has taken a foraging-inspired approach to analyse not only temporal features (i.e., IEIs), but also how individuals move about in acoustic space, thus beginning to address features of the quality of vocalizations in addition to their quantity \cite{ritwika2020exploratory}. 
However, we note that this prior work does not incorporate controls for the inherent burstiness of infant and caregiver vocal behaviors, i.e., the tendency for vocalizations occurring close together to be acoustically similar. 
Adapting the current analysis approach to aspects related to the quality of vocalizations, such as vocalization types (e.g., canonical babble, non-canonical babble, and meaningful words) and acoustic features, could also be a promising future direction.\\

While the current study focuses on infant-caregiver vocal interactions, the foraging framework can be flexibly applied to observations of infant talker bias, where infants exhibit a preference for (self or other) infant sounds \cite{polka2025infants}, as well as infants' strong endogenous motivations for vocal exploration, where large numbers of diverse protophones occur in non-social contexts \cite{long2020social, griebel2024emotional, long2022perspectives}.
Here, non-social protophones or more generally, infant vocal productions, may be conceptualised as rewards \cite{warlaumont2020infant} and hypothesised to be associated with shorter IEIs, providing utility in assessing endogenous reward circuits. 
Further, Fitness Signaling Theory \cite{griebel2024emotional, long2022perspectives} posits that caregivers are attuned to infants’ volubility and variety of sound types, implying that caregiver vocal engagement may be coupled to infant vocalization rate, salience, and/or variety. 
The interpersonal foraging framework could be adapted to test these hypotheses. 
Extending this, incorporating differential patterns of adult and infant engagement as a function of different vocalization types produced by the other agent \cite{warlaumont2014social,zhang2024developmental, reigado2011vocalizations, werker1994cross} is another avenue that would increase the explanatory power and theoretical value of the interpersonal foraging framework.
It would also be informative to run analyses separately for different individual interaction partners and to compare interpersonal foraging dynamics across individuals and/or demographic groups.\\

In the future, this approach could be used to compare child-caregiver vocal interactions across diverse cultural and socioeconomic contexts \cite{bergelson2023everyday}.
Moreover, a number of animal communication systems share the burstiness of communicative signals and the importance of social vocal engagement in infant vocal development \cite{gernat2018automated, burchardt2020comparison, price1979developmental, takahashi2015developmental, tomasello2008origins}.
As such, the use of correlations between successive IEIs to control for temporal structure in behavioral event series as well as the interpersonal foraging framework proposed in this study could be applied to understand interactive behaviors in other animal species \cite{abreu2022turn-taking,wegdell2025evolution}, such as marmosets, which also exhibit infant-caregiver vocal turn-taking \cite{takahashi2016early}, as well as in other communication modalities \cite{rohlfing2020multimodal,kosie2024infant-directed}, such as gestures in humans \cite{iverson2005gesture,burkhardt-reed2021origin,horton2022acquisition} and great apes \cite{tomasello2008origins}.\\

Finally, interpersonal foraging can be compared to other types of social foraging (e.g., cooperative or competitive foraging) \cite{santos2009can,martinez-garcia2013optimizing,golnaraghi2023optimal}. 
In both cooperative and competitive foraging, the occurrence of the foraged-for resource is independent of the foraging agents, even as other agents' behavior informs foraging strategy. 
In contrast, resources in interpersonal foraging are generated by the agents' foraging behaviors.
This is similar to some cases of foraging for mates in that the agents are searching for each other and interact non-antagonistically.
The diversity of social foraging processes notwithstanding, dynamic interactions between agents is a common thread shared by social foraging and interpersonal foraging as conceptualized here.  
As such, it could be fruitful to adapt computational models of social foraging to interpersonal foraging and assess implications for optimal foraging strategies.\\

In summary, the current study provides robust evidence for the facilitatory effects of social engagement on both infant and adult vocalizations in ecologically relevant contexts using a large (2,400+ hours of audio), longitudinal corpus of daylong, naturalistic audio data after controlling for the inherent burstiness of vocal behavior.
We validate our results using a human listener-labeled subset of the automatically labeled data, demonstrating that this pattern is robust to  labeling method.
These findings support the view that infant vocalization is an interpersonal foraging process.

\begin{center}
\noindent
--------------------------------------
\end{center}

\noindent
\textbf{Acknowledgments.} 
We thank the families who participated, the many research assistants who helped with data collection and annotation, and Gina Pretzer and Jimel Mutrie for coordinating, supervising, and performing data collection and annotation. 
This work was supported by the National Science Foundation (grant numbers 1529127 and 1539129/1827744) and by the James S. McDonnell Foundation (\url{https://doi.org/10.37717/220020507}).\\

\noindent
\textbf{Ethics information.}
All data for this study was collected in accordance with relevant guidelines and regulations. 
Informed consent was obtained from the infant participants’ legal guardians. 
Data collection protocols were approved by the University of California, Merced Institutional Review Board (protocol ID: UCM15-0031) and continuing data analysis was approved by the University of California, Los Angeles Institutional Review Board (protocol IDs: IRB-23-1317 and IRB\#23-001317).\\

\noindent
\textbf{Statement of Authorship.}
ASW, VPSR, CTK, and AG designed the research; SS coordinated, SS and LL performed, and ASW and CTK supervised data collection; JM performed, and SS, LL, and ASW supervised data annotation; VPSR and JM cleaned and pre-processed data; VPSR performed statistical analyses; VPSR and ASW wrote the paper with input from CTK, AG, LL, and SS.\\

\noindent
\textbf{Data and Code availability.}
All processing and analysis code is available on \href{https://github.com/Ritwikavps/BurstinessAndInterpersonalForagingPaper}{Github}. De-identified preprocessed data used in the present paper as well as all relevant results tables are available on \href{https://osf.io/5xp7z/}{OSF}. 
The raw audio data and LENA labels for participants who gave permission for us to share their data with other researchers are available within the San Joaquin Valley corpus \cite{warlaumont2024san} in HomeBank \cite{vandam2016homebank}.\\

\noindent
\textbf{Conflict of interest.}
The authors declare no conflict of interest.\\

\noindent
\textbf{AI declaration.}
No AI technologies were used in the creation of this manuscript.

%%%%%%%%%% Supplemental materials %%%%%%%%%%
\pagebreak
%\widetext
\begin{LARGE}
\noindent \textbf{Supplementary Information\\} 
\end{LARGE}

\setcounter{equation}{0}
\setcounter{figure}{0}
\setcounter{table}{0}
\setcounter{section}{0}
\setcounter{page}{1}
\makeatletter
\renewcommand{\theequation}{S\arabic{equation}}
\renewcommand{\thesection}{S\arabic{section}}
\renewcommand{\thefigure}{S\arabic{figure}}
\renewcommand{\thetable}{S\arabic{table}}
%\renewcommand{\bibnumfmt}[1]{[S#1]}
%\renewcommand{\citenumfont}[1]{S#1}
%%%%%%%%%% Prefix a "S" to all equations, figures, tables and reset the counter %%%%%%%%%%

\vspace{-4 mm}

%---SECTION: DataAbbrevs--------------------
\section{Abbreviations for the datasets}\label{sec:DataAbbrevs}
\begin{itemize}
\item \textbf{L (day)}: LENA daylong data consisting of daylong recordings collected, automatically segmented, and labeled using the LENA{\texttrademark} system. 
\item \textbf{H (5min)} and \textbf{H (5min; All Ad)}: Validation data consisting of human listener-annotated 5-minute sections of L (day) data, with all adult vocalizations included.
Note that H (5min) and H (5min; All Ad) abbreviations are used interchangeably.
In the main text, this dataset is referred to as \textbf{H: All Adult}.
\item \textbf{H (5min; T-Ad)}: Validation data consisting of human listener-annotated 5-minute sections of L (day) data, with only infant-directed adult vocalizations included.
The `T' refers to the annotation label for infant-directed adult vocalizations, specifying the adult vocalization as addressed `to' the infant wearing the recorder.
In the main text, this dataset is referred to as \textbf{H: Child-directed Adult}.
\item \textbf{L (5min)}: LENA-labeled counterpart of the human listener-labeled validation data, consisting of LENA segment onsets, offsets, and labels for the 5-minute sections annotated by human listeners.
In the main text, this dataset is referred to as \textbf{L: 5 min}.
\end{itemize}

%---SECTION: SuppMethods--------------------
\section{Extended methods}\label{sec:SuppMethods}

%---SUBSECTION: DataCollec--------------------
\subsection{Data collection}\label{subsec:DataCollec}

Data used consisted of daylong (10+ hour) naturalistic recordings (recorded on a typical day in the infant's home environment) collected longitudinally from 54 infants aged 3, 6, 9, and 18 months using the LENA\textsuperscript{TM} Pro system. 
45 infants had recordings at all four ages (see Section \ref{sec:DataOverview_FileNumAndAge} for details).
Infants resided in the San Joaquin Valley region of California, USA, and were exposed primarily to English and/or Spanish, although other languages were also spoken in some of the infants' homes.
Socioeconomic status (SES) varied: seven households reported annual income less than \$30,000, seven \$30,000-60,000, six \$60,000-90,000, and nine over \$90,000.
For reference, in 2015 (the year that the data collection started), the median household income in the United States was \$56,516 \cite{proctor2016income}.
The highest level of completed education within the families also varied: 2 reported some high school, 3 high school or GED, 10 some college, 3 associate’s degree, 18 bachelor’s degree, 11 master’s degree, and 7 doctoral degree. 
Infants' reported ethnicity/race were: 17 White, 13 Hispanic/Latino (race not specified), 13 Hispanic/Latino - White, 2 Hispanic/Latino - Native American, 2 Black/African American, 1 Asian, 4 Asian and White, and 2 Unknown.
Caregivers were instructed to turn recorders on and put them on the infants by 8 am and to finish recording no earlier than 7 pm. 
Caregivers were allowed to pause recording during the day for privacy, but total pause time was not to exceed 1 hour per recording day. 
Caregivers were also able to request that sections of the recording be deleted for privacy purposes. 
Monetary compensation was provided for the completion of each recording and associated questionnaires.\\

This is an expanded version of the dataset in a previous study \cite{ritwika2020exploratory_SI}, which can be referred to for further details regarding data collection methods.

%---SUBSECTION: VocIding--------------------
\subsection{Identification of infant and adult vocalizations}\label{subsec:VocIding}

Audio recordings were processed by the LENA Pro software, which provided vocalization onset and offset times and segment labels. 
Our analyses focus on segments labeled `CHN' (clear vocalizations likely produced by the infant wearing the recorder) containing at least one `Utt' (a speech-related vocalization, as opposed to cry, laugh, or vegetative vocalizations), referred to here as `ChSp' (\textbf{Ch}ild \textbf{Sp}eech-related); `FAN' (clear adult female vocalizations); and `MAN' (clear adult male vocalizations).
FAN and MAN vocalizations were grouped into an `Ad' (\textbf{Ad}ult) category for our analyses and collectively represent all adults in the infant's vicinity over the course of the day \cite{gilkerson2020guide_SI}.
We use `ChNsp' (short for \textbf{Ch}ild \textbf{N}on-\textbf{sp}eech-related) to refer to CHN segments that do not contain an Utt and thus are likely only cry, laugh, or vegetative.
ChNsp segments were not analyzed but were used to determine which inter-event intervals (IEIs) were eligible for analysis (see Section \ref{subsec:RespVsNoRespIeis}).
Unless otherwise specified, `infant vocalizations' refers to ChSp vocalizations in the context of this study. 
For ChSp, ChNsp, and Ad vocalizations, all instances of more than one vocalization of the same category separated by 0 s were merged to create a single vocalization (see Section \ref{subec:0IEIMerge}).

%---SUBSECTION: RespVsNoRespIeis--------------------
\subsection{Defining response (Resp) and no-response (NoResp) inter-event intervals (IEIs)}\label{subsec:RespVsNoRespIeis}

%Ritwika: The 0.5 response window is aimed to look at the case where the least data was excluded as NA and wanted to see what that looks like (cuz of the LENA IEI thresholds at 0.6 s cuz min segment length is generally 0.6 s). Note that because CHNNSP also triggers NA responses for the infant, there may still be some excluded IEIs for infants, cuz there may be a CHNNSP voc triggering an NA response. 
Infant IEIs were defined as the offset-to-onset time for a pair of sequential ChSp vocalizations, and similarly for adult IEIs (see Fig. \ref{fig:Schematic}A in the main text).  
Recorder pauses and deletions of portions of recorded audio created multiple sub-recordings. 
We did not compute IEIs that spanned sub-recordings.\\

ChSp and Ad IEIs were categorized as associated with a response (Resp IEIs) or not associated with a response (NoResp IEIs) for response window values ($T_{\rm resp}$) from 0.5 s to 10 s (with 1 s increments between 1 s and 10 s) based on the response computation scheme described in Fig. \ref{fig:Schematic} in the main text.
ChSp IEIs where the first vocalization offset was followed by a ChNsp or ChSp onset within time $T_{\rm resp}$ without an intervening Ad vocalization were categorized as IEIs associated with NA responses and excluded from response analyses, as were Ad IEIs $\leq T_{\rm resp}$ without an intervening ChSp vocalization.
Resp IEIs $\leq T_{\rm resp}$ were excluded from response analyses to avoid the confound of Resp IEIs being shorter due to not needing to wait the full $T_{\rm resp}$ time period to determine response receipt. 

%---SUBSECTION: StatTests--------------------
\subsection{Statistical tests relating IEI length to response receipt}\label{subsec:StatTests}

Statistical analyses assessing response effects were performed separately for 3, 6, 9, and 18 months.
All IEIs were log-transformed (base 10) and then normalized prior to subsequent analysis steps. 
Normalization was performed with respect to the age-block subset (e.g., infant IEIs at 3 months) being analyzed and did not take response receipt into account.\\

To test the effect of adult response receipt on infant IEIs, we used a two-step approach designed to control for the endogenous clustering of infant vocalizations in time (Fig. \ref{fig:Schematic}B in the main text as well as Figs. \ref{fig:ExtendedSchematics_Pt1}, \ref{fig:ExtendedSchematics_Pt2}).
Separate tests were conducted for each $T_{\rm resp}$ value.
First, a linear mixed effects model predicted current infant IEI with the previous infant IEI as a fixed effect and infant ID as a random effect (model structure: Current IEI $\sim$ Previous IEI + (1\textbar Infant ID)).
All computed IEIs were used in this analysis step.
Next, normalized residuals from this model were used as the dependent variable in a linear model that had a binary fixed effect representing whether the residual was for a Resp or a NoResp IEI (model structure: Residual IEI $\sim$ Response Receipt).
The binary response receipt variable was not normalized.
All IEIs $\leq T_{\rm resp}$ were excluded from this analysis step.
To test the effect of infant age on the association between response receipt and IEI residuals, the above protocol was applied for each recording day, with the only difference being that a linear model (as opposed to a linear mixed effects model) was used to regress current IEIs against previous IEIs, since infant ID is the same for all IEIs within a recording day.
This was  followed by a linear mixed effects model predicting recording day-level response effect $\beta$s, with infant age and age\textsuperscript{2} as fixed effects and infant ID as a random effect (model structure: Response Effect $\sim$ Infant Age + Infant Age\textsuperscript{2} + (1\textbar Infant ID)).
Infant age (in months) was not normalized for the age effects analyses.
Analogous analyses were done for adult IEIs and infant responses.\\

Because of the large number of statistical tests performed, we opted to only interpret effects with $p < 0.001$ as statistically significant.

%---SUBSECTION: ValDataMethods--------------------
\subsection{Validation using 5-minute excerpts labeled by human listeners}\label{subsec:ValDataMethods}

To validate the results based on automatically labeled LENA daylong data, we used a subset of the LENA daylong data from 45 infants, in which 420 5-minute sections from 148 daylong recordings (up to three 5-minute sections per recording) were independently segmented and labeled by trained human listeners (see Section \ref{sec:DataOverview_FileNumAndAge} for details).
These 5-minute sections were selected on the basis of having high numbers of child speech-related vocalizations (i.e., high child volubility) as suggested by the LENA software and representing different portions of the day. 
Section selection involved a criterion of the sections being 30 minutes apart, although some 5-minute sections were less than 30 minutes apart due to human error (see Section \ref{sec:DataOverview_FileNumAndAge} for details).
Research assistants used ELAN \cite{elan2024} to label the onsets and offsets of all target infant and adult vocalizations and to identify major sub-types of each:
canonical (C), non-canonical non-reflexive (X), cry-reflexive (R), and laugh-reflexive (L), for infant vocalizations; directed to the target child (T), not to the target child (N), or addressee unknown (U) for adult vocalizations.
The C and X sub-types together form an equivalent class to the CHN Utt label by LENA and therefore, our ChSp category, while the R and L sub-types together correspond to our ChNsp category.
%The R and L sub-types would be included within the CHN CryVfX label by LENA, corresponding to our ChNsp category.
Custom R and MATLAB scripts were written to check for potential annotation errors (e.g., annotations in the wrong tier, mistyped or missing annotations, or annotations with no (or negative) duration). 
All identified errors that could not be corrected automatically were corrected by one of the authors (JM), who was also one of the original trained human listeners.
We also identified two cases where the annotation file was linked to the incorrect audio file (see the \href{https://osf.io/5xp7z/wiki}{wiki}\footnote{\url{https://osf.io/5xp7z/wiki}} for the \href{https://osf.io/5xp7z/}{OSF repository}\footnote{\url{https://osf.io/5xp7z/}} associated with this study for further details).
Files with affected annotations are not included in the set of 420 5-minute sections analyzed in this study.\\

Unlike the LENA system's labels, infant and adult vocalizations identified by human listeners could overlap.
For consistency with the LENA system, these overlaps were excluded from further analyses (see Section \ref{sec:OverlapHumanData} for details). 
However, since human listeners only tagged adult and infant vocalizations, only infant-adult overlaps were excluded while (unlike the LENA system) cases of overlaps of either adults or the target infant with other children, electronic sounds, or environmental noise were not excluded.
In addition, unlike LENA labels\textemdash which have a minimum duration of 600 ms for infant vocalizations and a minimum duration of 1000 ms for adult vocalizations \cite{gilkerson2020guide_SI}\textemdash the human listener labels did not have prescribed duration minima  (see Sections \ref{subsec:LenaMeanSegDurAndNums}, \ref{subsec:ValDataMeanSegDurAndNums}, and \ref{subsec:DurDistribution}).
Nevertheless, we retained the human listener-tagged onsets and offsets\textemdash without thresholding vocalizations to match LENA's set minimum durations\textemdash in our analyses to enable assessment of how idiosyncrasies associated with LENA labeling (e.g., the presence of minimum durations) may bias analyses results.\\

To test the validity of the LENA-based analyses, we considered three different sets of labels for these 5-minute sections: (1) the automatically generated LENA labels for these sections (L (5min)), (2) the set of human-tagged ChSp and ChNsp vocalizations together with the set of all human-tagged adult vocalizations (H (5min; All Ad)), and (3) the set of human-tagged ChSp and ChNsp vocalizations together with the set of human-tagged infant-directed adult vocalizations (H (5min; T-Ad)).
Together, these three sets of labels allowed us to identify consistencies and differences in results across labeling methods, controlling for the portions of the day sampled.\\

IEIs from the three validation datasets (L (5min), H (5min; All Ad), and H (5min; T-Ad)) were analyzed following the same procedures for the full-day LENA-annotated dataset, except that data were not separated by infant age, and infant age (in months) was considered as a random rather than a fixed effect and was normalized prior to these analyses.
Infant ID and infant ID-infant age interaction were used as additional random effects (model structure: Current IEI $\sim$ Previous IEI + (1\textbar Infant ID) + (1\textbar Infant Age) + (1\textbar Infant ID:Infant Age)). 
Additional overall reliability metrics for the human versus LENA labeling are provided in Section \ref{sec:Reliability}.

%---SECTION: DataOverview_FileNumAndAge--------------------
\section[DataOverview]{An overview of the data: number of files\footnote{Note that the terms `file' and `recording' are used interchangeably in this document.} and contributing infant ages}\label{sec:DataOverview_FileNumAndAge}

The daylong data used in this study consists of 200 daylong recordings from 54 infants. Recordings were collected at ages 3, 6, 9, and 18 months; 45 infants have recordings at all four ages.\\
%Of these 45 infants, one infant (infant 583) had 4 recordings in the corpus at ages other than 3, 6, 9, and 18 months.
%These recordings were excluded from this study and did not contribute to the total of 200 recordings reported here. 

The validation data consists of 420 5-minute sections from 148 daylong recordings from 45 infants.
30 infants have validation data at all four ages. 
Up to three 5-minute sections were selected for annotation from each daylong recording based on the LENA system identifying them as having high child volubility.  
18 recordings in the validation dataset do not have three annotated sections.
Of these 18 files, one file (infant ID 225, 3 months) has four 5-minute sections while all others have fewer than three 5-minute sections annotated.
One section (infant ID 014, 3 months) chosen for annotation was 3 minutes long (instead of 5 minutes) due to human error. 
5-minute sections were assigned to human annotators in three batches, as annotation began prior to the completion of data collection. 
Within each batch, sections were assigned in random order.
Some of the assigned 5-minute sections were not included in the validation dataset because the human annotation phase of the project ended before these sections were completed. 
While most validation files have the 5-minute sections at least 30 minutes apart, 17 files have a pair of 5-minute sections less than 30 minutes apart due to human error. \\

\begin{figure}[H]
\centering
\includegraphics[width=\linewidth]{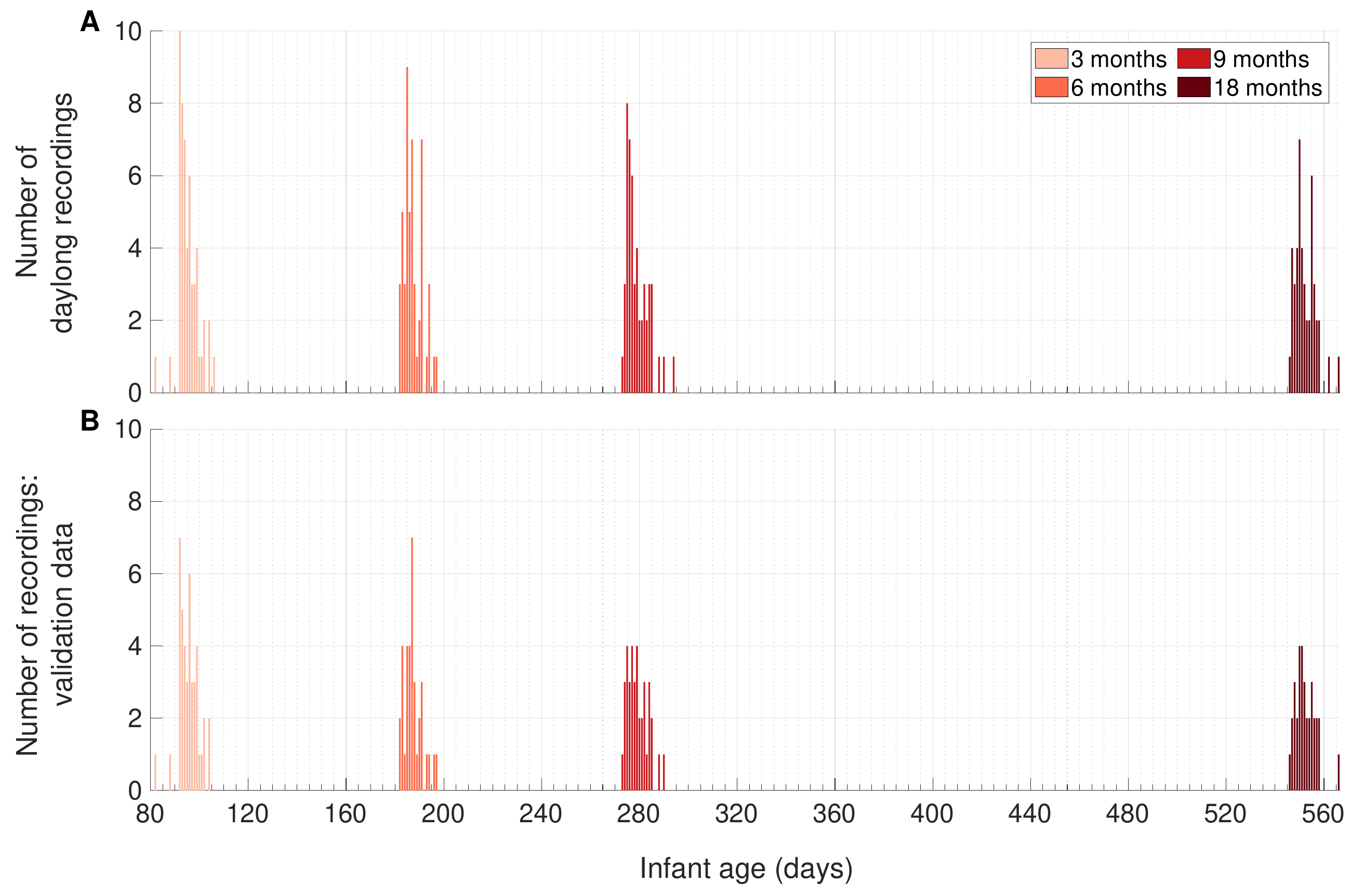}
\caption{\textbf{Distribution of infant ages (in days) in LENA daylong data and validation data.}
\textbf{(A)} shows the number of recordings at each unique infant age (in days) in the LENA daylong data. 
The bars are color-coded by the month-level age group (3, 6, 9, or 18 months; see legend) the recordings belong to.
\textbf{(B)} shows the number of recordings at each unique infant age (in days) in the validation data, color coded by month-level age group.
A and B share a legend.}
\label{fig:DataSummaryAgeDistribution}
\end{figure}

For a breakdown of the number of recordings in the LENA daylong dataset and the validation dataset by infant age, see Table \ref{tab:RecNumsByAge}. 
Each month-level infant age group corresponds to a distribution of infant ages at the day-level.
These day-level infant age distributions for LENA daylong data and validation data are shown in Fig. \ref{fig:DataSummaryAgeDistribution}.

%----Tab: RecNumsByAge-------------
\begin{table}[H]
\centering
\begin{tabular}[t]{C{3 cm} C{3.5 cm} C{3.5 cm}}
\hline
Infant age (months) & Number of recordings: LENA daylong data & Number of recordings: Validation data\\
\hline
3 & 54 & 43\\
6 & 51 & 35\\
9 & 50 & 37\\
18 & 45 & 33\\
\hline
\textbf{Total} & \textbf{200} & \textbf{148}\\
\hline
\end{tabular}
\caption{\textbf{Breakdown of number of recordings by infant age for LENA daylong data and validation data}.
The last row of the table is the total number of recordings (summed across infant age) by dataset type.}
\label{tab:RecNumsByAge}
\end{table}

%----Fig: Infant age distributions for data (DataSummaryAgeDistribution)-------------

%---SECTION: SegmentNumAndDuration--------------------
\section{Numbers and durations of audio segments by type}\label{sec:SegmentNumAndDuration}

%---SUBSECTION: SegmentNumsAndDurOverview--------------------
\subsection{Overview}\label{subsec:SegmentNumsAndDurOverview}
The LENA daylong data amounts to a total of 2418.10 hours of recorded audio, of which 571.12 hours (23.62\% of total recorded time) are silence (SIL), while key segments of interest\textemdash \textbf{Ch}ild \textbf{Sp}eech-related (ChSp), \textbf{Ch}ild \textbf{N}on-\textbf{sp}eech-related (ChNsp), and \textbf{Ad}ult (Ad)\textemdash cumulatively amount to 329.25 hours (13.62\% of total recorded time).  
Mean recording length is 12.09 hours, with a standard deviation of 1.78 hours and a range of 9.33--16.00 hours.\\

The validation data amounts to a total of 34.97 hours of recorded audio from a total of 420 annotated sections.
All sections are 5 minutes long except for one which was only 3 minutes long due to human error in generating the annotation assignments.
For L (5min) data, key segments make up 11.21 hours (32.60\% of total recorded time) while for H (5min) data, key segments make up 11.87 hours (33.91\% of total recorded time), after removing overlaps (see Sections \ref{subsec:ValDataMethods} and \ref{sec:OverlapHumanData}).
For H (5min; T-Ad) data, key segments make up 9.25 hours (26.44\% of total recorded time) after removing overlaps, reflecting the removal of adult vocalizations that are not directed towards the key infant (U and N annotations), which make up 2.61 hours (7.47\% of total recorded time).

%---SUBSECTION: LenaMeanSegDurAndNums--------------------
\subsection{Summary statistics: LENA daylong data}\label{subsec:LenaMeanSegDurAndNums}

%----Table: Lena mean segment durations for all segments-------------
\begin{table}[H]
\centering
\begin{tabular}[t]{l l c}
\hline
Segment label & Description & Mean segment duration (s)\\
\hline
\textbf{ChSp} & \textbf{Key child, speech-related} & \textbf{1.14}\\
\textbf{ChNsp} & \textbf{Key child, non-speech-related} & \textbf{1.43}\\
CHF & Key child, faint& 0.80\\
\textbf{Ad} & \textbf{Female or male adult} & \textbf{1.71}\\
FAF & Female adult, faint & 1.19\\
MAF & Male adult, faint & 1.42\\
CXN & Other child & 1.13\\
CXF & Other child, faint & 1.01\\
OLN & Overlap & 2.28\\
OLF & Overlap, faint & 1.60\\
TVN & Electronic & 2.14\\
TVF & Electronic, faint & 1.56\\
NON & Noise & 4.31\\
NOF & Noise, faint & 1.78\\
SIL & Silence & 3.18\\
\hline
\end{tabular}
\caption{\textbf{Mean segment duration for LENA daylong data}.
Mean durations (in seconds) for all segment  labels as identified by LENA{\texttrademark} for the LENA daylong data are shown.
Key segment labels (infant speech-related, ChSp; infant non-speech-related, ChNsp; Adult, Ad) are indicated in bold.
Segment labels are organized such that faint sounds of each segment label category follows the `near' segment label for that category.
For example, CHF is the segment label for very quiet instances of vocalizations by the child wearing the recorder. 
Similarly, FAF and MAF are respectively faint female and faint male adult, while Ad is combined female or male adult vocalizations that are louder from the infant's perspective (LENA's `FAN' and `MAN' categories, respectively, combined into a single adult category). 
ChSp, ChNsp, and Ad summary statistics are computed after merging segments with the same label separated by 0 s IEIs (see Sections \ref{subsec:VocIding} and \ref{subec:0IEIMerge} for details).
For more details on the segment labels and descriptions, see LENA technical reports \cite{gilkerson2020guide_SI} and \cite{xu2008lena}. 
The numbers in this table correspond to the closed circles in Fig. \ref{fig:LdaySegsDurAndNumsSummaryStatswErrBars_ChnAd0IviMerged}B.}
\label{tab:MeanSegDur}
\end{table}

%----Fig: Lena segment durations and number summary stats for all segments-------------
\begin{figure}[H]
\centering    \includegraphics[width=\linewidth]{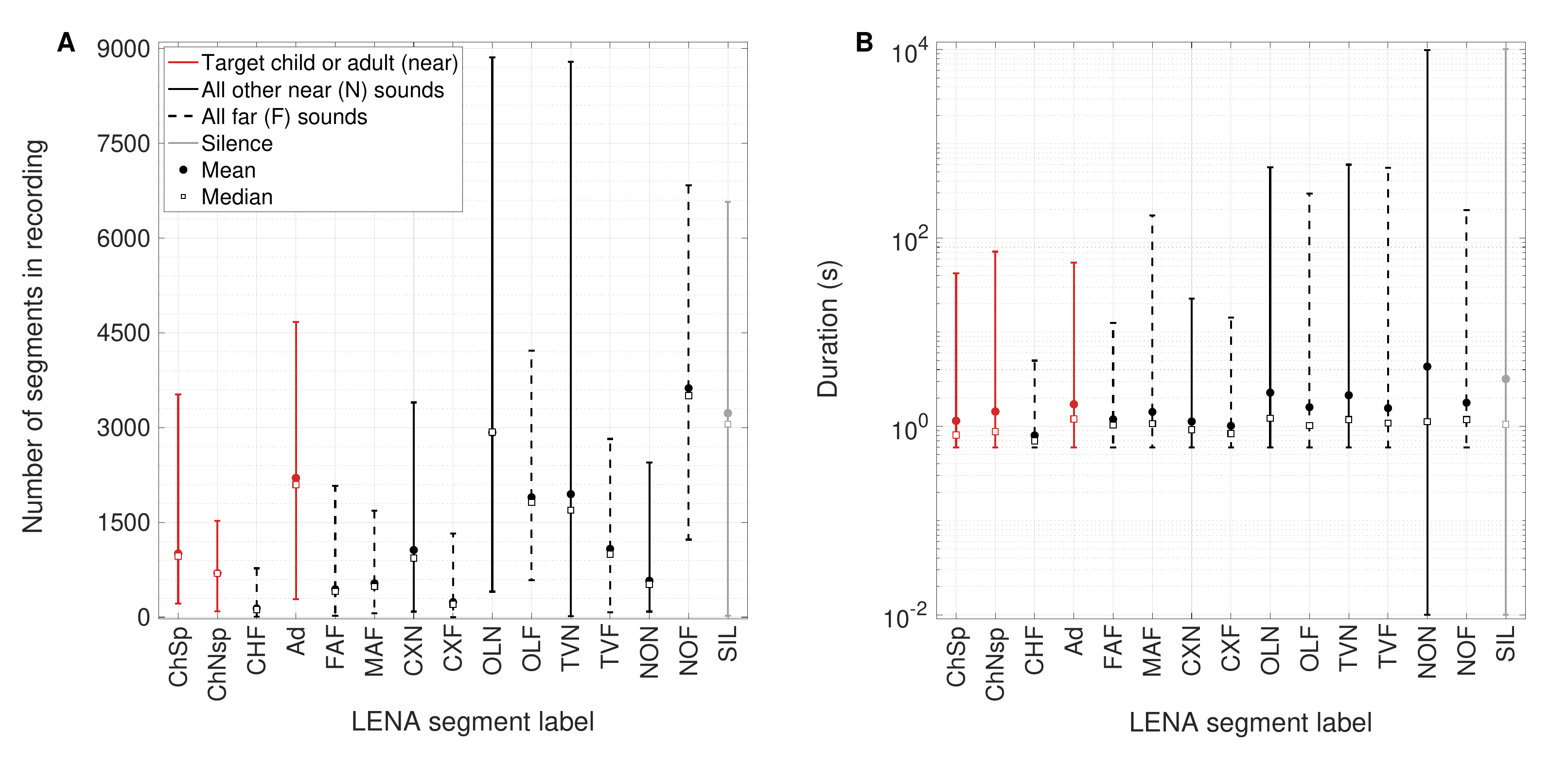}
\caption{\textbf{Summary statistics for number of segments in a recording and segment duration for LENA daylong data.}
\textbf{(A)} Mean (closed circle) and median (open square) number of segments in a recording day for all segment labels identified by LENA are shown.
Error bars indicate minimum and maximum values across all recordings.
Key segment labels (infant speech-related, ChSp; infant non-speech-related, ChNsp; Adult, Ad) are shown in red, silence (SIL) is shown in grey, and all other segment labels are in black. 
For sound segment labels, all `near' (N) sounds are indicated by solid lines while all faint (F) sounds are indicated by dashed lines (see \cite{gilkerson2020guide_SI} and \cite{xu2008lena} for definitions of near and faint).
Note that ChSp, ChNsp, and Ad are all near sounds.
\textbf{(B)} shows the mean (closed circle) and median (open square) individual segment durations (in seconds) for all segment labels identified by LENA. 
A and B share a legend.
%For reference: summary statistics for A and B are computed at the dataset level (excluding recordings at ages other than 3, 6, 9 and 18 months). 
For A and B, ChSp, ChNsp, and Ad summary statistics are computed after merging segments with the same label separated by 0 s IEIs (see Section \ref{subec:0IEIMerge} for details).
Summary statistics for all other segment labels are presented without similar merging.
We make this distinction because the 0 s IEI merging is part of the processing pipeline for ChSp, ChNsp, and Ad segments with respect to the results presented in this paper.
As such, we wanted to preserve that information here.} \label{fig:LdaySegsDurAndNumsSummaryStatswErrBars_ChnAd0IviMerged}
\end{figure}

%---SUBSECTION: ValDataMeanSegDurAndNums--------------------
\subsection{Summary statistics: Validation data}\label{subsec:ValDataMeanSegDurAndNums}

%----Table: Validation data mean segment duration for all segments-------------
\begin{table}[H]
\centering
\begin{tabular}[t]{c C{3.1cm} C{3.2cm}}
\hline
Segment label & Mean duration (s): L (5min) & Mean duration (s): H (5min)\\
\hline
ChSp &  1.42 & 0.78 \\
ChNsp & 1.53 & 1.38\\
Ad & 1.55 & 1.10\\
\hline
\end{tabular}
\caption{\textbf{Mean segment duration for validation data}.
Mean durations (in seconds) for key segment labels (infant speech-related, ChSp; infant non-speech-related, ChNsp; Adult, Ad) for L (5min) data and H (5min) data are shown.
Means are computed after merging segments with the same label separated by 0 s IEIs (see Section \ref{subec:0IEIMerge} for details) and, for H (5min) data, after overlaps have been removed (see Sections \ref{subsec:ValDataMethods} and \ref{sec:OverlapHumanData} for details).}
\label{Tab:MeanSegDurValData}
\end{table}

\begin{figure}[H]
\centering
\includegraphics[width=\linewidth]{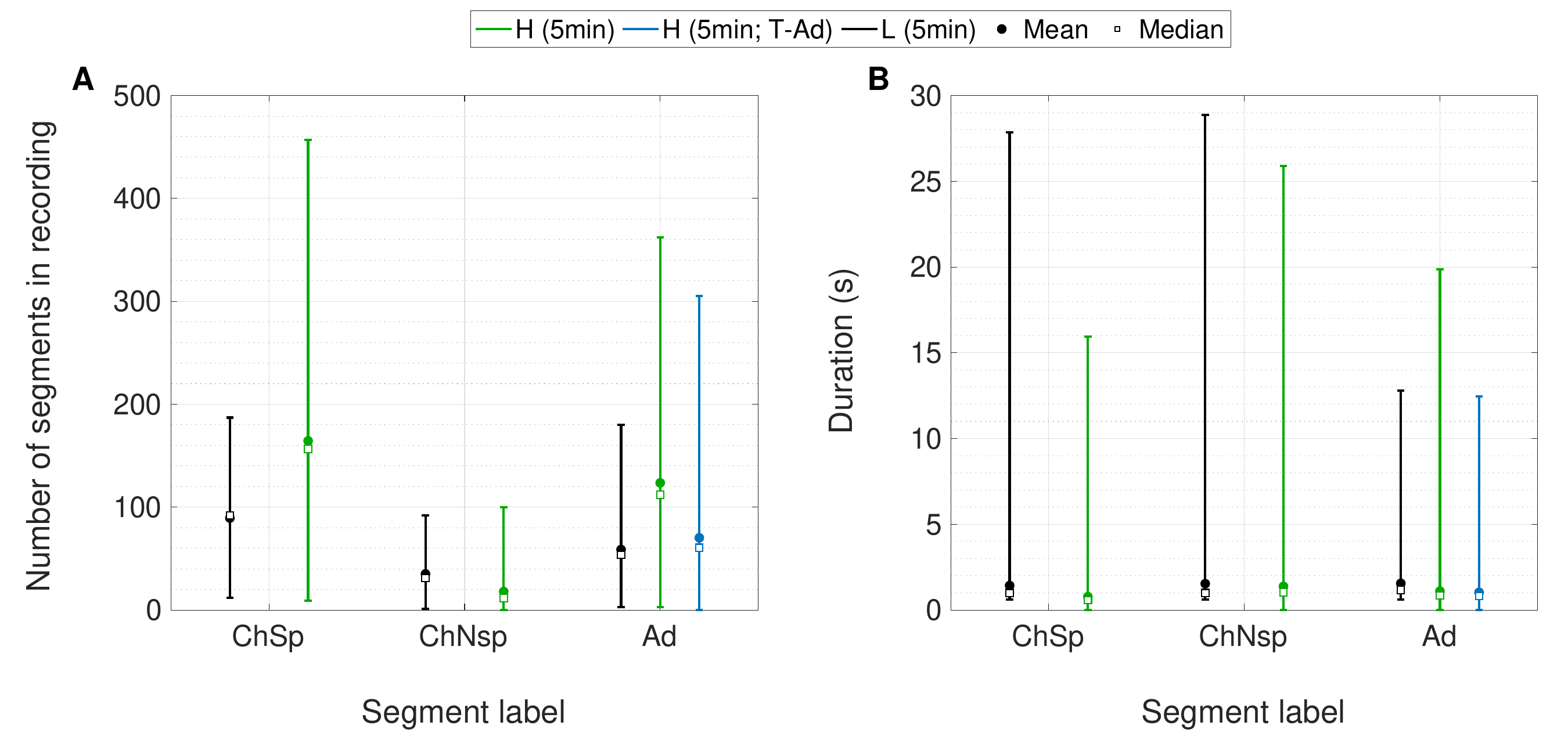}
\caption{\textbf{Summary statistics for number of segments in a file and segment duration for validation data.}
\textbf{(A)} Mean (closed circle) and median (open square) number of segments in a validation data file for key segment labels (ChSp, ChNsp, Ad) are shown for L (5min) data (black), H (5min) data (green), and H (5min; T-Ad) data (blue).
Summary statistics from H (5min; T-Ad) data is only shown for Ad segments because all infant vocalizations\textemdash and therefore, associated summary statistics\textemdash are the same for H (5min; T-Ad) and H (5 min) datasets.  
Error bars indicate minimum and maximum  values across all files and each file typically contains up to three 5-minute sections (see Section \ref{sec:DataOverview_FileNumAndAge} for details). 
%for reference: Summary statistics for A are computed at the dataset level, where each data point represents one file in the relevant validation dataset, which contains up to four 5-minute sections. 
\textbf{(B)} Mean (closed circle) and median (open square) individual segment durations (in seconds) for key segment labels are shown for L (5min) data (black), H (5min) data (green), and H (5min; T-Ad) data (blue).
A and B share a legend.
Summary statistics in A and B are computed after merging segments with the same label separated by 0 s IEI (see Section \ref{subec:0IEIMerge} for details), and in the case of human listener-labeled data, after removing overlaps (see Sections \ref{subsec:ValDataMethods} and \ref{sec:OverlapHumanData} for details).}
\label{fig:ValDataSegsDurAndNumsSummaryStatswErrBars}
\end{figure}

For human listener-labeled data with only child-directed adult vocalizations included, i.e., H (5min; T-Ad), the mean duration of adult vocalizations is 1.03 s\textemdash slightly lower than the mean adult vocalization duration when all adult vocalizations are included (Table \ref{Tab:MeanSegDurValData}). 
Mean ChSp and ChNsp durations are the same for H (5min; T-Ad) and H (5min) data since all infant vocalization segments are the same for these two datasets (see also Fig. \ref{fig:ValDataSegsDurAndNumsSummaryStatswErrBars}).

%----Fig: Validation data mean segment durations and numbers for key segments------------- 

%---SUBSECTION: LdayAndValDataSegNumsAndDurCompared--------------------
\subsection{LENA daylong and validation data compared}\label{subsec:LdayAndValDataSegNumsAndDurCompared}

%---Fig: TotnumsAndDur--------------------
\begin{figure}[H]
\centering  \includegraphics[width=\linewidth]{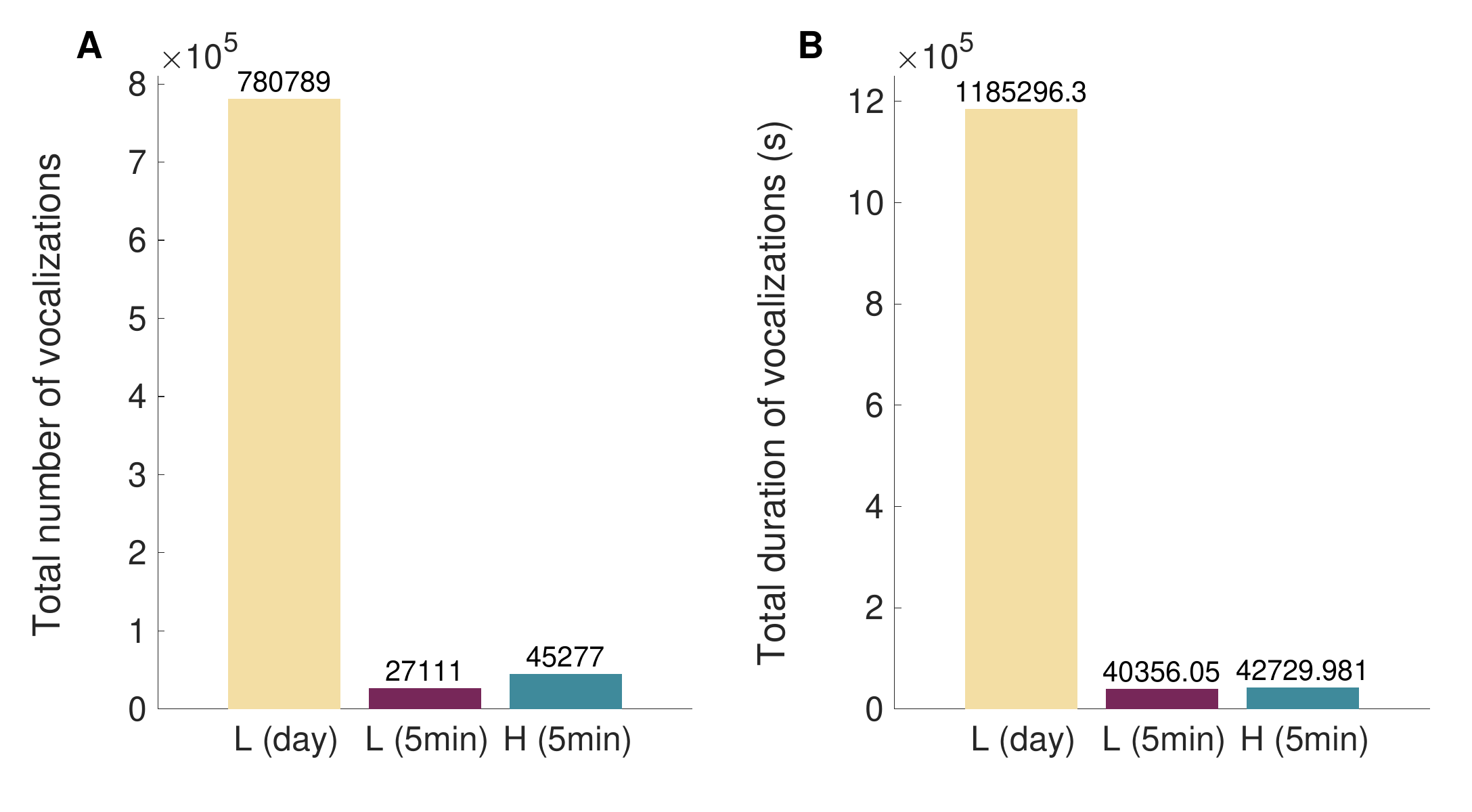}
\caption{\textbf{Vocalization number and duration totals for LENA daylong data and validation data.} 
\textbf{(A)} Total number of vocalizations (sum of the number of adult and infant---speech-related \textit{and} non-speech-related---vocalizations; $n_{\rm{tot}} = n_{\rm{Ad}} + n_{\rm{ChSp}} + n_{\rm{ChNsp}}$) are shown for LENA daylong data (L (day), yellow), human listener-labeled 5-minute sections (H (5min), green), and corresponding 5-minute sections as labeled by LENA (L (5min), maroon).
\textbf{(B)} Total duration of vocalizations (sum of the duration of adult and infant vocalizations; $d_{\rm{tot}} = d_{\rm{Ad}} +d_{\rm{ChSp}} + d_{\rm{ChNsp}}$) are shown for L (day), H (5min), and L (5min) data. 
For A and B, the bars for H (5min) data represent totals after overlaps have been processed. 
For all three datasets, totals are  presented after 0 s IEIs have been merged. 
}
\label{fig:TotnumsAndDur}
\end{figure}
%The OLP processing, resulting in more voc nums, could account for the increase in the number of total vocalisation for H (5min) data compared to L (5min) data vs. comparable total durations for both.

For human listener-labeled data with only infant-directed adult vocalizations included (H (5min; T-Ad)), the total number of key segments (ChSp, ChNsp, and Ad) is 37377, while the total duration of key segments is 33319.553 s. 
See Figs. \ref{fig:ValdataVocNumRecDayTots_BarAndBoxChts} and \ref{fig:ValdataVocDurRecDayTots_BarAndBoxChts} for details. %see also SI Figs 6 and 8

%---Fig: LdayVocNumRecDayTots_BarAndBoxChts--------------------    
\begin{figure}[H]
\centering
\includegraphics[width = \linewidth]{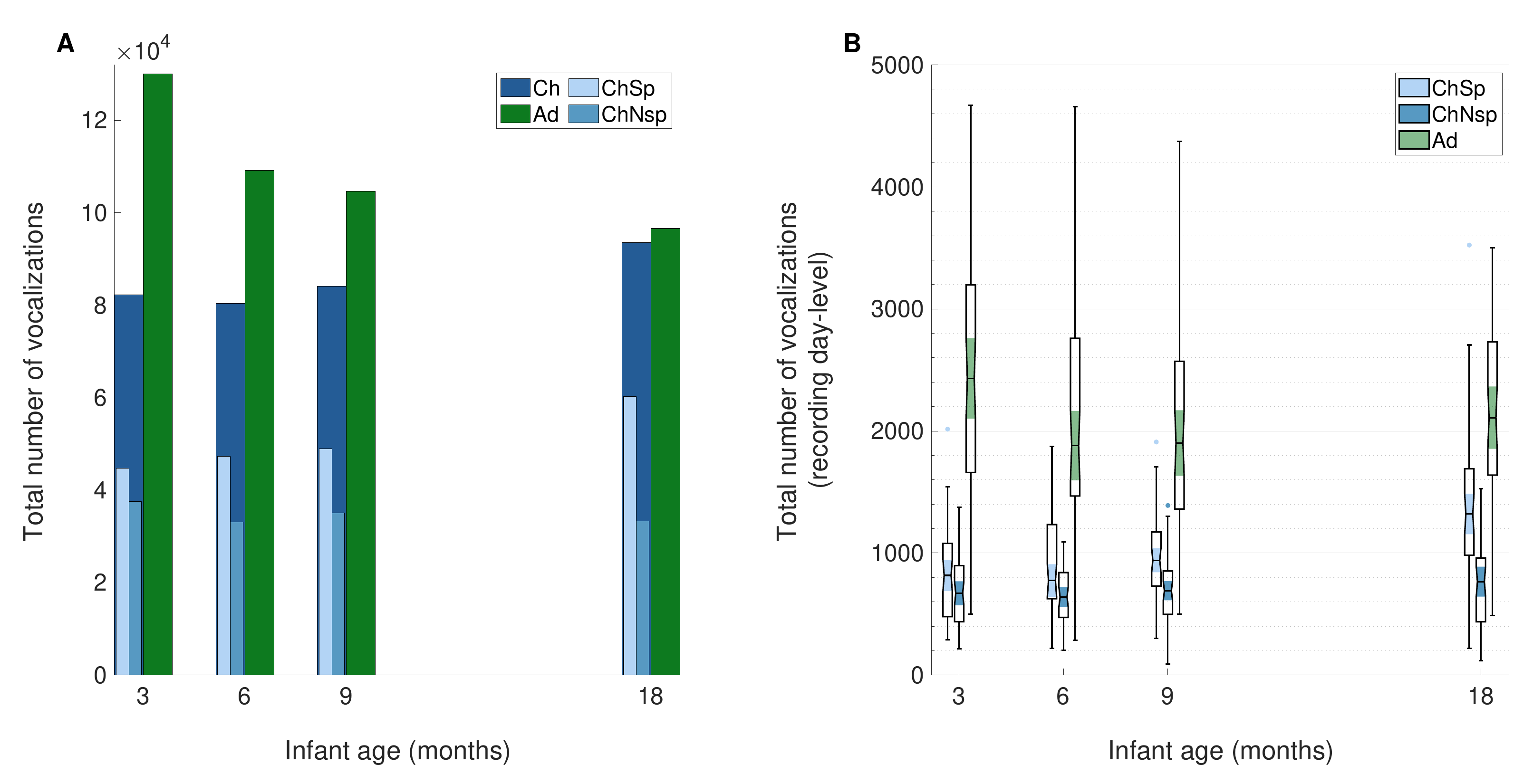} 
\caption[DummyCaption]{\textbf{Number of vocalizations in LENA daylong data by infant age and vocalization type, expressed as totals at the dataset-level and as box plots for totals at the recording day-level. (A)} 
Total numbers of infant (Ch; dark blue) and adult (Ad; green) vocalizations in the L (day) dataset for each infant age (months; x-axis) are shown as a bar plot. 
Bars for infant speech-related (ChSp) and infant non-speech-related (ChNsp) totals are shown within the infant (Ch) bars, indicating that these are sub-categories.
\textbf{(B)} Box plots summarizing total numbers of infant speech-related (ChSp), infant non-speech-related (ChNsp), and adult (Ad) vocalizations at the recording day-level are shown as a function of infant age (months; x-axis). 
Box plots whose notches do not overlap have different medians at the 0.05 significance level\footnotemark.
Number of vocalizations are reported after vocalizations with the same label are merged when separated by 0 s IEIs. 
Note that total vocalization numbers shown in A do not account for the fact that different ages have different numbers of daylong recordings (see Table \ref{tab:RecNumsByAge}).
%Note that while there appears to be fewer adult vocalizations at 18 months in A, it is confounded by the fact that there are fewer daylong recordings at 18 months than other ages (see Table \ref{tab:RecNumsByAge}).
As such, B provides a more accurate representation of how the total number of each vocalization type in a daylong recording changes with infant age.
However, data plotted in B have not been normalized to account for differences in total number of vocalizations at the recording-level due to variations in recording duration across daylong recordings (see Section \ref{subsec:SegmentNumsAndDurOverview}).}
\label{fig:LdayVocNumRecDayTots_BarAndBoxChts}
\end{figure}

\footnotetext{Per the documentation of MATLAB's \href{https://www.mathworks.com/help/matlab/ref/boxchart.html\#mw_f1510deb-1332-4210-8cb0-cb2706d47198}{\texttt{boxchart} function}, which was used to generate box plots, ``the significance level is based on a normal distribution assumption, but the median comparison is reasonably robust for other distributions".}

%---Fig: ValdataVocNumRecDayTots_BarAndBoxChts--------------------
\begin{figure}[H]
\centering
\includegraphics[width = \linewidth]{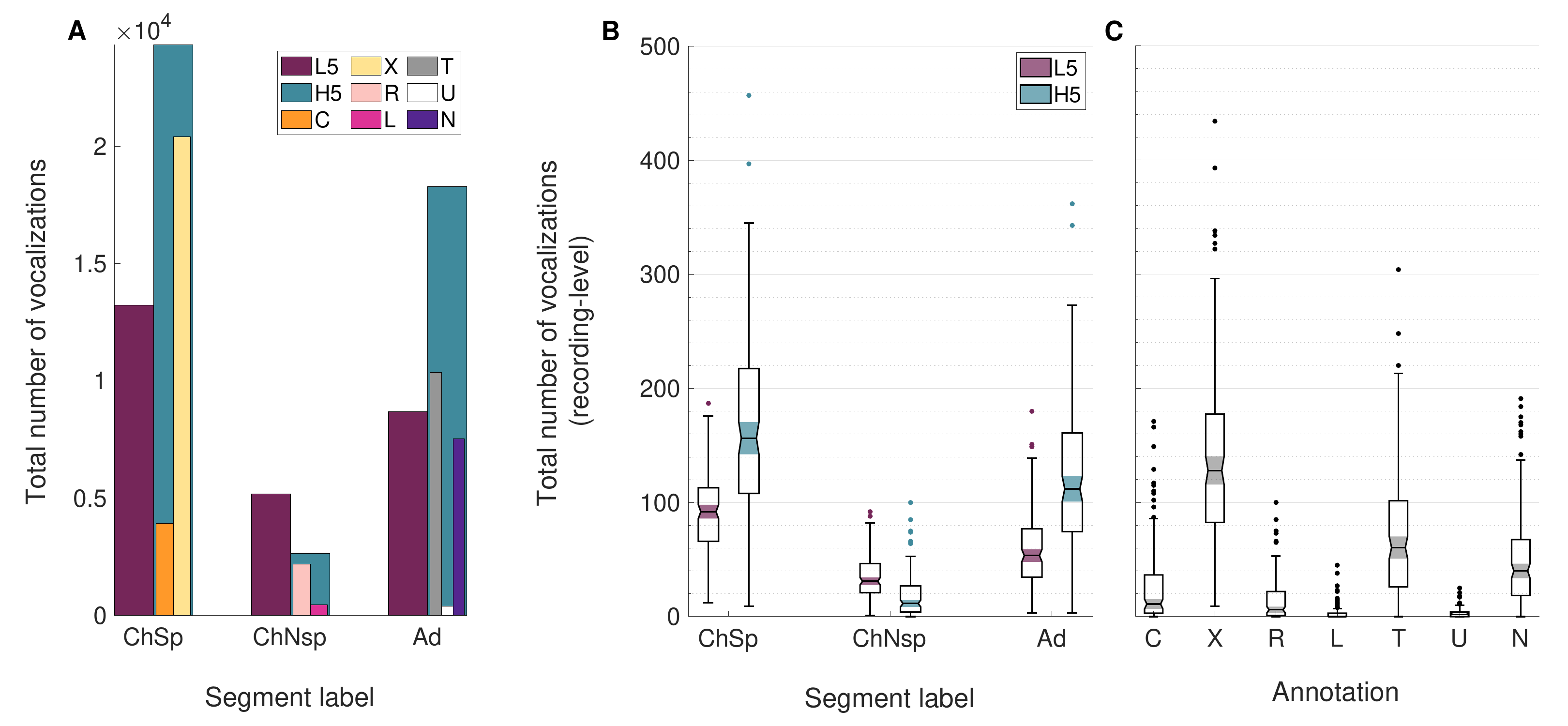}
\caption{\textbf{Number of vocalizations in the validation datasets by vocalization type, expressed as totals at the dataset-level and as box plots for totals at the recording-level. (A)} 
Total numbers of infant speech-related (ChSp), infant non-speech-related (ChNsp), and adult (Ad) vocalizations in the validation datasets (L (5min), maroon; H (5min), green) are shown as a bar plot.
Bars for annotation-level totals for the human listener-labeled data are shown within the bar for the corresponding segment label, indicating that these are sub-categories.
ChSp annotation types are C (canonical) and X (non-canonical non-reflexive), ChNsp annotation types are R (cry) and L (laugh), and Ad annotation types are T (infant-directed), U (addressee unknown), and N (not infant-directed).
\textbf{(B)} Box plots summarizing total numbers of infant speech-related (ChSp), infant non-speech-related (ChNsp), and adult (Ad) vocalizations at the recording-level are shown for L (5min) and H (5min) data.
\textbf{(C)} 
Box plots summarizing total vocalization numbers by annotation type (x-axis) at the recording-level are shown for H (5min) data.
B and C share a y-axis. 
For H (5min) data, vocalization numbers are estimated after overlaps have been processed and for all datasets, number of vocalizations are reported after vocalizations with the same label are merged when separated by 0 s IEIs.
For B and C, box plots whose notches do not overlap have different medians at the 0.05 significance level.
Data plotted in B and C have not been normalized to account for differences in vocalization numbers at the recording-level due to variations in the number of annotated sections across recordings (see Section \ref{sec:DataOverview_FileNumAndAge}).
}
\label{fig:ValdataVocNumRecDayTots_BarAndBoxChts}
\end{figure}

% While we do not report a similar breakdown of number of vocalizations by vocalization and annotation type for H (5min; T-Ad) data, we note that the only difference between H (5min) and H (5min; T-Ad) data is that the latter excluded non-child-directed adult vocalizations (U and N annotations).
% As such, the breakdown by annotation type shown in Fig. \ref{fig:ValdataVocNumRecDayTots_BarAndBoxChts} A and C contains the relevant information. 

%---Fig: LdayVocDurRecDayTots_BarAndBoxChts--------------------
\begin{figure}[H]
\centering
\includegraphics[width = \linewidth]{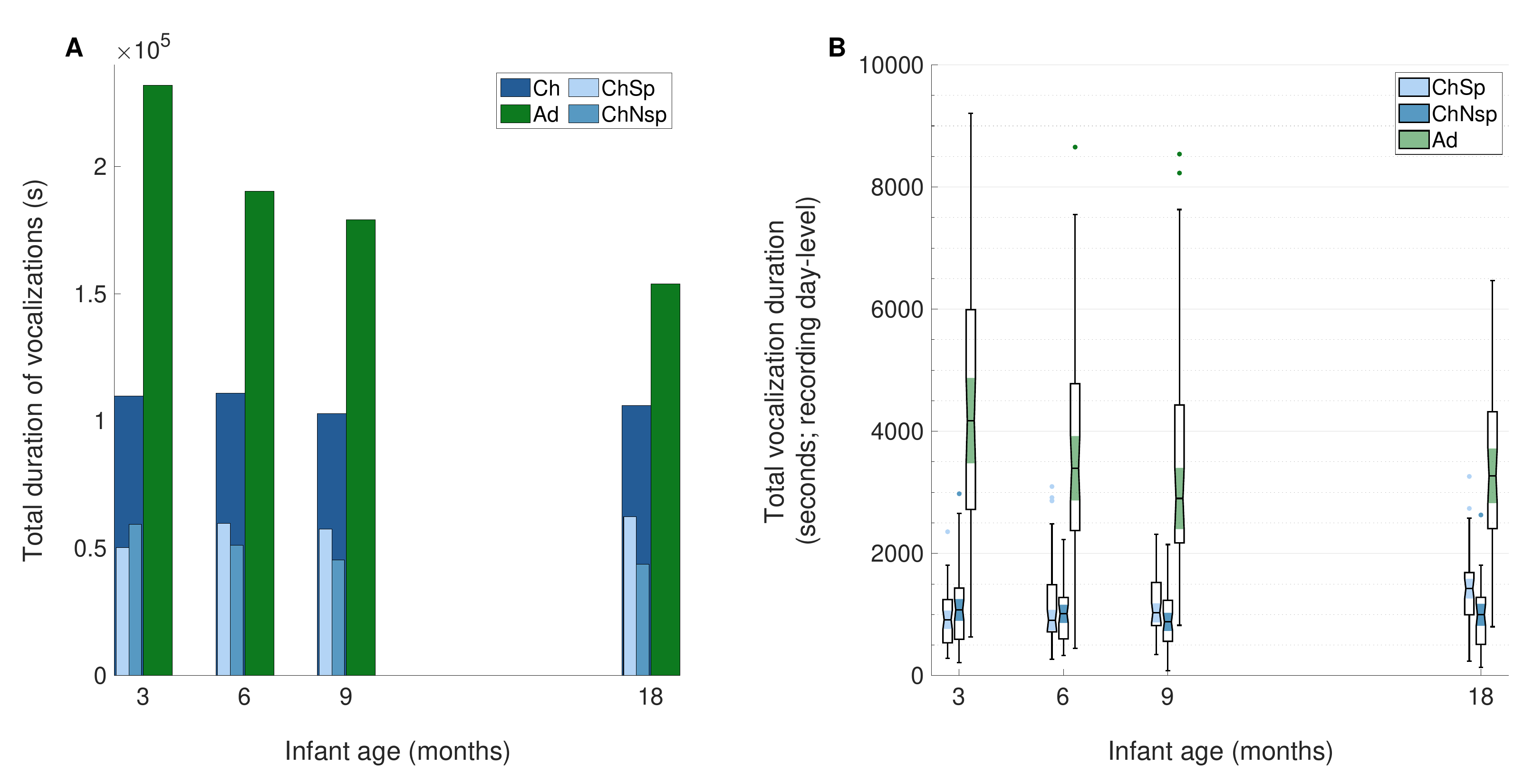}
\caption{\textbf{Vocalization durations in LENA daylong data by infant age and vocalization type, expressed as totals at the dataset-level and as box plots for totals at the recording day-level. (A)} 
Total durations of infant (Ch; dark blue) and adult (Ad; green) vocalizations in the L (day) dataset for each infant age (months; x-axis) are shown as a bar plot. 
Bars for infant speech-related (ChSp) and infant non-speech-related (ChNsp) total durations are shown within the infant (Ch) bars, indicating that these are sub-categories.
\textbf{(B)} Box plots summarizing total durations of infant speech-related (ChSp), infant non-speech-related (ChNsp), and adult (Ad) vocalizations at the recording day-level are shown as a function of infant age (months; x-axis). 
Box plots whose notches do not overlap have different medians at the 0.05 significance level.
Vocalization durations are reported after vocalizations with the same label are merged when separated by 0 s IEIs.
%However, this does not matter since duration totals don't change due to IEI merges
Note that total vocalization durations shown in A do not account for the fact that different ages have different numbers of daylong recordings (see Table \ref{tab:RecNumsByAge}).
As such, B provides a more accurate representation of how the total duration of each vocalization type in a daylong recording changes with infant age.
However, data plotted in B have not been normalized to account for differences in total vocalization durations at the recording-level due to variations in recording duration across daylong recordings (see Section \ref{subsec:SegmentNumsAndDurOverview}).}
\label{fig:LdayVocDurRecDayTots_BarAndBoxChts}
\end{figure}

%---Fig: ValdataVocDurRecDayTots_BarAndBoxChts--------------------
\begin{figure}[H]
\centering
\includegraphics[width = \linewidth]{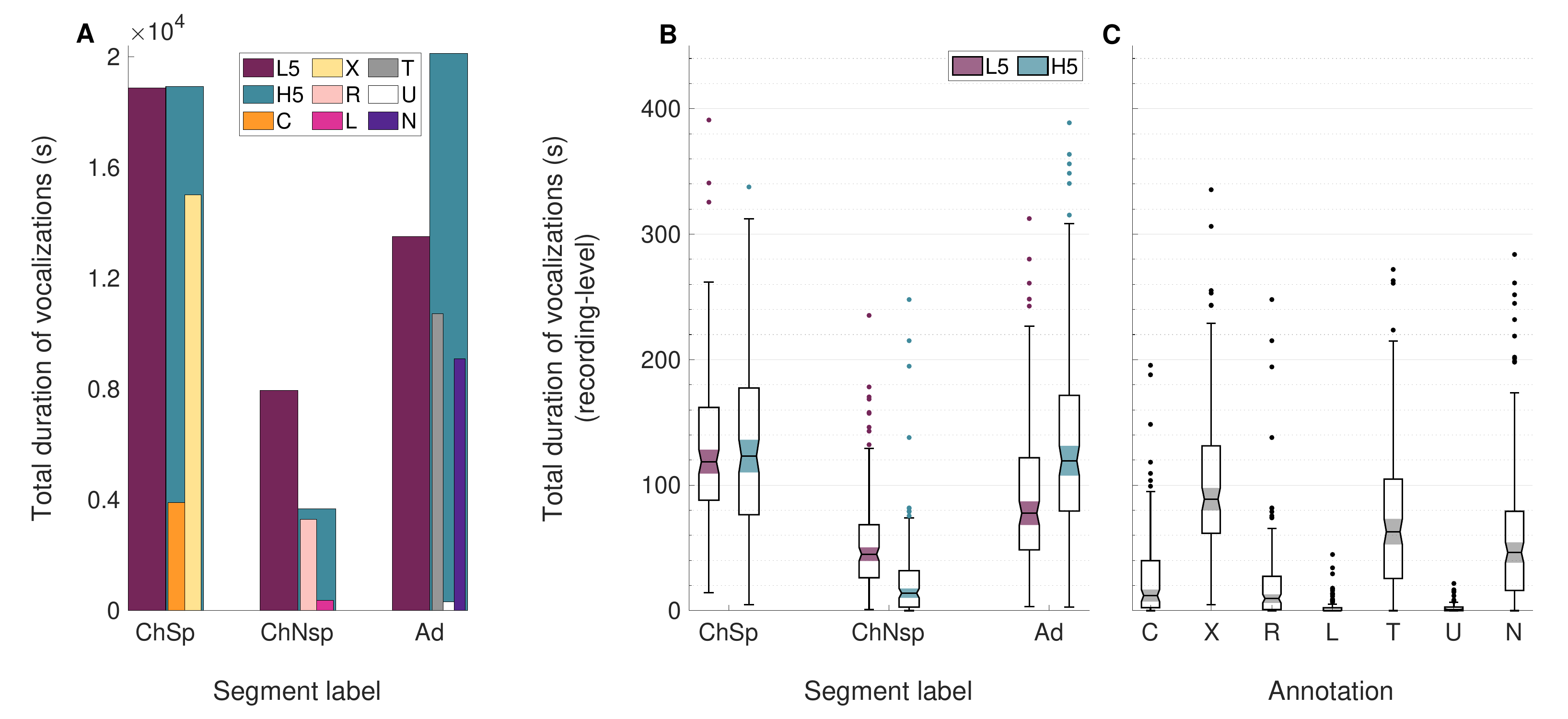}
\caption{\textbf{Vocalization durations in the validation datasets by vocalization type, expressed as totals at the dataset-level and as box plots for totals at the recording-level. (A)} 
Total durations of infant speech-related (ChSp), infant non-speech-related (ChNsp), and adult (Ad) vocalizations in the validation datasets (L (5min), maroon; H (5min), green) are shown as a bar plot.
Bars for annotation-level totals for the human listener-labeled data are shown within the bar for the corresponding segment label, indicating that these are sub-categories.
ChSp annotation types are C (canonical) and X (non-canonical non-reflexive), ChNsp annotation types are R (cry) and L (laugh), and Ad annotation types are T (infant-directed), U (addressee unknown), and N (not infant-directed).
\textbf{(B)} Box plots summarizing total durations of infant speech-related (ChSp), infant non-speech-related (ChNsp), and adult (Ad) vocalizations at the recording-level are shown for L (5min) and H (5min) data.
\textbf{(C)} 
Box plots summarizing total vocalization durations by annotation type (x-axis) at the recording-level are shown for H (5min) data.
B and C share a y-axis. 
For H (5min) data, vocalization durations are estimated after overlaps have been processed and for all datasets, vocalization durations are reported after vocalizations with the same label are merged when separated by 0 s IEIs.
For B and C, box plots whose notches do not overlap have different medians at the 0.05 significance level.
Data plotted in B and C have not been normalized to account for differences in total vocalization durations at the recording-level due to variations in the number of annotated sections across recordings (see Section \ref{sec:DataOverview_FileNumAndAge}).
}
\label{fig:ValdataVocDurRecDayTots_BarAndBoxChts}
\end{figure}

\newpage
%---SECTION: IEIAndDurationDistributions--------------------
\section{IEI and duration distributions of ChSp and Ad segments}\label{sec:IEIAndDurationDistributions}
%One reason for showing IEI and duration distributions vs box plots is because the data is long-tailed. 

Both IEI and duration distributions have long tails, with a large number of data points occurring in a relatively small range compared to the range of the data.
We use a recursive binning approach based on percentile values (adapted from \citenum{jiang2013head}) to estimate (empirical) distributions for both IEI and duration data.\\

We first partition the data at a specified percentile value.
For illustrative purposes, let us set this to the 5\textsuperscript{th} percentile. 
Then, this process divides the data into two sets: the first consisting of all data less than or equal to the 5\textsuperscript{th} percentile value (the `head') and the second consisting of all data greater than the 5\textsuperscript{th} percentile value (the `tail').
Since this is the first of several partitions, we refer to the head and tail obtained from this partition as the first head and the first tail. 
Next, we partition the first tail at \emph{its} 5\textsuperscript{th} percentile value to get the second tail.
We repeat this process till partitioning is no longer possible\textemdash this is enforced by stopping partitioning when the resultant tail has fewer than 2 data points and discarding the corresponding partition value.
Thus, we get a series of partitions for the data given by the set of 5\textsuperscript{th} percentile values of the recursive tails.
We use the partitions obtained using this method to bin the data and estimate the empirical distribution.
The first bin ranges from the minimum value in the data to the 1\textsuperscript{st} partition value (corresponding to the 5\textsuperscript{th} percentile of the data), the second bin ranges from the 1\textsuperscript{st} partition value to the 2\textsuperscript{nd} partition value (corresponding to the 5\textsuperscript{th} percentile of the first tail), and so on.
The last bin ranges from the last partition value to the maximum value in the data.
Finally, the bin centers are computed to plot the distributions. \\

To best represent the long-tailed nature of the data while also showcasing how the data is distributed in the tail, we present the data as a probability distribution as well as an empirical complementary cumulative distribution function (CCDF).
The probability distribution is estimated by normalizing the bin counts such that the bin counts sum to 1.
%Note that this is distinct from an empirical probability distribution function where the normalization is such that the area under the curve is equal to 1 (the area under the curve summing to 1 is equivalent to bin counts summing to 1 in the limit of bin number tending to infinity) 
The CCDF is derived from the cumulative distribution function (CDF).
The CDF estimates the probability that the value of a random variable, $X$, is less than or equal to a given number, $x$. 
This is expressed mathematically as $F_X(x) = P(X \leq x)$, where $F_X(x)$ is the CDF. 
The CCDF, then, is given by $1-F_X(x) = P(X > x)$, and provides an easy way to visualize distribution tails \cite{clauset2009power}.
The empirical CCDFs presented in Sections \ref{subsec:IEIdistribution} and \ref{subsec:DurDistribution} are derived from the probability distributions estimated using the method described above.\\

We use 5\textsuperscript{th} percentile values to partition LENA daylong data, and 10\textsuperscript{th} percentile values to partition validation data because of the much lower sample size of the latter. 
We also truncate the CCDFs such that probability values less than 10\textsuperscript{-6} are discarded.\\
%Since CCDFs are necessarily monotonically decreasing functions, this means that the truncation only affects the tails

IEI and duration distributions from L (day) and L (5min) data have artifacts resulting from minimum duration thresholds for different segment types as set by LENA{\texttrademark} \cite{gilkerson2020guide_SI}. 
Minimum duration thresholds for infant (speech-related and non-speech-related) and adult segments are 600 ms and 1 s, respectively.
However, we find that roughly 8\% of the adult vocalizations (after merging vocalizations of the same type separated by 0 s IEIs) in the L (day) data have durations ranging from 600 ms to 1 s. 
Accordingly, infant speech-related (ChSp) and adult (Ad) duration distributions from both L (day) and L (5min) data begin at or around 600 ms (see Section \ref{subsec:DurDistribution}).
Adult duration distributions peak at 1 s\textemdash corresponding to the 1 s minimum\textemdash while ChSp duration distributions peak at 600 ms. 
Duration distributions for the human listener-labeled data, however, begin at lower duration values.\\

While LENA{\texttrademark} also sets minimum durations for all other segment labels (ranging from 800 ms to 1 s, depending on the segment type; see \cite{gilkerson2020guide_SI}), 
we note that the observed minimum duration for most segment types in the L (day) data\textemdash and by extension, L (5min) data\textemdash is 600 ms (Fig. \ref{fig:LdaySegsDurAndNumsSummaryStatswErrBars_ChnAd0IviMerged}B).
As a result, the smallest possible IEIs for infant speech-related (ChSp) vocalizations (after merging vocalizations of the same type separated by 0 s IEIs) are necessarily 600 ms long\textemdash set by the minimum segment duration across all LENA segment labels\textemdash since at least one segment type other than ChSp must intervene between any two ChSp vocalizations. 
The same considerations also apply for adult (Ad) IEIs\footnote{While IEIs shorter than 600 ms are possible as a result of very short NON or SIL segments\textemdash which have observed minimum durations less than 600 ms (Fig. \ref{fig:LdaySegsDurAndNumsSummaryStatswErrBars_ChnAd0IviMerged}B)\textemdash intervening between consecutive vocalizations of the same type, we do not observe ChSp or Ad IEIs shorter than 600 ms in our data.}. 
These lower limits are reflected in the IEI distributions for L (day) and L (5min) data but do not apply to human listener-labeled data (see Section \ref{subsec:IEIdistribution}). 

%Finally, ChSp data are identical for human listener-labeled data that includes all adult vocalizations (H (5min; All Ad)) and human listener-labeled data that only includes infant-directed adult vocalizations (H (5min; T-Ad)). As such, we show distributions for H (5min; T-Ad) data only for adult vocalizations.
%For LENA day long duration, there are *so* many 600 ms vocalizations that the first few 5 percentile values were 600 ms and those bins were merged together. This is explained in the code. 

%---SUBSECTION: IEIdistribution--------------------
\subsection{IEI distributions}\label{subsec:IEIdistribution}

%---Fig: IEIDist_ChnspLday--------------------
\begin{figure}[H]
\centering
\includegraphics[width=\linewidth]{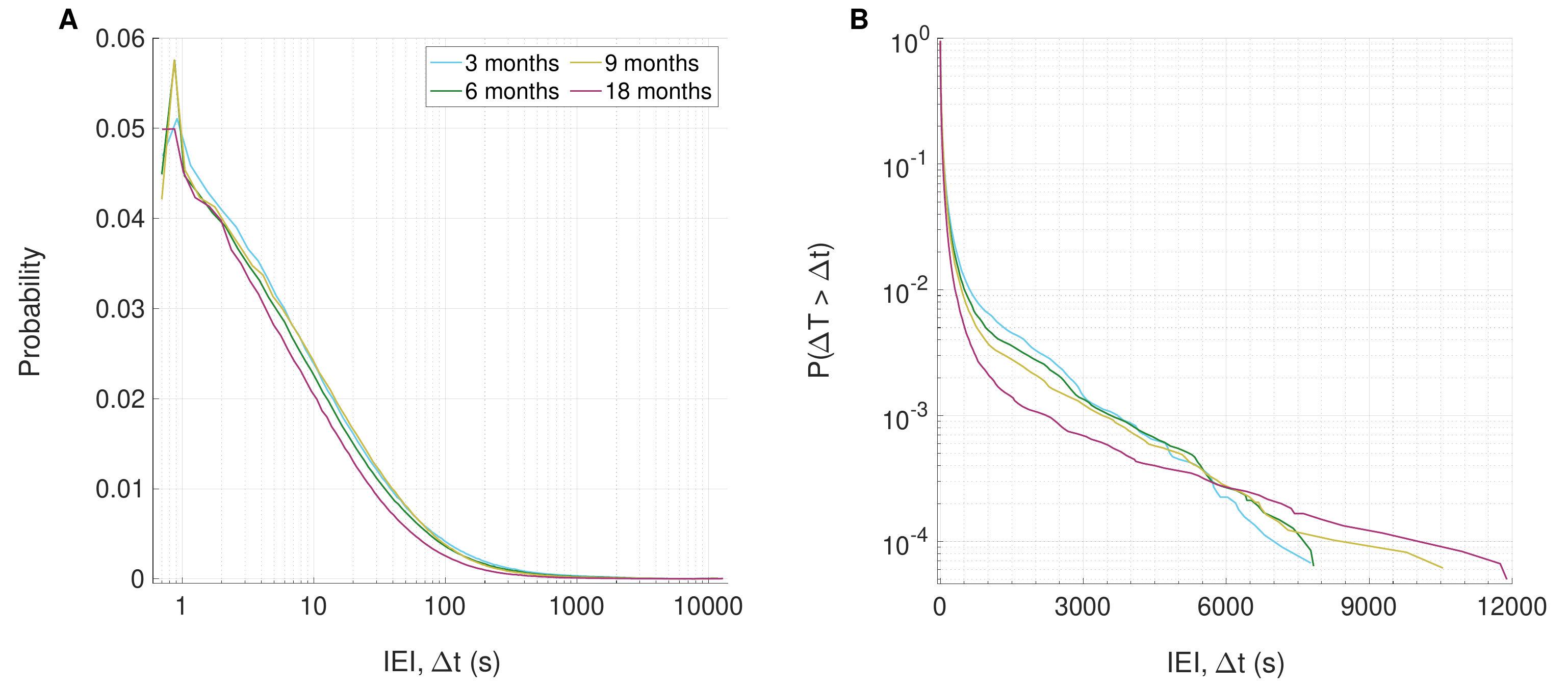} 
\caption{\textbf{Infant speech-related IEI distributions for LENA daylong data. (A)} 
Empirical probability distributions of infant speech-related (ChSp) IEIs ($\Delta t$) for 3, 6, 9, and 18 months (see legend).
\textbf{(B)} Empirical complementary cumulative distribution functions (CCDF) of ChSp IEIs for 3, 6, 9, and 18 months.
Here, $P(\Delta T > \Delta t)$ indicates the probability that the random variable representing IEIs, $\Delta T$ is greater than the value of the IEI on the x-axis, $\Delta t$.
A and B share a legend.}
\label{fig:IEIDist_ChnspLday}
\end{figure}

%---Fig: IEIDist_ChnspValdata--------------------
\begin{figure}[H]
\centering
\includegraphics[width=\linewidth]{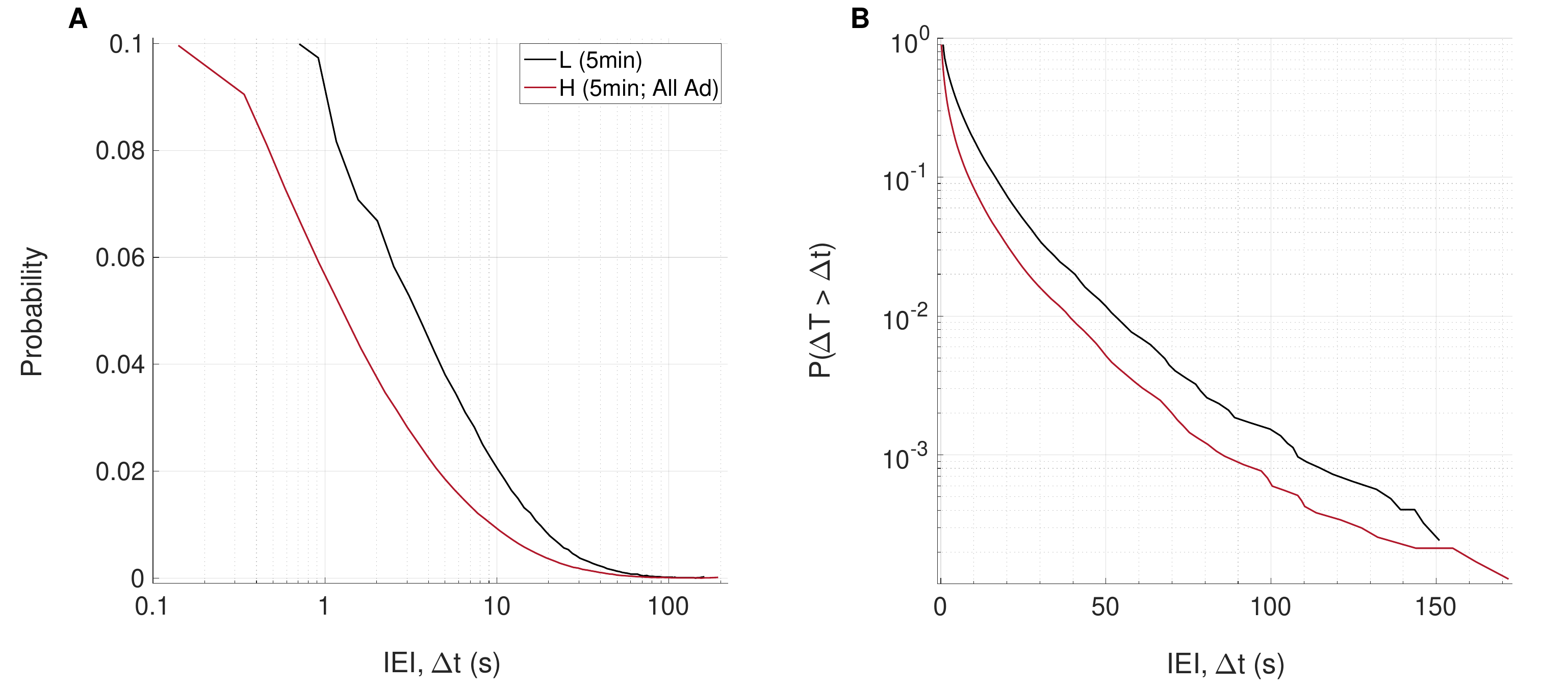}
\caption{\textbf{Infant speech-related IEI distributions for validation data. (A)} 
Empirical probability distributions of infant speech-related (ChSp) IEIs ($\Delta t$) for H (5min; All Ad) and L (5min) data (see legend). 
\textbf{(B)} Empirical complementary cumulative distribution functions (CCDF) of ChSp IEIs for H (5min; All Ad) and L (5min) data.
Here, $P(\Delta T > \Delta t)$ indicates the probability that the random variable representing IEIs, $\Delta T$ is greater than the value of the IEI on the x-axis, $\Delta t$.
A and B share a legend.}
\label{fig:IEIDist_ChnspValdata}
\end{figure}

%---Fig: IEIDist_AnLday--------------------
\begin{figure}[H]
\centering
\includegraphics[width=\linewidth]{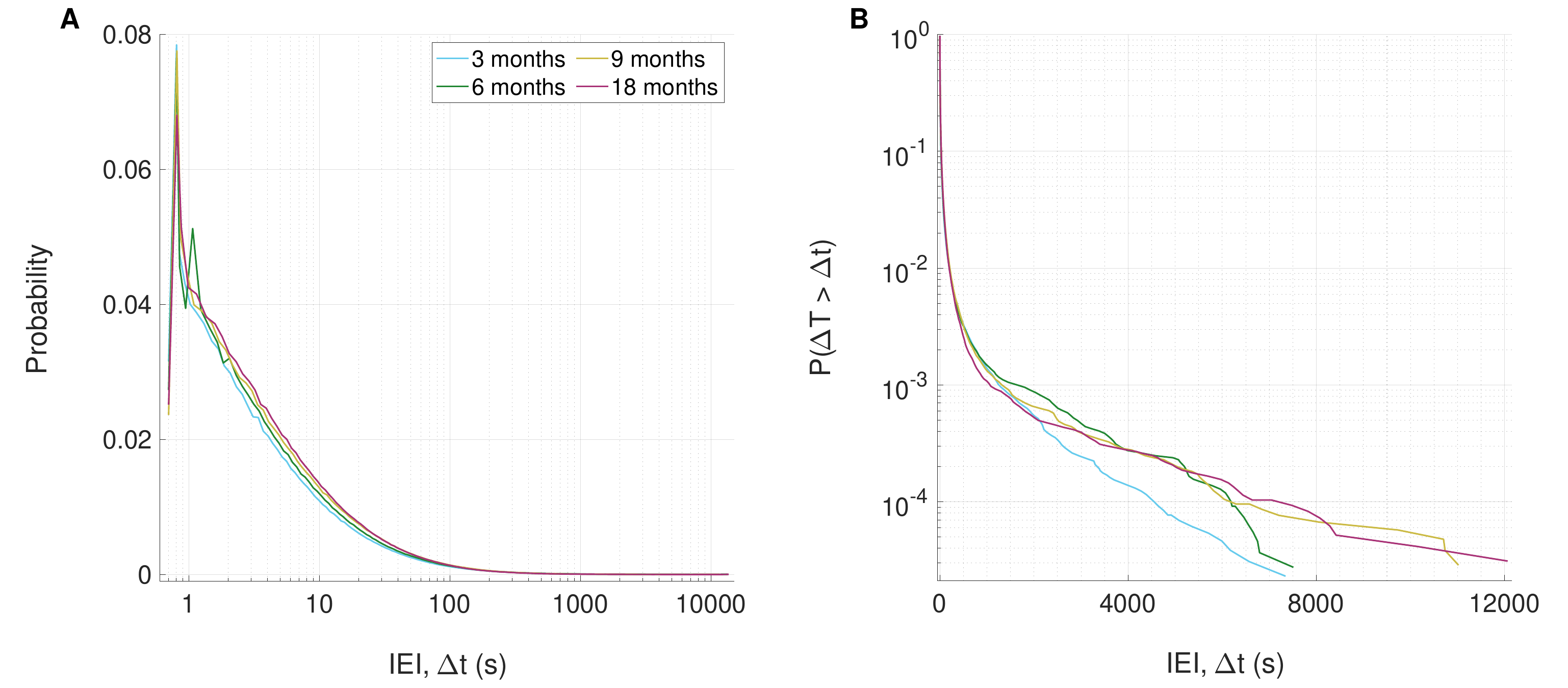} 
\caption{\textbf{Adult IEI distributions for LENA daylong data. (A)} 
Empirical probability distributions of adult (Ad) IEIs ($\Delta t$) for 3, 6, 9, and 18 months (see legend).
\textbf{(B)} Empirical CCDFs of Ad IEIs for 3, 6, 9, and 18 months.
Here, $P(\Delta T > \Delta t)$ indicates the probability that the random variable representing IEIs, $\Delta T$ is greater than the value of the IEI on the x-axis, $\Delta t$.
A and B share a legend.}
\label{fig:IEIDist_AnLday}
\end{figure}

%---Fig: IEIDist_AnValdata--------------------
\begin{figure}[H]
\centering
\includegraphics[width=\linewidth]{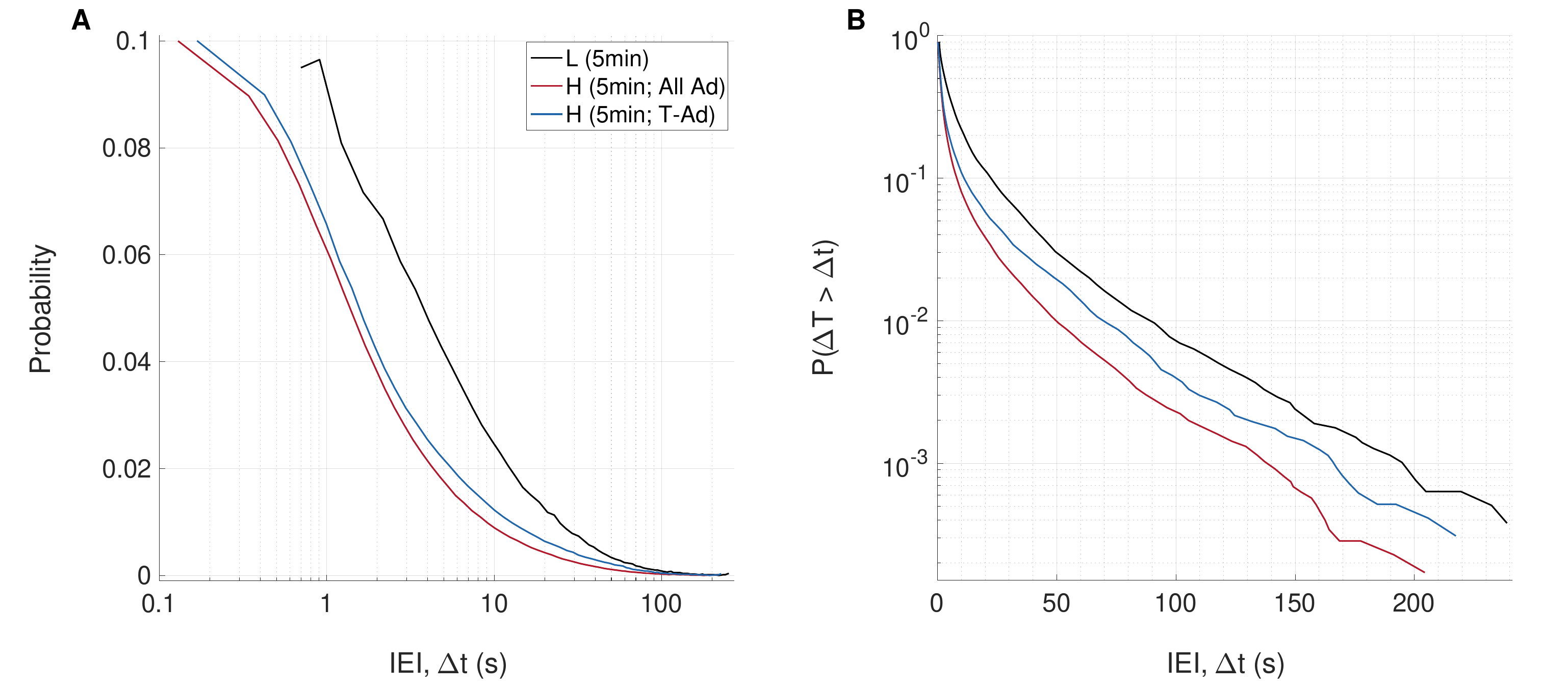}
\caption{\textbf{Adult IEI distributions for  validation data. (A)} 
Empirical probability distributions of adult (Ad) IEIs ($\Delta t$) for H (5min; All Ad), H (5min; T-Ad), and L (5min) data (see legend).
\textbf{(B)} Empirical CCDFs of Ad IEIs for H (5min; All Ad), H (5min; T-Ad), and L (5min) data.
Here, $P(\Delta T > \Delta t)$ indicates the probability that the random variable representing IEIs, $\Delta T$ is greater than the value of the IEI on the x-axis, $\Delta t$.
A and B share a legend.}
\label{fig:IEIDist_AnValdata}
\end{figure}

%---SUBSECTION: DurDistribution--------------------
\subsection{Duration distributions}\label{subsec:DurDistribution}

%---Fig: DurDist_ChnspLday--------------------
\begin{figure}[H]
\centering
\includegraphics[width=\linewidth]{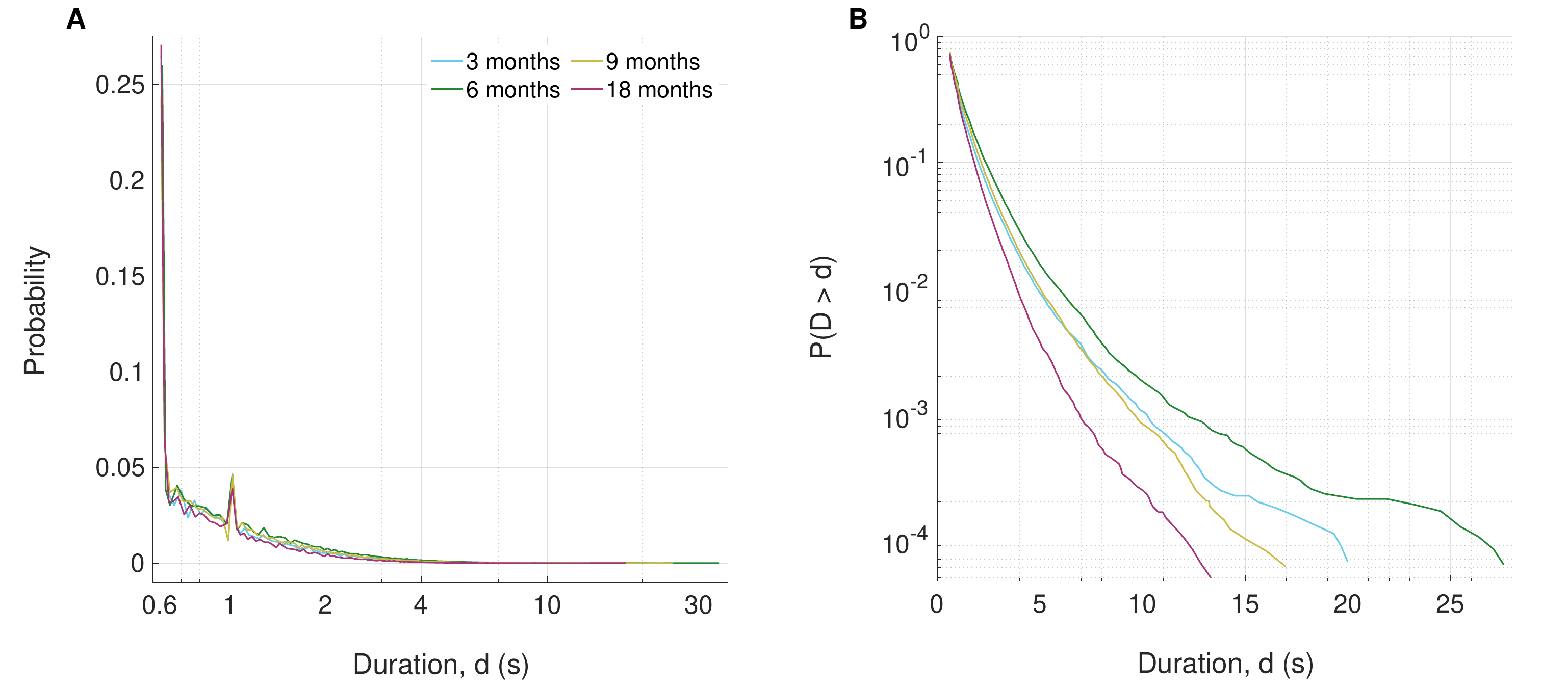}
\caption{\textbf{Infant speech-related duration distributions for LENA daylong data. (A)} 
Empirical probability distributions of infant speech-related (ChSp) durations ($d$) for 3, 6, 9, and 18 months (see legend).
%Aadditional details: Note that infant vocalizations have a large number of instances where the duration is equal to the minimum duration value, 0.6 s. As a result, the recursive 5 percentile binning method resulted in the first few bins having both end points equal to 0.6 s, because vocalizations with durations equal to 0.6 s made up more than the first 5 percentile of the data. These bins were merged into a single bin (details are available in the Github repository).
\textbf{(B)} Empirical CCDFs of ChSp durations for 3, 6, 9, and 18 months.
Here, $P(D > d)$ indicates the probability that the random variable representing durations, $D$ is greater than the value of the duration on the x-axis, $d$.
A and B share a legend.}
\label{fig:DurDist_ChnspLday}
\end{figure}

%---Fig: DurDist_ChnspValdata--------------------
\begin{figure}[H]
\centering
\includegraphics[width=\linewidth]{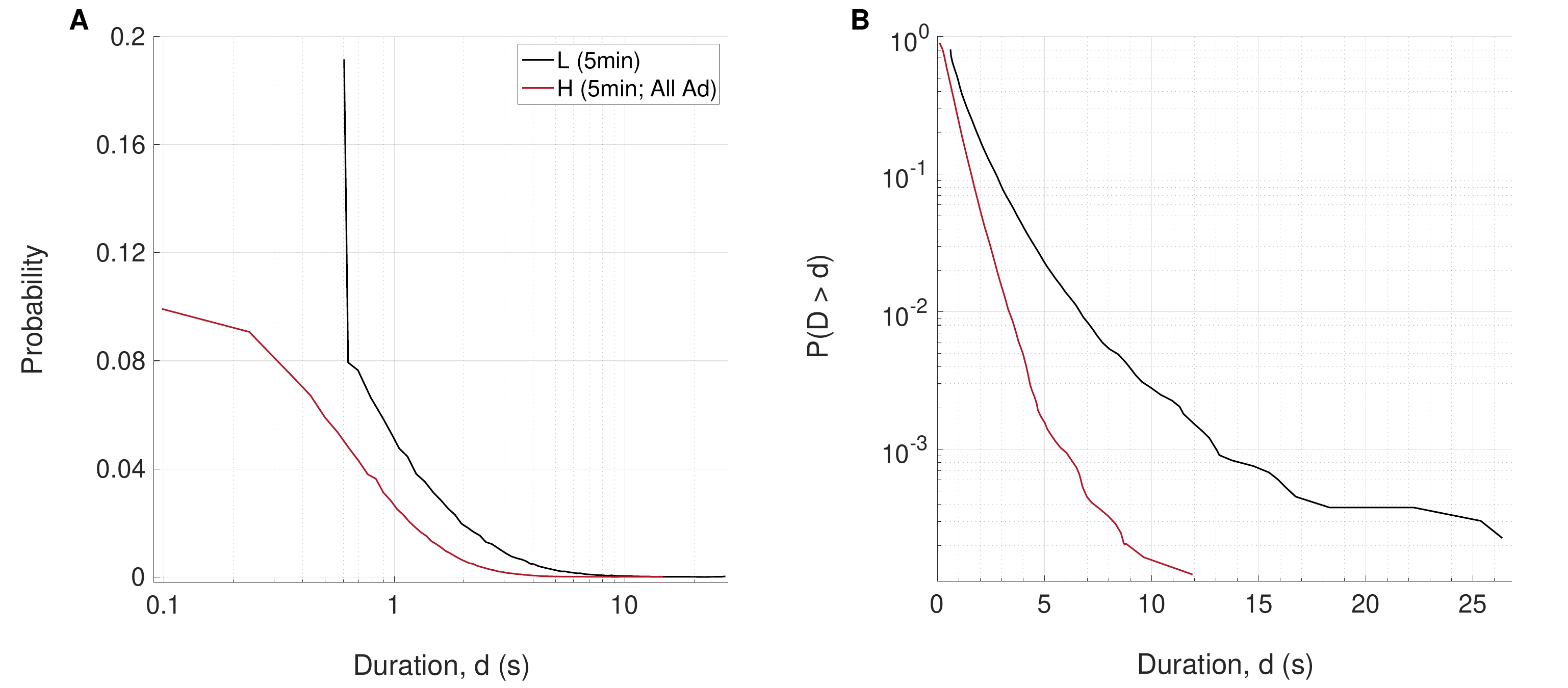}
\caption{\textbf{Infant speech-related duration distributions for validation data. (A)} 
Empirical probability distributions of infant speech-related (ChSp) durations ($d$) for H (5min; All Ad) and L (5min) data (see legend).
\textbf{(B)} Empirical CCDFs of ChSp durations for H (5min; All Ad) and L (5min) data.
Here, $P(D > d)$ indicates the probability that the random variable representing durations, $D$ is greater than the value of the duration on the x-axis, $d$.
A and B share a legend.}
\label{fig:DurDist_ChnspValdata}
\end{figure}

%---Fig: DurDist_AnLday--------------------
\begin{figure}[H]
\centering
\includegraphics[width=\linewidth]{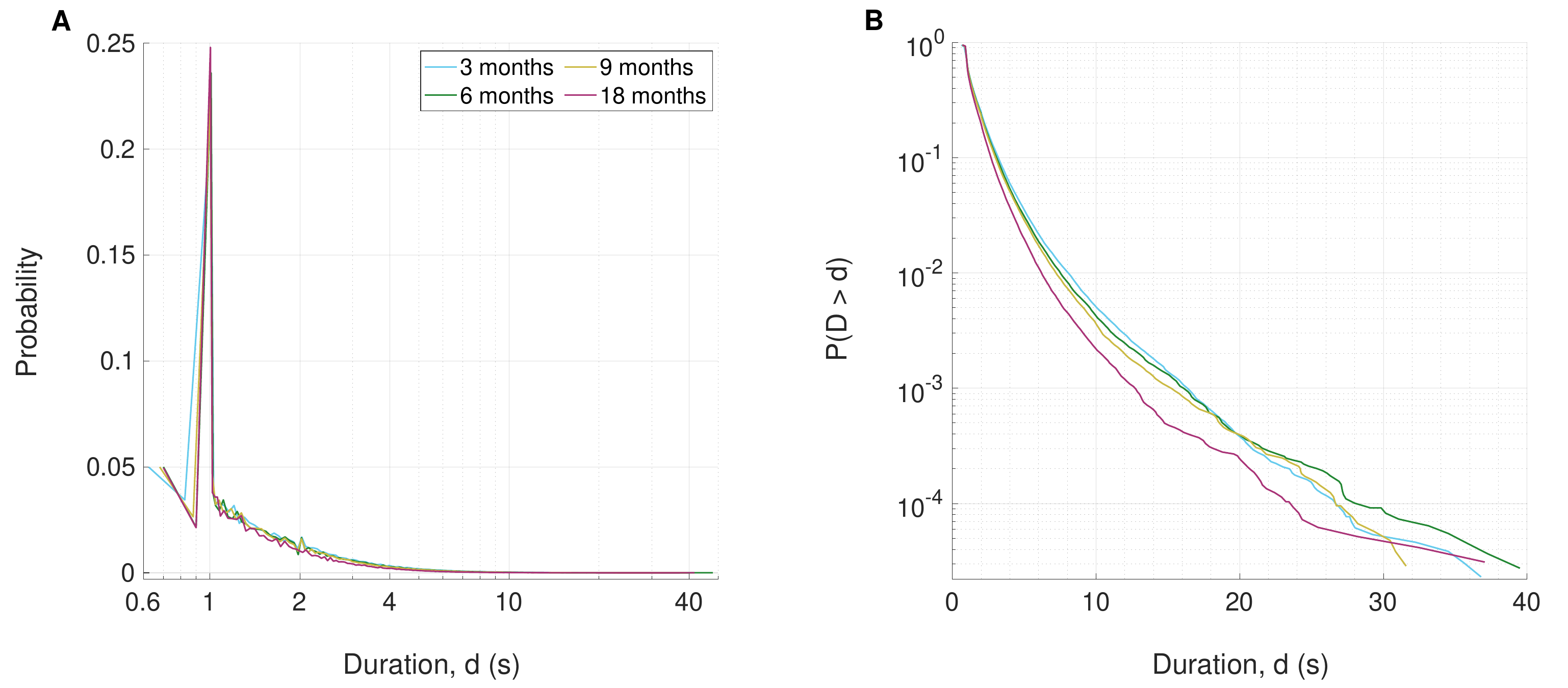}
\caption{\textbf{Adult duration distributions for LENA daylong data. (A)} 
Empirical probability distributions of adult (Ad) durations ($d$) for 3, 6, 9, and 18 months (see legend).
\textbf{(B)} Empirical CCDFs of Ad durations for 3, 6, 9, and 18 months.
Here, $P(D > d)$ indicates the probability that the random variable representing durations, $D$ is greater than the value of the duration on the x-axis, $d$.
A and B share a legend.}
\label{fig:DurDist_AnLday}
\end{figure}

%---Fig: DurDist_AnValdata--------------------
\begin{figure}[H]
\centering
\includegraphics[width=\linewidth]{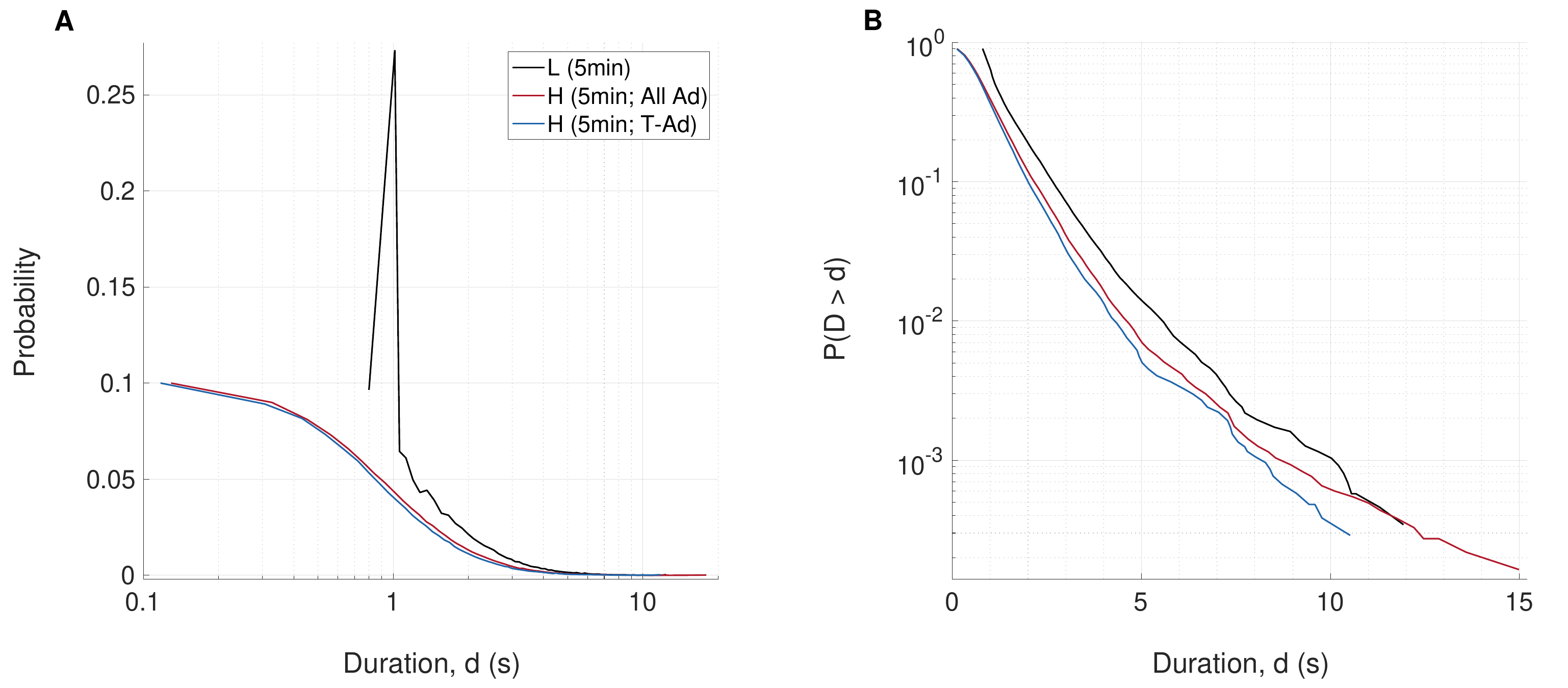}
\caption{\textbf{Adult duration distributions for validation data. (A)} 
Empirical probability distributions of adult (Ad) durations ($d$) for H (5min; All Ad), H (5min; T-Ad), and L (5min) data (see legend).
\textbf{(B)} Empirical CCDFs of Ad durations for H (5min; All Ad), H (5min; T-Ad), and L (5min) data.
Here, $P(D > d)$ indicates the probability that the random variable representing durations, $D$ is greater than the value of the duration on the x-axis, $d$.
A and B share a legend.
}
\label{fig:DurDist_AnValdata}
\end{figure}

%---SECTION: Burstiness--------------------
\section{Burstiness and hierarchical clustering of vocalization event series}\label{sec:burstiness}

%---Fig: BurstinessFig--------------------
\begin{figure}[H]
    \centering
    \includegraphics[width=\linewidth]{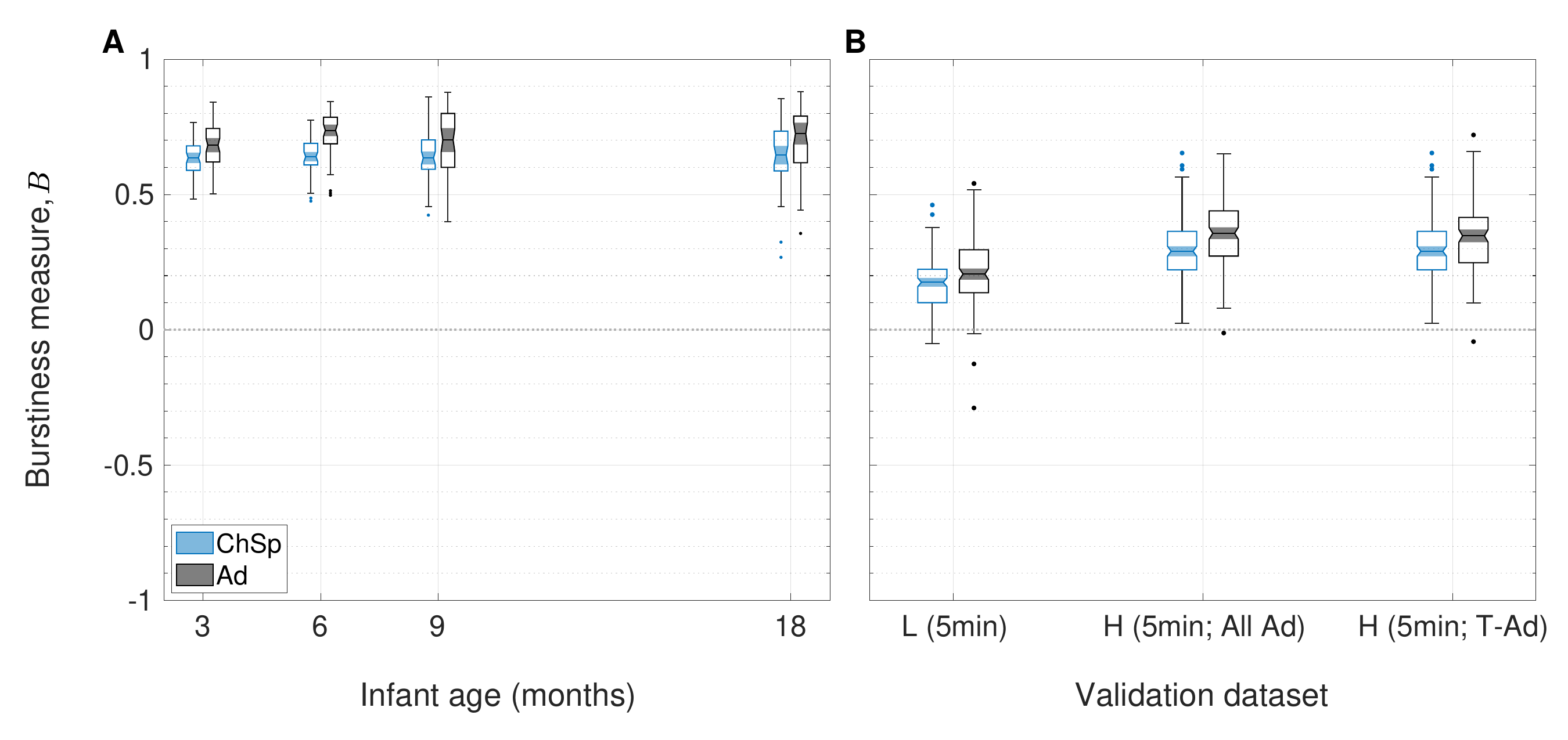}
    \caption{\textbf{Burstiness measures ($B$) for LENA daylong data and validation data at the recording-level, expressed as  box plots.
    (A)} Box plots summarizing burstiness measures, $B$, for infant speech-related (ChSp; blue) and adult (Ad; black) vocalizations, computed at the recording day-level are shown as a function of infant age (months; x-axis) for the LENA daylong data. 
    Box plots whose notches do not overlap have different medians at the 0.05 significance level.
    The dotted horizontal line indicates $B$ = 0, representing a neutral time series while the y-axis limits are set to $B$ = 1 (maximally bursty time series) and $B$ = -1 (periodic time series) \cite{goh2008burstiness_SI}.
    \textbf{(B)} Box plots summarizing burstiness measures for ChSp (blue) and Ad (black) vocalizations, computed at the recording-level are shown for the validation datasets: human listener-labeled 5-minute sections with all adult vocalizations included (H (5min; All Ad)), human listener-labeled 5-minute sections with only infant-directed adult vocalizations included (H (5min; T-Ad)), and corresponding 5-minute sections as labeled by LENA (L (5min)). 
    Validation data has not been separated into age blocks.
    IEIs are estimated for the human listener-labeled datasets after processing overlaps and for all datasets, after merging vocalizations with the same label when separated by 0 s IEIs.
    A and B share a legend as well as the y-axis.
    }
    \label{fig:BurstinessFig}
\end{figure}

To quantify the burstiness of infant and caregiver vocalizations in the daylong and validation datasets, we computed a burstiness measure, $B$ \cite{goh2008burstiness_SI, abney2025advancing} for infant speech-related (ChSp) and adult (Ad) vocalization time series (Fig. \ref{fig:BurstinessFig}).
$B$ is defined as 

\begin{equation}
    B = \frac{\sigma - \mu}{\sigma + \mu}
\end{equation}

where $\sigma$ is the standard deviation of the set of IEIs and $\mu$ is the mean IEI.
Here, $B$ = 1 corresponds to maximally bursty time series, $B$ = -1 corresponds to a perfectly periodic time series, and $B$ = 0 corresponds to a neutral (i.e., random) time series. 
$B$ was computed using the set of IEIs at the recording-level and only for recordings that had at least 10 IEIs for the relevant vocalizer category (i.e., ChSp or Ad).
For both daylong and validation data, IEIs were computed within each sub-recording or 5-minute section as applicable, and were then pooled together to estimate $B$ at the recording level.\\

\begin{figure}[H]
    \centering
    \includegraphics[width = \linewidth]{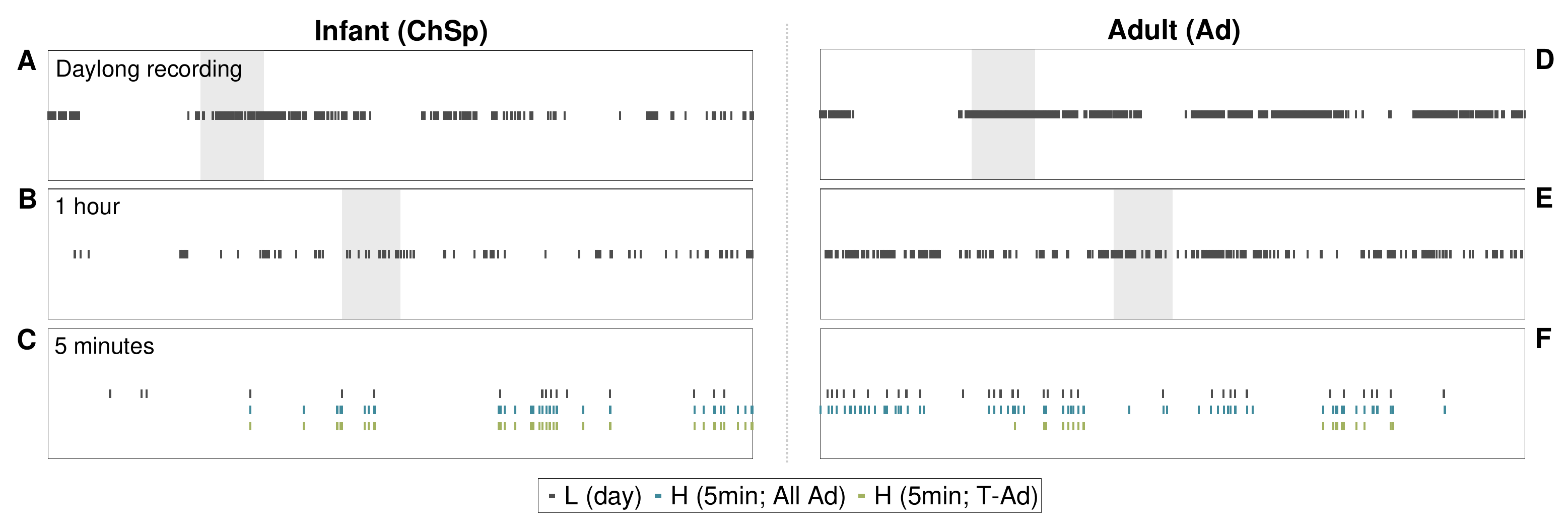}
    \caption{\textbf{Hierarchical clustering in infant speech-related (ChSp) and adult (Ad) vocalizations over the course of a day.
    (A)} The onsets of ChSp vocalizations as identified by LENA over the course of a randomly selected daylong recording ($\sim$ 11 hours; infant age = 9 months) are shown (black vertical dashes), from earlier in the day (left) to later in the day (right). 
    A one hour period within this recording is highlighted in gray. 
    \textbf{(B)} The onsets of ChSp vocalizations as identified by LENA in the one-hour period highlighted in A are shown.
    A 5-minute period within this hour (which corresponds to a randomly selected human listener-labeled 5-minute section within the daylong recording) is highlighted in gray. 
    \textbf{(C)} The onsets of ChSp vocalizations as identified by LENA (black) in the 5-minute period highlighted in B are shown along with ChSp onsets as identified by human listeners for the H (5min; All Ad) (human listener-labeled 5-minute sections with all adult vocalizations included; dark cyan) and H (5min; T-Ad) (human listener-labeled 5-minute sections with only infant-directed adult vocalizations included; green) datasets.
    To ensure that onsets are continuous in time, the daylong recording was selected from the subset of recordings that have no recorder pauses or deletions. 
    \textbf{(D), (E),} and \textbf{(F)} show similar depictions of the adult (Ad) vocalization onsets for the same daylong recording, hour, and 5-minute section as in A, B, and C, respectively. 
    Note that while the ChSp onsets are the same for the H (5min; All Ad) and H (5min; T-Ad) datasets, this is not the case for the Ad onsets. 
    }
    \label{fig:HierarchicalClusteringDemo}
\end{figure}

Burstiness values for the validation datasets tend to be lower than those for the LENA daylong data for both ChSp and Ad vocalizations, likely due to the limited temporal and contextual ranges of the former. 
Nevertheless, values of \textit{B} for the human listener-labeled datasets and the corresponding LENA-labeled 5-minute subset span comparable ranges (Fig. \ref{fig:BurstinessFig}B). 
As previous work has demonstrated \cite{abney2017multiple_SI, warlaumont2022daylong_SI}, the temporal clustering of vocalizations evidenced by the burstiness measure extends across multiple timescales over the course of a day (see Fig. \ref{fig:HierarchicalClusteringDemo} for an illustrative example), resulting in IEI distributions with long tails (see Section \ref{subsec:IEIdistribution}).
Burstiness values for the data shown in Fig. \ref{fig:HierarchicalClusteringDemo} at the various temporal scales represented are given in Table \ref{tab:FigS18BurstinessVals}.

%---Fig: HierarchicalClusteringDemo--------------------

%---Tab: FigS18BurstinessVals--------------------
\begin{table}[H]
    \centering
    \begin{tabular}{lccc}
    \hline
    Data type & Fig. \ref{fig:HierarchicalClusteringDemo} panels &$B$ (ChSp) & $B$ (Ad)\\
    \hline
    L (day)          & A, D & 0.64 & 0.77\\
    L (day), 1 hour  & B, E & 0.29 & 0.25\\
    L (5min)         & C, F & 0.03 & 0.17\\
    H (5min; All Ad) & C, F & 0.28 & 0.31\\
    H (5min; T-Ad)   & C, F & 0.28 & 0.47\\
    \hline
    \end{tabular}
    \caption{\textbf{Burstiness values ($B$) for data at various temporal scales as shown in Fig. \ref{fig:HierarchicalClusteringDemo}.}
    Burstiness values for the LENA daylong data at the day, hour, and 5-minute scales, as well as the human listener-labeled datasets  (H (5min; All Ad), H (5min; T-Ad)) are shown for infant speech-related (ChSp) and adult (Ad) vocalization time series.
    The corresponding panels in Fig. \ref{fig:HierarchicalClusteringDemo} are indicated in the relevant column.
    }
    \label{tab:FigS18BurstinessVals}
\end{table}

\newpage
%---SECTION: Reliability--------------------
\section{Reliability between LENA and human listener labels}\label{sec:Reliability}

\begin{figure}[H]
\centering
\includegraphics[width = \linewidth]{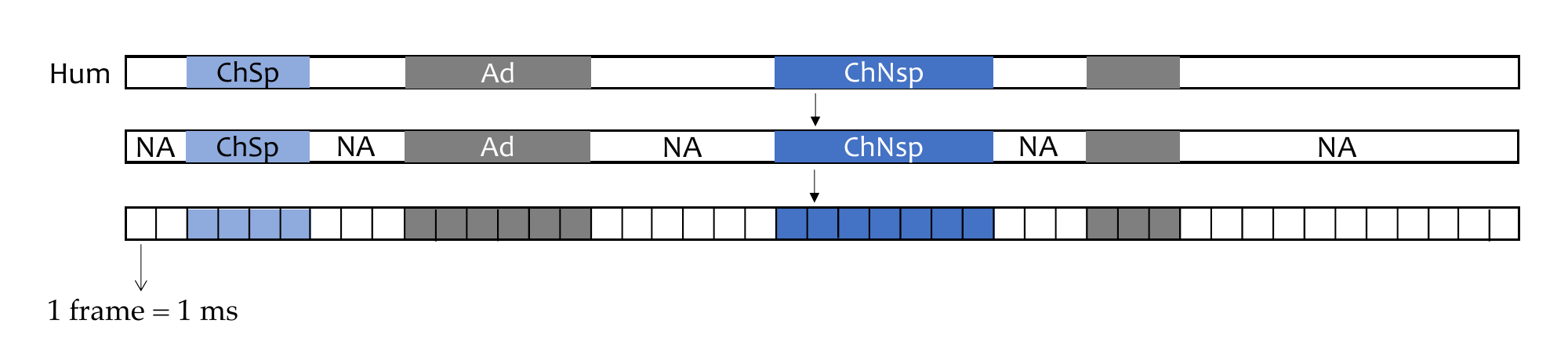}
\vspace{0.5 mm}
\caption{\textbf{Schematic describing labeling error rate estimation.}
A hypothetical sequence of human listener labels (indicated by `Hum' to the left of the sequence) is shown to demonstrate processing steps prior to computing labeling error rates.
Solid-tipped arrows indicate the sequence of processing steps. 
Human listeners identify onsets and offsets of key segments\textemdash infant speech-related (ChSp, light blue), infant non-speech-related (ChNsp, dark blue), and adult (Ad, gray)\textemdash and assign corresponding labels (see Section \ref{subsec:ValDataMethods} for details). 
Note that the label sequence presented here is understood to have already gone through processing to remove overlaps and merge vocalizations of the same type when separated by 0 s IEIs (see Sections \ref{subsec:ValDataMethods}, \ref{sec:OverlapHumanData}, and \ref{subec:0IEIMerge} for details). 
Before estimating error rates, we assign `NA' labels to all segments spanning times between labeled vocalizations (ChSp, ChNsp, Ad) as shown.
While LENA provides exhaustive labels spanning the entire length of recorded audio, human listeners only code for key segments, and there is no information to compare against LENA labels that are not ChSp, ChNsp, or Ad.
Therefore, for both LENA and human listener-labeled data, the NA label serves as a catch-all to indicate that the segment does not carry a key segment label. 
Segments labeled NA are indicated by the white spaces in the sequence.
In the next step, the sequence is divided into `frames' that are 1 ms long. 
The processing steps applied to human listener-labeled data  are also applied to LENA-labeled data before estimating error rates.  
\label{fig:ReliabSchematicPt1}}
\end{figure}

\begin{figure}[H]
\centering
\includegraphics[width = \linewidth]{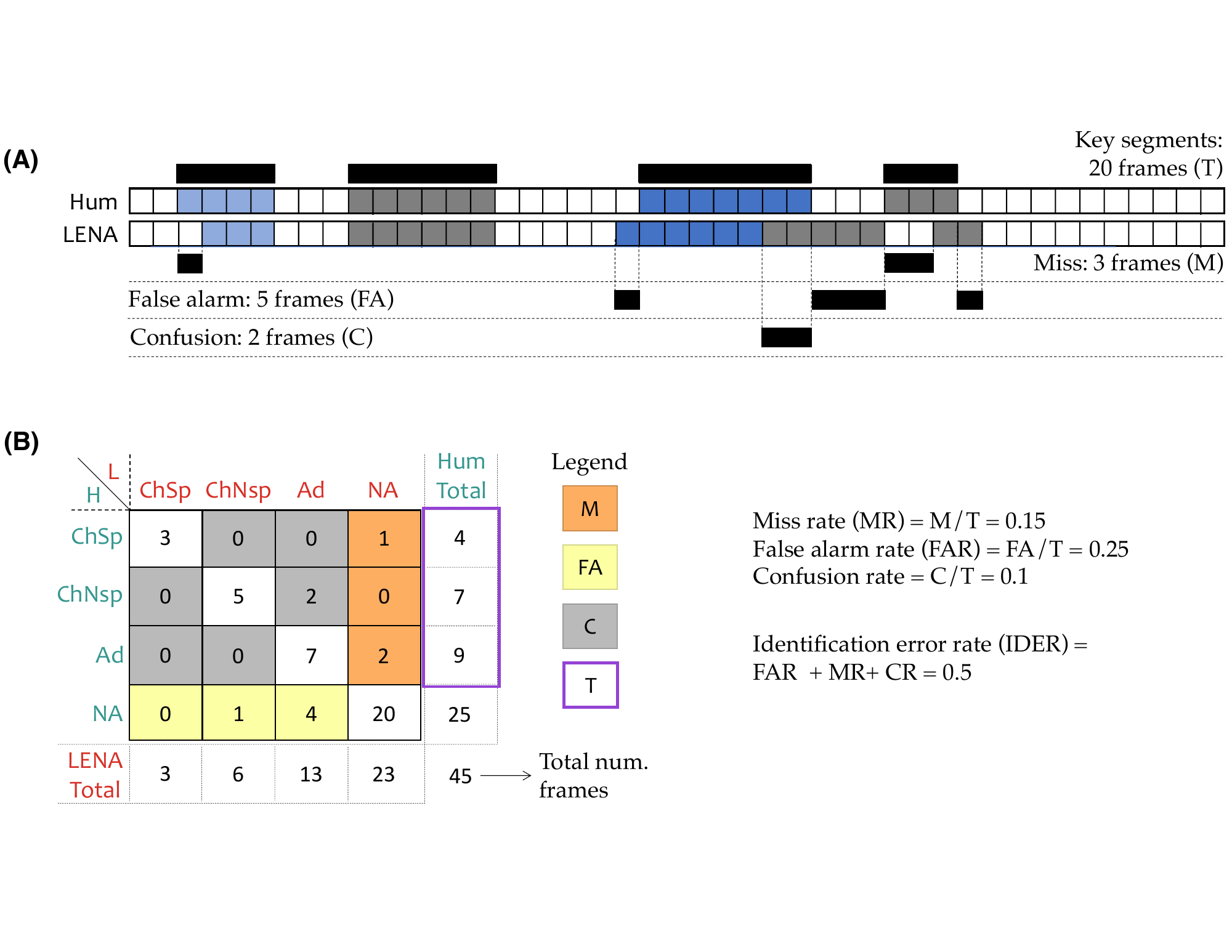} %\vspace{1.5 mm}
\caption{\textbf{Schematic describing labeling error rate estimation (continued). (A)} 
The human listener-labeled sequence from Fig. \ref{fig:ReliabSchematicPt1} (indicated by `Hum' to the left of the sequence) and the corresponding hypothetical LENA-labeled sequence (indicated by `LENA' to the left of the sequence) are presented next to each other with 1 ms frames indicated. 
Frames with ChSp labels are light blue, frames with ChNsp labels are dark blue, frames with Ad labels are gray, and NA frames are white.
The total number of frames identified by the human listener as containing key segments (T) is indicated above the human listener-labeled sequence.
The number of frames with false alarms (FA), where LENA detected key segments but the human listener did not; misses (M), where LENA did not detect a key segment but the human listener did; and confusions (C), where LENA and the human listener detected key segments but disagreed on the key segment label, are indicated below the LENA sequence. 
\textbf{(B)} The confusion matrix for the human listener-labeled sequence and the corresponding LENA labels in A is shown (left side of panel B). 
The row indices, $i$, correspond to human listener labels (green, H) while column indices, $j$, correspond to LENA labels (red, L), both in order of ChSp, ChNsp, Ad, and NA.
The $(i,j)$\textsuperscript{th} element of the confusion matrix is the number of frames that were assigned the $i$\textsuperscript{th} label by the human listener and the $j$\textsuperscript{th} label by LENA.
The diagonal, therefore, gives the number of frames that LENA and the human listener agreed on the label. 
The total number of frames labeled a certain type by the human listener is the sum of the corresponding row, while the total number of frames labeled a certain type by LENA is the sum of the corresponding column. 
The totals for each label type are shown as a separate column (human listener labels) and row (LENA labels) next to the confusion matrix. 
The overall total number of frames is the sum of the human total column or of the LENA total row, and is also indicated. 
The number of misses (M; orange), false alarms (FA; yellow), and confusions (C; gray), as well as the total number of frames identified by the human listener as key segments (T; purple outline) are indicated in or adjacent to the confusion matrix as applicable (see legend next to the matrix).
On the left side of panel B, the miss rate (MR), false alarm rate (FAR), and confusion rate (CR) are computed for the label sequences presented in A.
The identification error rate, which is the sum of MR, FAR, and CR, is also shown. 
\label{fig:ReliabSchematicPt2}}
\end{figure}

To get reliability estimates, we divided human and LENA-labeled 5-minute sections into 1 ms frames following \cite{cristia2020thorough} (see also Fig. \ref{fig:ReliabSchematicPt1}).
Human listener-labeled sections were compared against the corresponding LENA-labeled sections by lining up the start and end times of the 5-minute sections (Fig. \ref{fig:ReliabSchematicPt2}A).
We used 1 ms frames as opposed to the choice of 10 ms frames in \cite{cristia2020thorough} to accommodate the resolution of human listener onset and offsets. 
We computed false alarm rates (FAR), miss rates (MR), confusion rates (CR), and identification error rates  (IDER) as defined in \cite{cristia2020thorough} as well as simple percent agreement and Cohen's kappa values using frame-by-frame comparisons between human listener labels and LENA labels.
We also computed precision and recall estimates as outlined in \cite{cristia2020thorough} and associated confusion matrices (Fig. \ref{fig:ReliabSchematicPt2}B).
These reliability measures were computed for human listener-labeled data with all adult vocalizations included (H (5min; All Ad)), and human listener-labeled data with only child-directed adult vocalizations included (H (5min; T-Ad)).
All reliability measures were computed at the recording level, i.e., aggregated over all 5-minute validation sections in a daylong recording. \\
% When there were multiple recording files b/c a deletion had been made and human annotation had been performed on subrecording level wav and eaf files, corresponding eafs were merged into a single file for reliability estimates (and all other analyses)

%---Fig: ReliabSchematicPt1--------------------

%---Fig: ReliabSchematicPt2--------------------

%---Fig: TotConfMat--------------------
%Note: I simply divided the total number of frames by 1000 (and rounded to the nearest number) to get the number of seconds here, just because the number of frames--being at 1 ms resolution--showed much bigger numbers that are a little harder to look at and make sense of. Also, these matrices are totals across the entire validation data set. 
\begin{figure}[H]
\centering
\includegraphics[width=\linewidth]{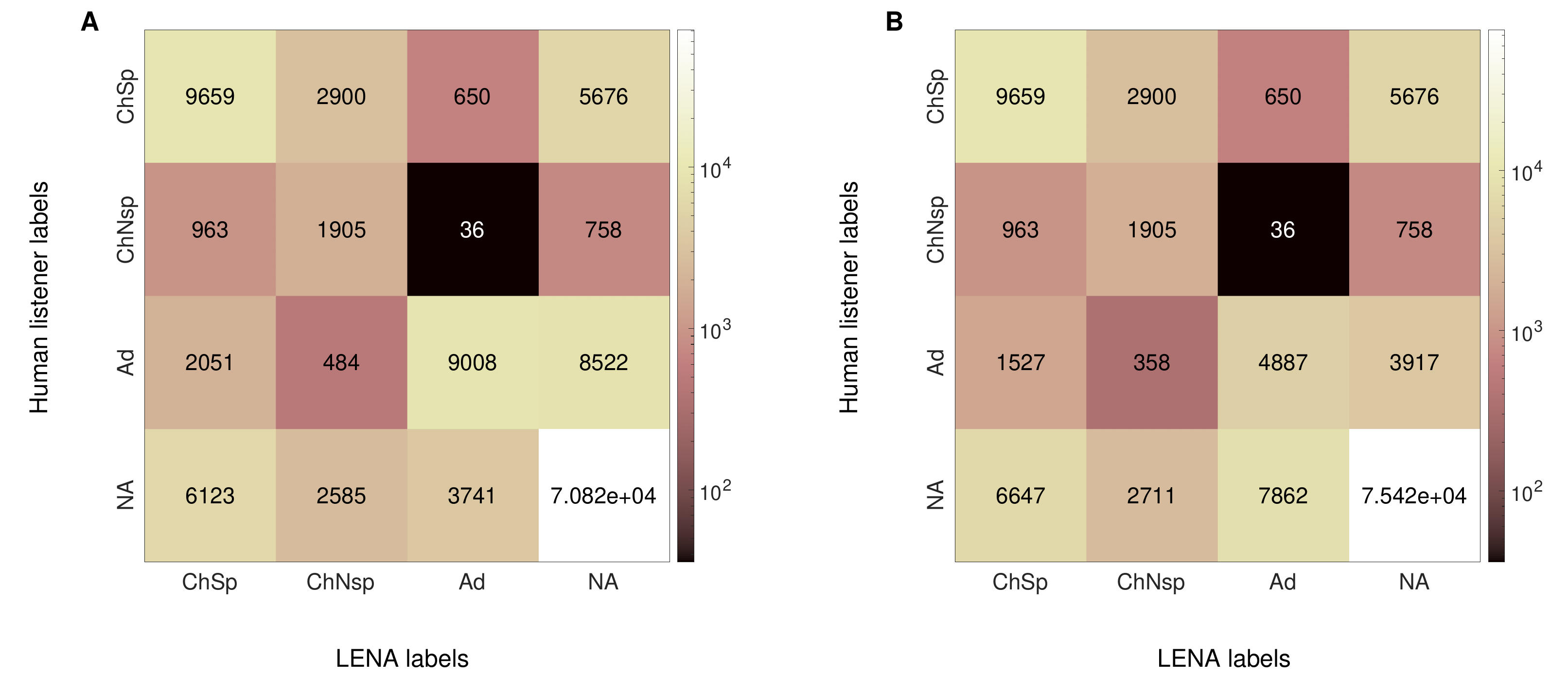}
\caption{\textbf{Confusion matrices for validation data. (A)} 
Confusion matrix (expressed in seconds; see color bar) between LENA (x-axis) and human listener labels (y-axis) when all adult vocalizations are included in the human listener-labeled dataset, H (5min; All Ad).
For example, the first row of the confusion matrix can be summarized as follows: LENA and human listeners agree on infant speech-related (ChSp) labels for 9659 s of audio, while 2900 s of audio labeled as ChSp by human listeners are labeled as infant non-speech-related (ChNsp) by LENA, 650 s of audio labeled as ChSp by human listeners are labeled as adult (Ad) by LENA, and 5676 s of audio labeled as ChSp by human listeners are identified as containing no key segment audio (corresponding to NA labels for reliability estimation purposes)  by LENA. 
Note that the color bar is on a log scale.
\textbf{(B)} Confusion matrix (expressed in seconds; see color bar) between LENA and human listener labels when only child-directed adult vocalizations are included in the human listener-labeled dataset, H (5min; T-Ad).
As expected, only the last two rows in the confusion matrix (corresponding to human listener Ad and NA labels) differ from the confusion matrix in A. 
Further, the number of seconds of audio lost from the human listener-labeled Ad category (with respect to the confusion matrix in A) is gained by the NA category (also with respect to the confusion matrix in A). 
This is as expected, since only considering child-directed adult vocalizations results in an increase in seconds of audio that do not contain any key segment.  
}
\label{fig:TotConfMat}
\end{figure}

%---Tab: Reliability--------------------
\begin{table}[H]
\centering
\begin{tabular}[t]{p{3 cm}C{1.4 cm}C{1.4 cm}cC{1.4 cm}C{1.4 cm}}
\hline
& \multicolumn{2}{c}{H (5min; All Ad)}& &\multicolumn{2}{c}{H (5min; T-Ad)}  \\
\cline{2-3} \cline{5-6}
Reliability measure & Mean & Total & & Mean & Total\\
\hline
Percent Agreement & 72.61 & 72.61 & & 72.94 & 72.99 \\
Cohen's Kappa & 0.44 & 0.46 && 0.40 & 0.42 \\
FAR  & 0.34 & 0.29 && 0.72 & 0.52 \\
MR  & 0.36 & 0.35 && 0.32 & 0.31 \\
CR  & 0.16 & 0.17 && 0.19 & 0.19 \\
IDER  & 0.87 & 0.81 && 1.23 & 1.02 \\
\hline
\end{tabular}
\caption{\textbf{Reliability measures for validation data.} 
Mean and total percent agreements, Cohen's kappa values, false alarm rates (FAR), miss rates (MR), confusion rates (CR) and identification error rates (IDER) are shown when comparing human-listener data with all adult vocalizations included (H (5min, All Ad)) and only child-directed adult vocalizations included (H (5min; T-Ad)) against corresponding LENA-labeled 5-minute sections (L (5min)).
The means are computed over recording day-level reliability measures where each file typically has up to three 5-minute sections (see Sections \ref{subsec:ValDataMethods} and \ref{sec:DataOverview_FileNumAndAge} for details).}
\label{Tab:Reliability}
\end{table}

%---Fig: TotPrecRecallKeySegsFileLvl--------------------
\begin{figure}[H]
\centering
\includegraphics[width=\linewidth]{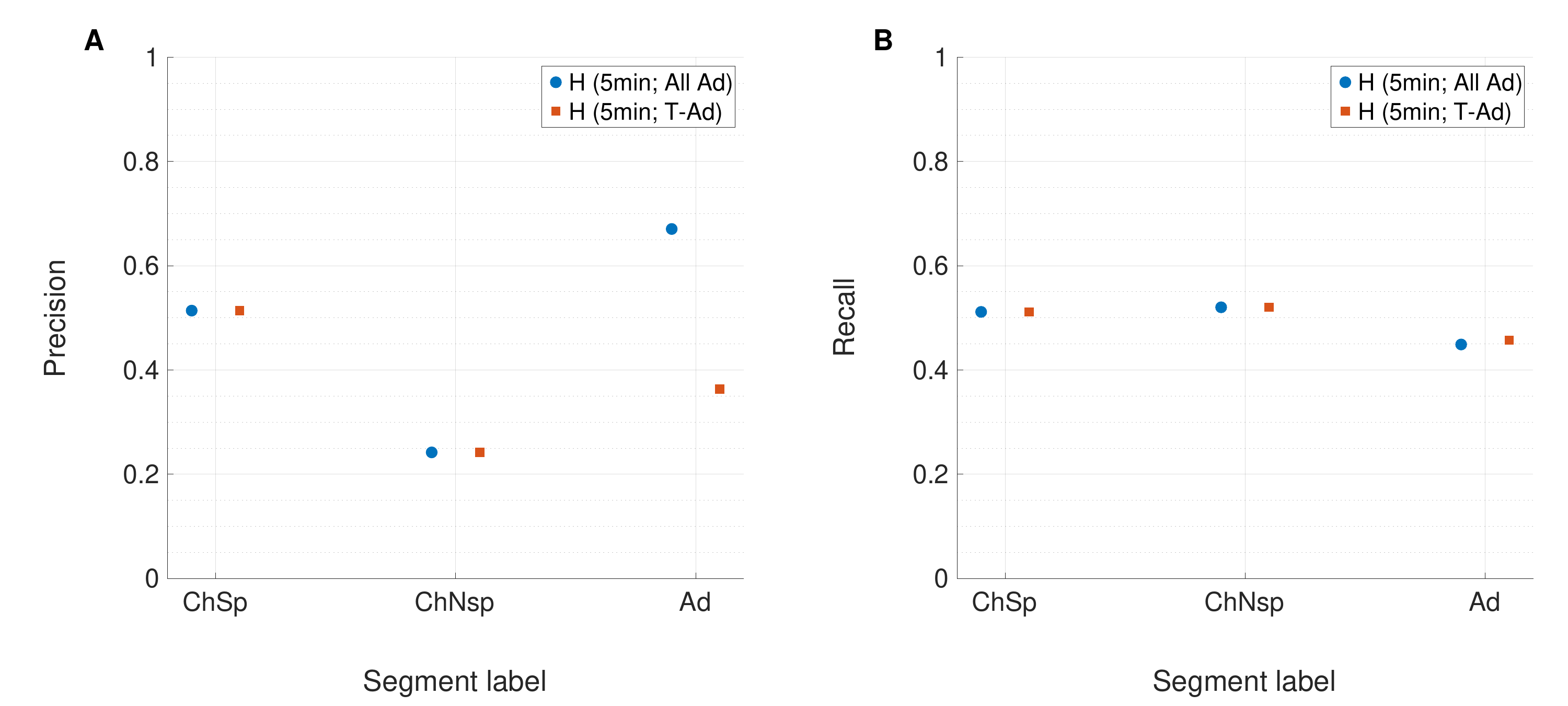}
\caption{\textbf{Precision and recall for key segments for validation data. (A)}
Total precision for key segments (infant speech-related, ChSp; infant non-speech-related, ChNsp; adult, Ad) between LENA and human listener labels are shown when all adult vocalizations are included in the human listener-labeled dataset (H (5min; All Ad); blue) vs. when only child-directed adult vocalizations are included (H (5min; T-Ad); red).
Precision is the number of frames LENA and human listeners agree on a label divided by the total number of frames LENA identifies as that label.
High precision means that frames LENA identifies as a certain category are likely to have also been identified by human annotators as belonging to that category.
\textbf{(B)} Total recall for key segments (ChSp, ChNsp, Ad) between LENA and human listener labels for H (5min; All Ad) and H (5min; T-Ad) data are shown.
Recall is the number of frames LENA and human listeners agree on a label divided by the total number of frames human listeners identify as that label. 
High recall means that the frames human annotators identify as a category are likely to have also been labeled by LENA as that category. 
Precision and recall numbers for the two human listener-labeled datasets differ only for Ad vocalizations, since ChSp and ChNsp labels are the same for both human-labeled datasets.}
\label{fig:TotPrecRecallKeySegsFileLvl}
\end{figure}

%---SECTION: MethDetails--------------------
\section{Methodological details}\label{sec:MethDetails}

%---SUBSECTION: OverlapHumanData--------------------
\subsection{Overlap processing for human listener-labeled data} \label{sec:OverlapHumanData}

%---Fig: OlpProcessSchem--------------------
\begin{figure}[H]
\centering
\includegraphics[width=\linewidth]{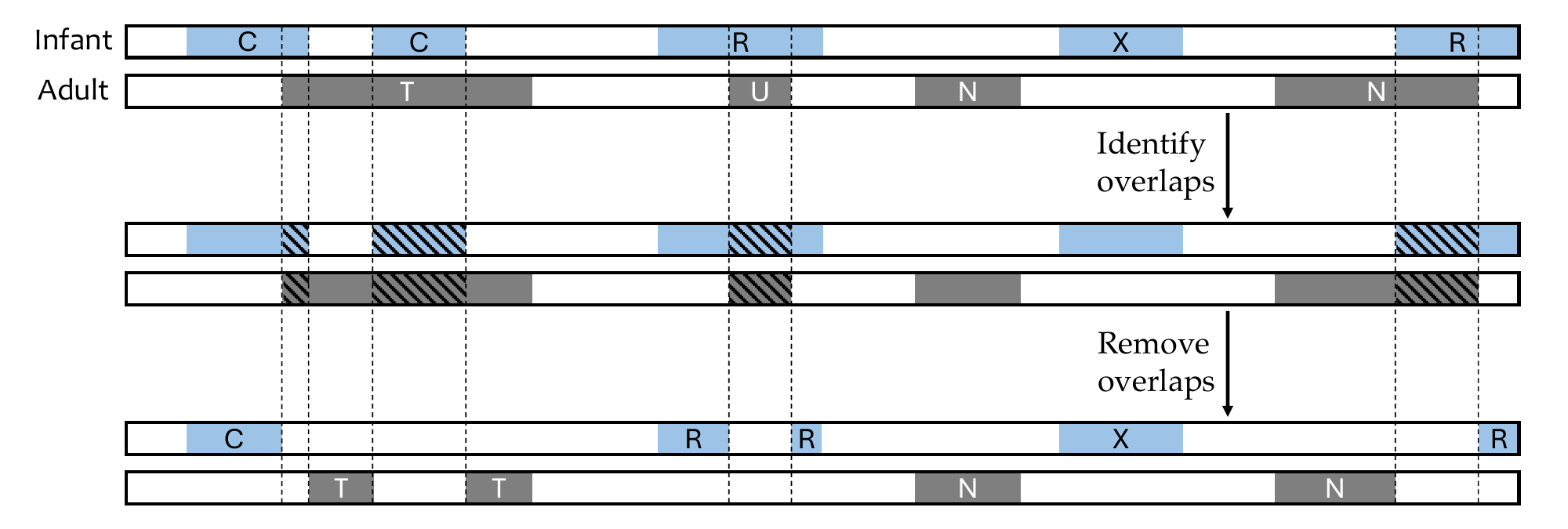}
\caption{\textbf{Schematic describing overlap processing for human listener-labeled data.} 
A hypothetical sequence of infant (speech-related (ChSp) and non-speech-related (ChNsp); blue) and adult (Ad; gray) vocalizations as labeled by human listeners are shown, prior to removing overlaps.
%Note that infant speech-related (ChSp) and infant non-speech related (ChNsp) utterances are treated as part of the infant utterance category here, because these two categories do not overlap.
Annotation labels for infant vocalizations (canonical, C; non-canonical non-reflexive, X; cry, R; or laugh, L) and adult vocalizations (infant-directed, T; not infant-directed, N; or addressee unknown, U) are indicated. 
The sequence of steps in the overlap processing protocol are indicated by the solid arrows.
In the first step, overlaps (crosshatched portions) are identified based on onsets and offsets of vocalizations.
In the second step, these overlaps are removed to obtain non-overlapping infant and adult vocalization time series. 
%Overlaps may be `full overlaps', where an adult utterance fully overlaps with a longer infant utterance (infant utterance 3 and adult utterance II) or vice-versa (as is the case for infant utterance 2 and adult utterance I).
This results in vocalizations being shortened (e.g., first infant vocalization with annotation label C), removed (if they overlap fully with another vocalization; e.g., adult vocalization with annotation label U), or split into multiple vocalization events (e.g., adult vocalization with annotation label T).
For vocalizations that are split into multiple vocalizations or in cases where part of a vocalization is removed as a result of overlap processing, the resultant non-overlapping or truncated vocalization(s) retain the segment (i.e., ChSp, ChNsp, or Ad) and annotation labels of the `parent' vocalization, as shown. 
Overlap processing was performed on the human listener-labeled dataset with all adult vocalizations included (H (5min; All Ad)) prior to extracting the human listener-labeled dataset with only child-directed adult vocalizations included (H (5min; T-Ad)).}
\label{fig:OlpProcessSchem}
\end{figure}

%In some cases, an overlap results in a shorter utterance for one or both overlapping utterances. An example of a case where both overlapping utterances become shorter as a result of overlap processing is shown for infant utterance 5 and adult utterance IV. Finally, there may be overlaps involving more than two utterances, as is the case with overlaps involving infant utterances 1 and 2, and adult utterance I. Note that this example is only one realisation of a more complex overlap scenario. Such complex overlap scenarios can result in there being fewer utterances of one or more utterance types than there were before overlap processing.

%---Tab: OlpSummary--------------------
\begin{table}[H]
\centering
\begin{tabular}[t]{llll}
\hline
 & ChSp & ChNsp & Ad\\
\hline
Number of full overlaps & 1471 & 124 & 534\\
Number of partial overlaps & 2666 & 562 & 3668\\
Total vocalization number before overlap removal & 25522 & 2600 & 17458\\
Total vocalization number after overlap removal & 24341 & 2665 & 18437\\
Total vocalization duration (s) before overlap removal & 21050.18 & 4355.82 & 22937.70\\
Total vocalization duration (s) after overlap removal & 18928.32 & 3670.82 & 20130.84\\
%Total duration (s) removed as overlaps  & 2121.85 & 685 & 2806.85\\
\hline
\end{tabular}
\caption{\textbf{A summary of overlap processing: vocalization numbers and durations.}
The number of full and partial overlaps as well as total numbers and durations of vocalizations before and after removing overlaps are shown for infant speech-related (ChSp), infant non-speech-related (ChNsp), and adult (Ad) vocalizations. 
%Because of the way overlaps work, removed vocalization durations are double counted. For eg. if an Adult and CHNSP vocalization overlap, the removed portion gets counted for both vocalization types, as reported in this table. Also, there are different kinds of overlaps, as demonstrated in teh schematic, which is why the number of vocs before and after olp processing isn't non-intuitive, but hopefully, the number of full and partial overlaps help. Essentially there are far more full ChSp overlaps than Ad (even after accounting for the number of these types of vocs)
A full overlap is when the entire vocalization in question is removed as part of overlap processing (e.g., adult vocalization with annotation label U in Fig. \ref{fig:OlpProcessSchem}). 
%This is counted towards the number of full overlaps for adult vocalizations.
A partial overlap is when part of the vocalization in question is removed during overlap processing (e.g., both infant vocalizations with annotation label R in Fig. \ref{fig:OlpProcessSchem}).
%This is counted towards the number of partial overlaps for infant vocalizations.
}
\label{Tab:OlpSummary}
\end{table}
%Overlap processing was done BEFORE extracting the child-directed utterances only adult data

\newpage
%---SUBSECTION: 0IEIMerge--------------------
\subsection{Merging vocalizations separated by 0 s IEIs}\label{subec:0IEIMerge}

It was possible for both LENA and human listener labels to have vocalizations of the same type perfectly adjacent to each other, with no time between the offset of one vocalization and the onset of the next vocalization of the same type. 
These were identifiable in our analysis pipeline as cases of 0 s IEIs. 
Prior to all statistical analyses (and in the case of human listener-labeled data, after overlap processing), we merged all vocalizations of the same type separated by 0 s IEIs.
While most of these merges involved a pair of vocalizations being merged into one vocalization, there were also cases where a sequence of more than two vocalizations separated by 0 s IEIs was merged into one vocalization.
We refer to the latter as a merge `chain'.

%---Tab: 0IEIMerge--------------------
\begin{table}[H]
\centering
\begin{tabular}[t]{llll}
\hline
 & ChSp & ChNsp & Ad\\
\hline
L (day) & 10950	& 5897	& 47197 \\
L (5min) & 1083 &	257	& 816\\
H (5min: All Ad) & 8& 5	& 141\\
H (5min; T-Ad) & 8 & 5 & 15\\
\hline
\end{tabular}
\caption{\textbf{A summary of 0 s IEI merges.}
The number of instances where 2 or more vocalizations of the same type separated by 0 s IEI(s) were merged into one vocalization are shown by vocalization type (infant speech-related, ChSp; infant non-speech-related, ChNsp; adult, Ad) for LENA daylong data (L (day)) as well as validation data (human listener-labeled data with all adult vocalizations included, H (5min; All Ad); human listener-labeled data with only infant-directed adult vocalizations included, H (5min; T-Ad); and corresponding LENA-labeled 5-minute sections, L (5min)).
For a summary of the number of such vocalization merges involving more than 2 vocalizations, see below.
}
\label{Tab:0IEIMerge}
\end{table}

%---Fig: VocMergeChainsLdayL5min--------------------
\begin{figure}[H]
\centering
\includegraphics[width=\linewidth]{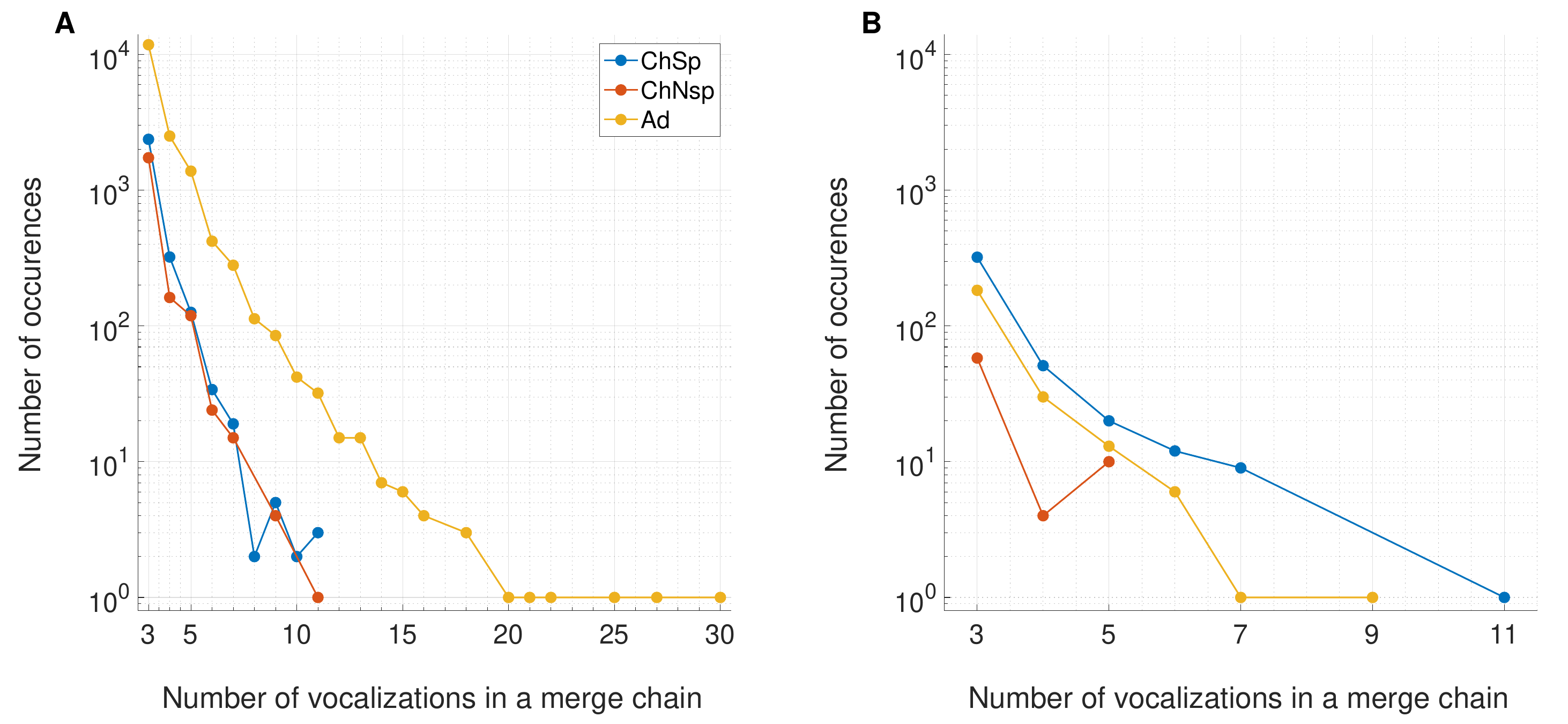}
\caption{\textbf{Frequency distributions of vocalization merge chains. (A)} 
The number of instances (y-axis) of vocalization merge chains\textemdash where more than 2 vocalizations of the same type separated by 0 s IEIs were merged into one vocalization\textemdash are shown as a function of the number of vocalizations in a merge chain (x-axis), for infant speech-related (ChSp, blue), infant non-speech-related (ChNsp, red), and adult (Ad, yellow) vocalizations in the LENA daylong  (L (day)) data. 
\textbf{(B)} The number of vocalization merge chains in the LENA-labeled validation subset  (L (5min) data) are shown as a function of the number of vocalizations in a merge chain for ChSp, ChNsp, and Ad vocalizations.
A and B share a legend.}
\label{fig:VocMergeChainsLdayL5min}
\end{figure}
%These plots start at 3 vocs in a merge chain because the shortest chain is 3 vocs long. Also, A and B have the same Y axis scale and limits. 

\vspace{-2 mm}
\noindent
Unlike both LENA-labeled datasets (L (day) and L (5min)), all vocalization merge chains for human listener-labeled data had three vocalizations.
For H (5min; All Ad) data, there were 10 such merge chains for adult vocalizations, while for H (5min; T-Ad) data, there was only one merge chain for adult vocalizations.
For both H (5min; All Ad) and H (5min; T-Ad) datasets, there was one vocalization merge chain for ChSp vocalizations and one merge chain for ChNsp vocalizations. 

%Note: a merge chain of n vocs has n-1 individual 'merge events'. A merge event as defined here may or not result in a voc that has non-0 IEIs with preceding or succeedings vocs. However, in the summary table, one merge chain is counted as one merge, resulting in a single merged voc

%---SUBSECTION: ResponseAnalysis--------------------
\subsection{Response computation and analysis}\label{subsec:ResponseAnalysis}

\vspace{-4 mm}
%---Fig: ExtendedSchematics_Pt1--------------------
\begin{figure}[H]
\centering
\includegraphics[width = 0.82\textwidth]{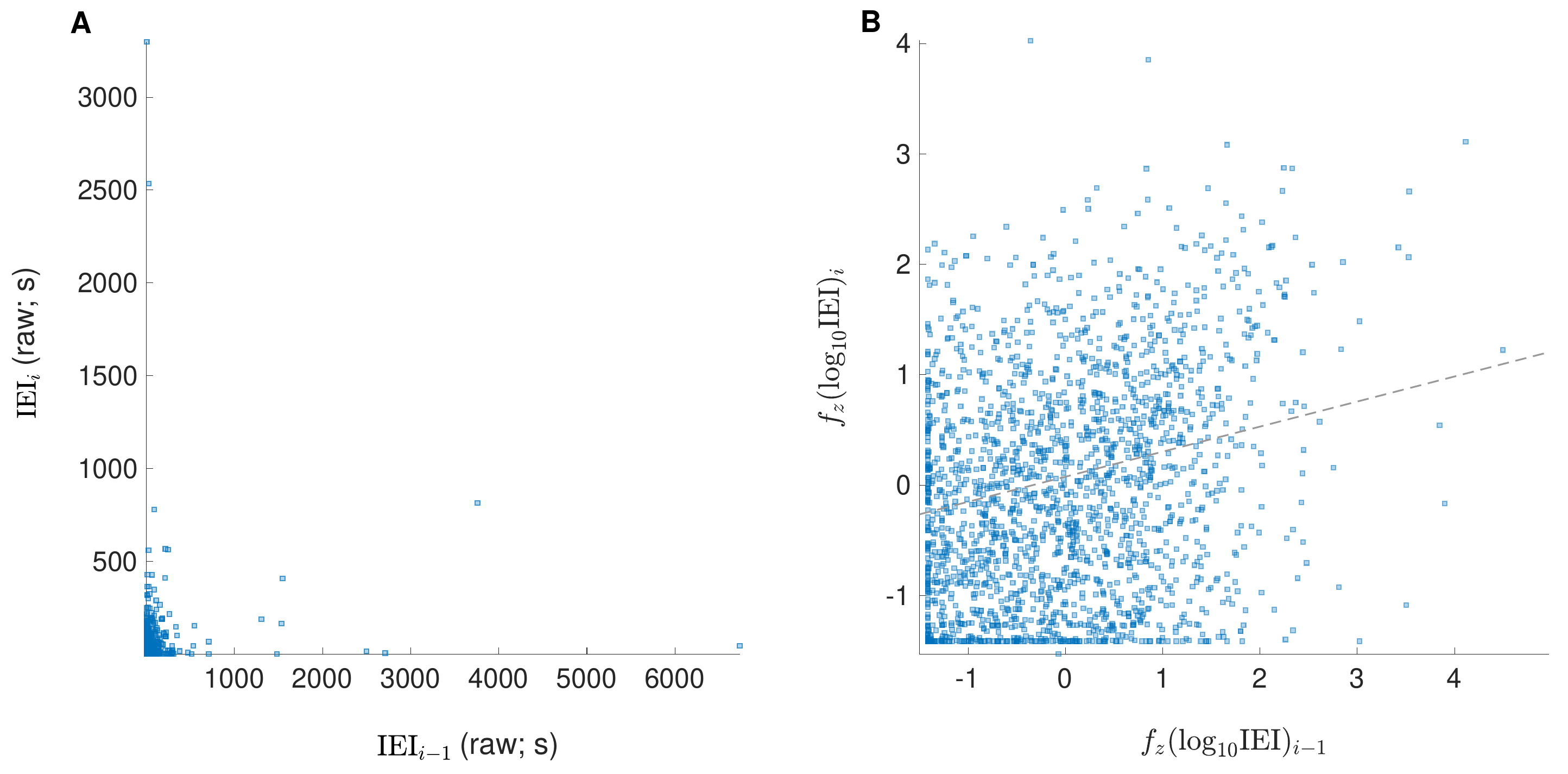} 
\caption{\textbf{A demonstration of the effect of IEI transformation on data distribution using infant IEIs at 18 months. (A)}   
A scatter plot of current infant speech-related (ChSp) IEIs (IEI$_i$) plotted against previous ChSp IEIs (IEI$_{i-1}$) is shown using LENA daylong data (L (day)) from infants at 18 months.
The infant age represented here was chosen randomly. 
To facilitate visualization, the plot displays 2000 randomly chosen points from the full set of daylong LENA infant IEIs at 18 months.
Note that the IEIs depicted here are `raw' IEIs, prior to log-transformation and normalization.
\textbf{(B)} is a scatter plot showing the positive correlation between successive ChSp IEIs for infants at 18 months, after IEIs are log-transformed and normalized with respect to the full 18 month L (day) infant IEI dataset. 
The normalization is indicated by the $f_z$ function.
To remain consistent with panel A, the plot displays raw IEIs corresponding to the same 2000  data points in A.
The dashed line represents $\overline{\rm IEI}$ (see Fig. \ref{fig:Schematic}B in the main text) and is obtained by regressing current IEIs over previous IEIs for all IEIs within the 18 month daylong infant data. 
Note that association with response is not part of this analysis step.
Thus, the choice of $T_{\rm resp}$ is not relevant to this analysis step. 
This regression is used to compute current IEI residuals which go into our response analyses (see Section \ref{subsec:StatTests} and Fig. \ref{fig:ExtendedSchematics_Pt2} for details).
}
\label{fig:ExtendedSchematics_Pt1}
\end{figure}

\vspace{-3 mm}
%---Fig: ExtendedSchematics_Pt2--------------------
\begin{figure}[H]
\centering
\includegraphics[width=0.82\linewidth]{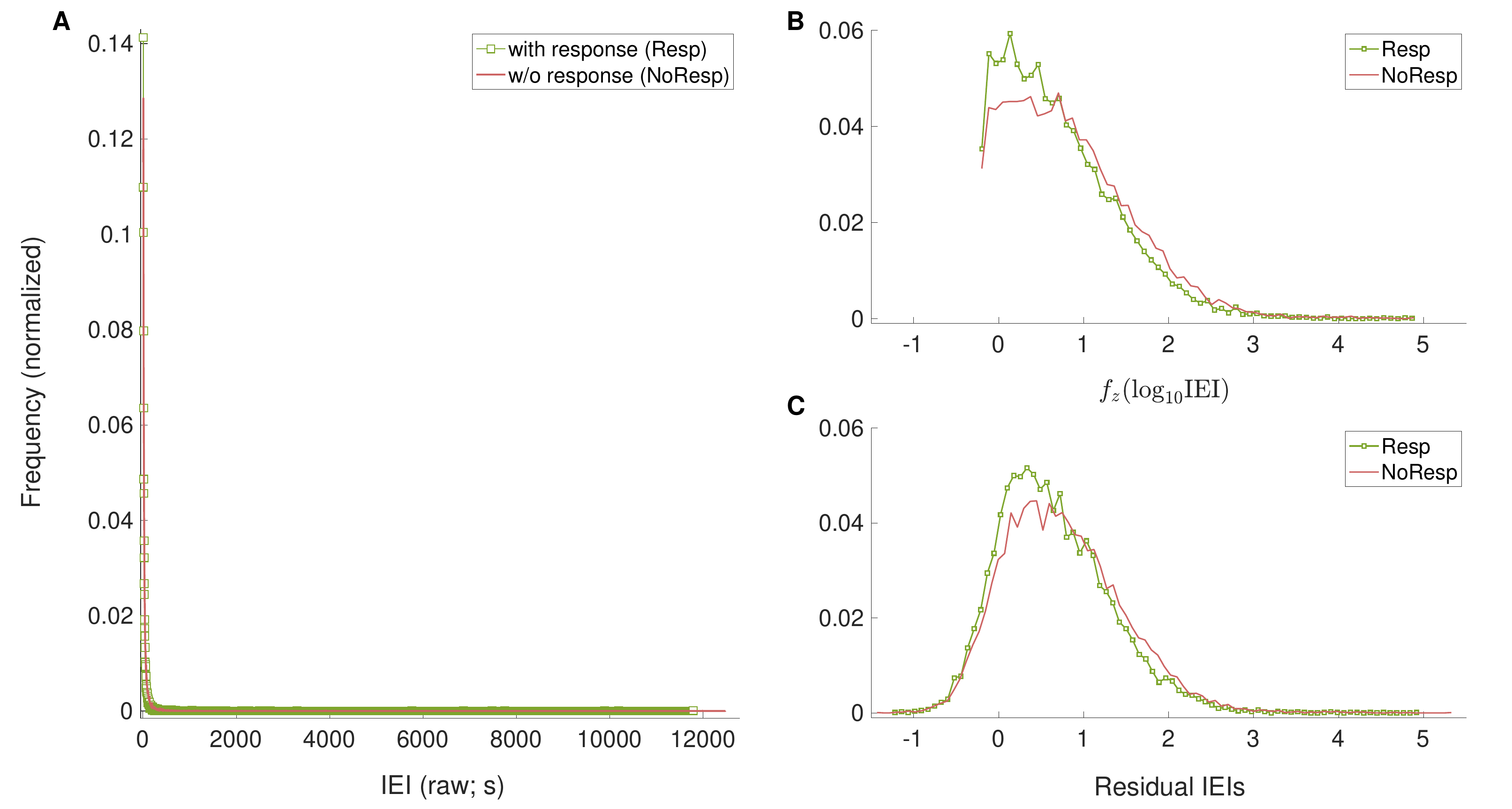}
\caption{\textbf{A demonstration of the effect of residual estimation on data distribution for Resp and NoResp IEIs ($T_{\rm resp} = 5$ s) using infant IEIs at 18 months. (A)}
shows infant speech-related (ChSp) IEI distributions at 18 months for LENA daylong data, prior to log-transformation and normalization. 
Data has been separated into IEIs associated with an adult response (Resp; green) and IEIs not associated with an adult response (NoResp; red)  for a response window ($T_{\rm{resp}}$) value of 5 s as an illustrative example.
IEIs associated with NA responses are not represented.
\textbf{(B)} uses the same data to show the distributions of Resp (green) and NoResp (red) IEIs after log-transformation and normalization.
\textbf{(C)} uses the same data to show the distributions of Resp (green) and NoResp (red) IEI residuals.
All IEIs were log-transformed and then normalized with respect to the full 18 month daylong infant IEI dataset (see Section \ref{subsec:StatTests} and Fig. \ref{fig:ExtendedSchematics_Pt1}B) prior to residual computation.
In A, B, and C, each distribution has been normalized independently.
The number of bins for each distribution was determined separately using the Freedman-Diaconis rule \cite{freedman1981histogram}.
As such, not all distributions have the same number of bins.
%B and C have the same x-axis scale and have the same x-axis limits.
}
\label{fig:ExtendedSchematics_Pt2}
\end{figure}

%---Fig: InfIEI_WrWorResidsByAge--------------------
\begin{figure}[H]
\centering
\includegraphics[width=\linewidth]{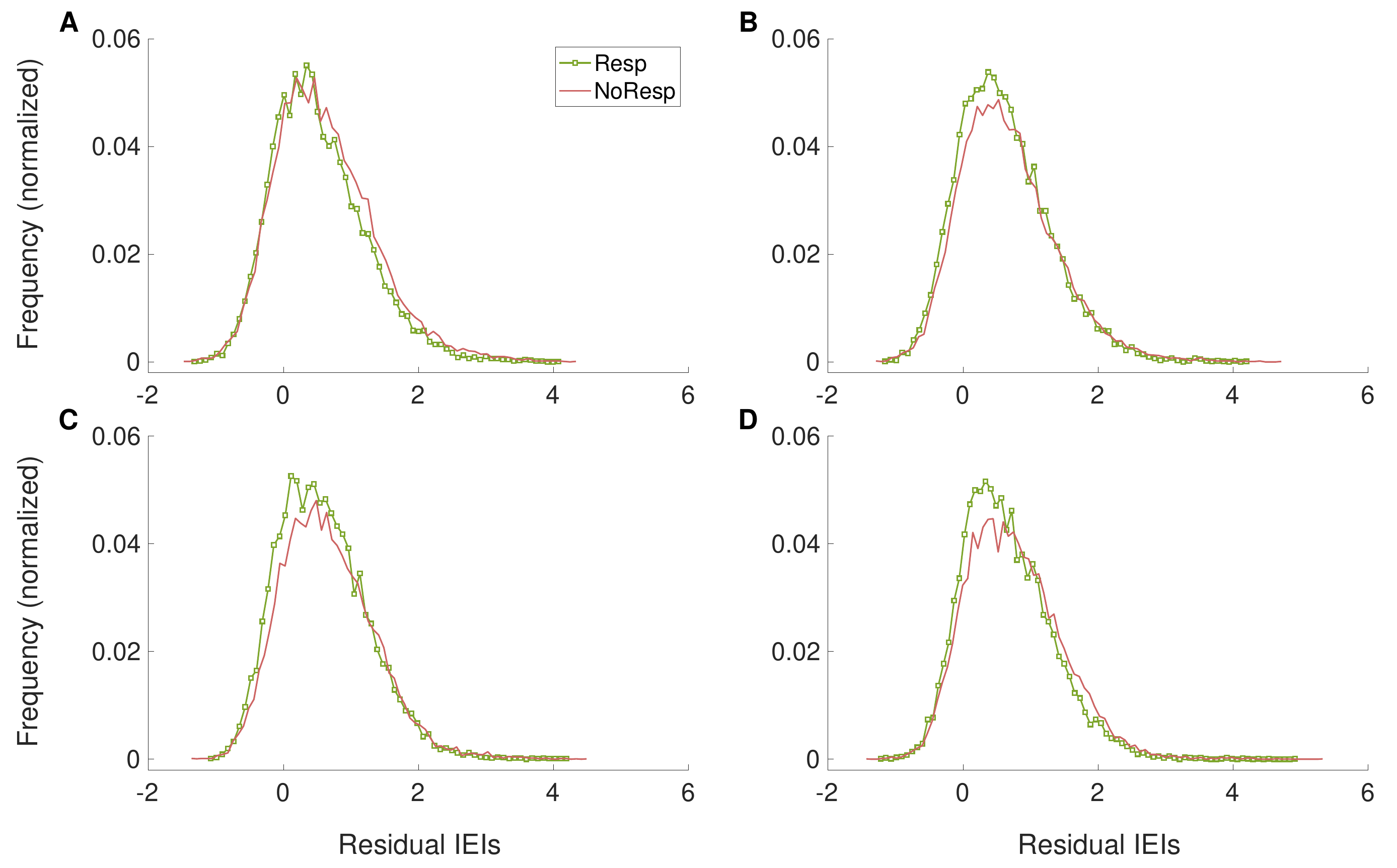}
\caption{\textbf{Infant Resp and NoResp IEI residual distributions for LENA daylong data for $T_{\rm resp}$ = 5 s. (A)} 
Distributions of infant speech-related (ChSp) IEI residuals separated into IEIs associated with an adult response (Resp; green) and IEIs not associated with an adult response (NoResp; red) are shown for LENA daylong data using data from infants at 3 months, for $T_{\rm resp}$ = 5 s. 
Residuals for IEIs associated with NA responses are not shown.
Resp and NoResp distributions have been normalized independently. 
\textbf{(B)}, \textbf{(C)}, and \textbf{(D)} show similar plots for Resp and NoResp ChSp IEI residuals for infants at 6, 9, and 18 months, respectively.
All plots share a legend.
The number of bins for each distribution was determined separately using the Freedman-Diaconis rule.
As such, not all distributions have the same number of bins.}
\label{fig:InfIEI_WrWorResidsByAge}
\end{figure}

%---Fig: AdultIEI_WrWorResidsByAge--------------------
\begin{figure}[H]
\centering
\includegraphics[width=\linewidth]{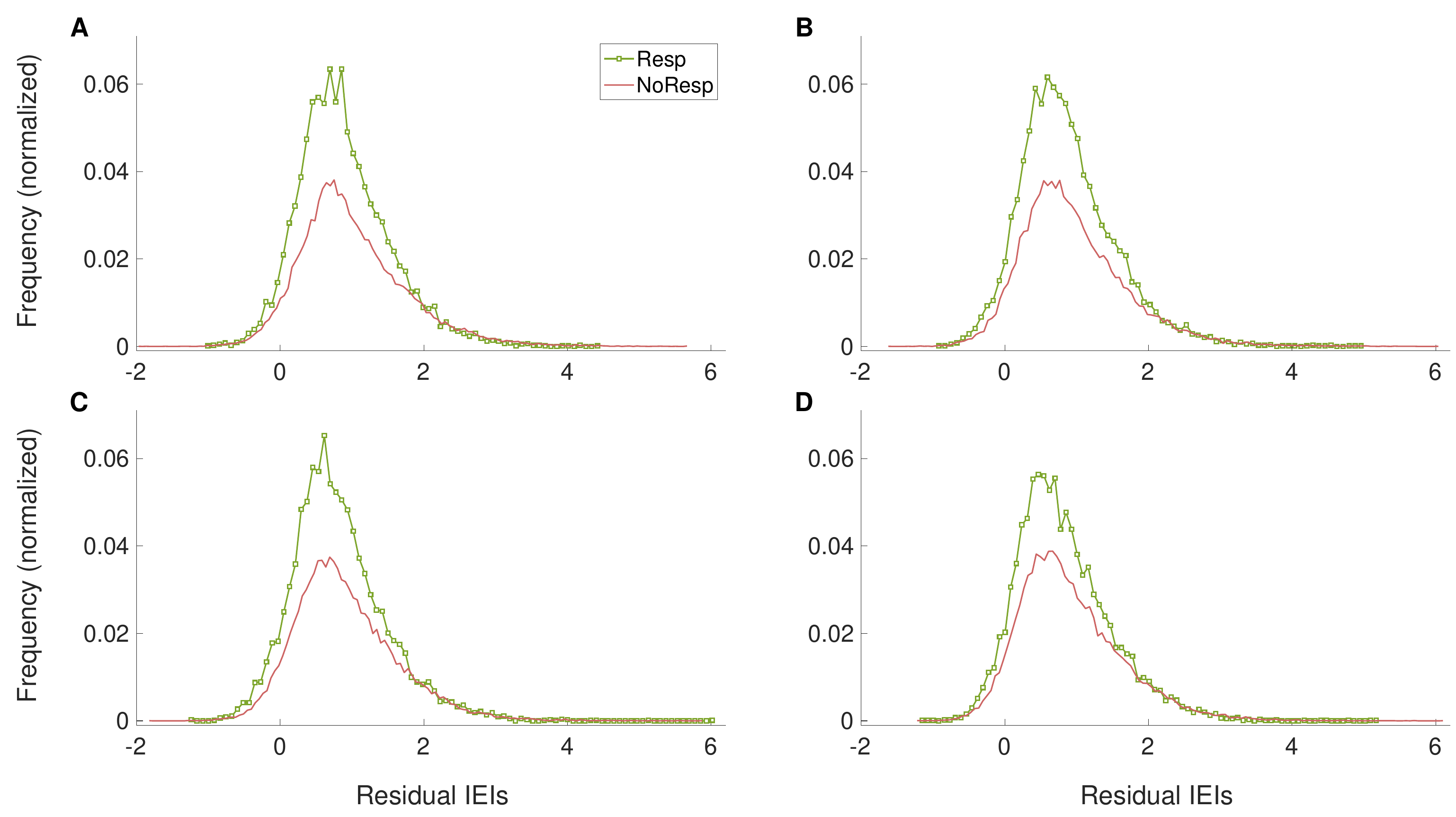}
\caption{\textbf{Adult Resp and NoResp IEI residual distributions for LENA daylong data for $T_{\rm resp}$ = 5 s. (A)} 
Distributions of adult IEI residuals separated into IEIs associated with an infant response (Resp; green) and IEIs not associated with an infant response (NoResp; red) are shown for LENA daylong data using data when infants are 3 months old, for $T_{\rm resp}$ = 5 s. 
Residuals for IEIs associated with NA responses are not shown.
Resp and NoResp distributions have been normalized independently. 
\textbf{(B)}, \textbf{(C)}, and \textbf{(D)} show similar plots for Resp and NoResp adult IEI residuals when infants are 6, 9, and 18 months old, respectively.
All plots share a legend.
The number of bins for each distribution was determined separately using the Freedman-Diaconis rule.
As such, not all distributions have the same number of bins.}
\label{fig:AdultIEI_WrWorResidsByAge}
\end{figure}

%---Fig: NumResponses_Ldaylong--------------------
\begin{figure}[H]
\centering
\includegraphics[width=\linewidth]{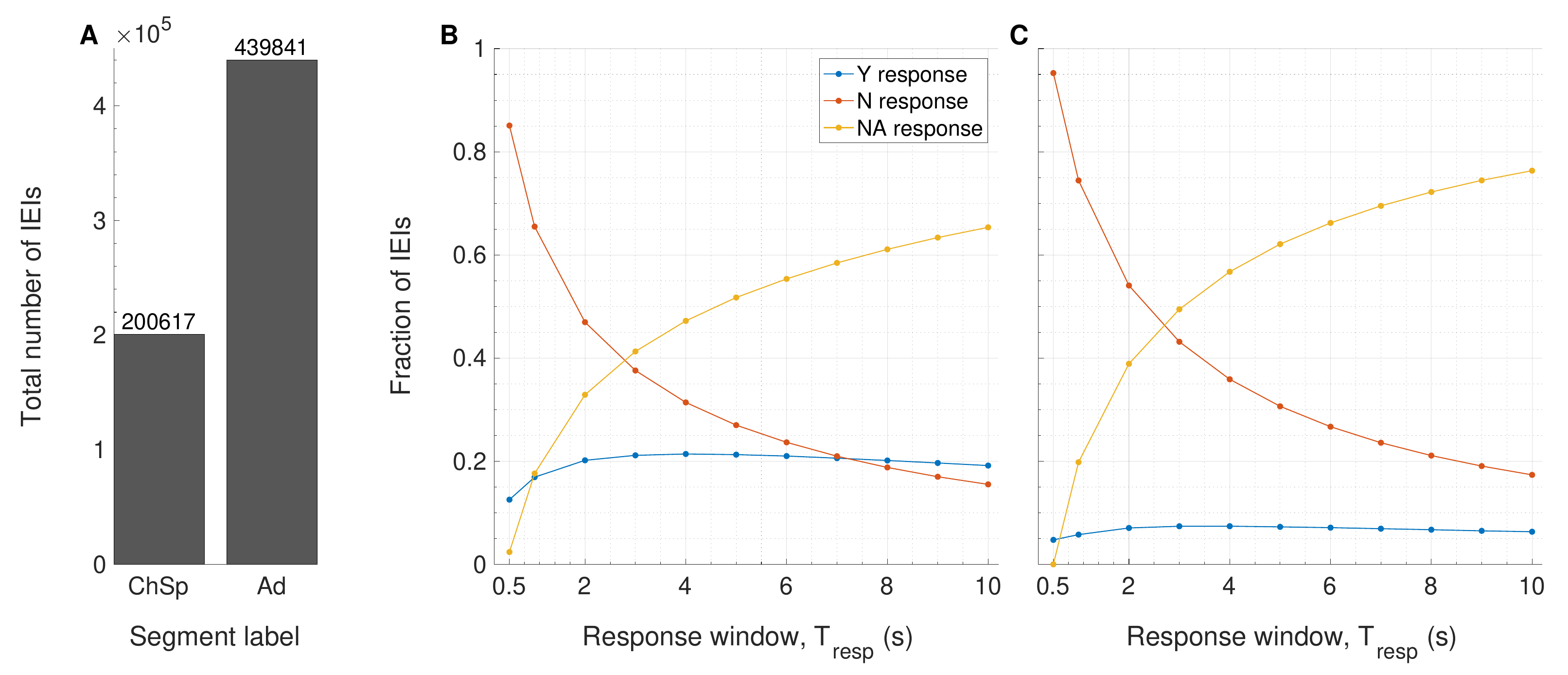}
\caption{\textbf{Number of IEIs associated with different response types for LENA daylong data. (A)} 
Total number of infant speech-related (ChSp) and adult (Ad) IEIs are shown as a bar plot.
%The total number of IEIs are also provided as labels for the relevant bars.
\textbf{(B)} Fractions of ChSp IEIs associated with an adult response (blue, Y response), not associated with an adult response (red, N response), and associated with NA response (yellow) are shown as a function of response window, $T_{\rm{resp}}$ (x-axis, seconds).
All fractions are computed with respect to the total number of ChSp IEIs.
\textbf{(C)} Fractions of Ad IEIs associated with a ChSp response (blue, Y response), not associated with a ChSp response (red, N response), and associated with NA response (yellow) are shown as a function of response window, $T_{\rm{resp}}$.
All fractions are computed with respect to the total number of Ad IEIs.
B and C share the y-axis and legend.}
\label{fig:NumResponses_Ldaylong}
\end{figure}
%Note: Re: Min durations of vocs/sound types and how that translates as number of NA responses here vs how that isn't the case for H5min data fig below--essentially, the fact that min sound duration is around 600 to 1000 ms for LENA means that 0.5 response window has practically no NA reponses. Also note that NA responses include ALL IEIs less than or equal to response window and for large response windows, we still get sig results even with 50-60 percent of data excluded. 

%---Fig: NumIEIsValdata--------------------
\begin{figure}[H]
\centering
\includegraphics[width=0.5\linewidth]{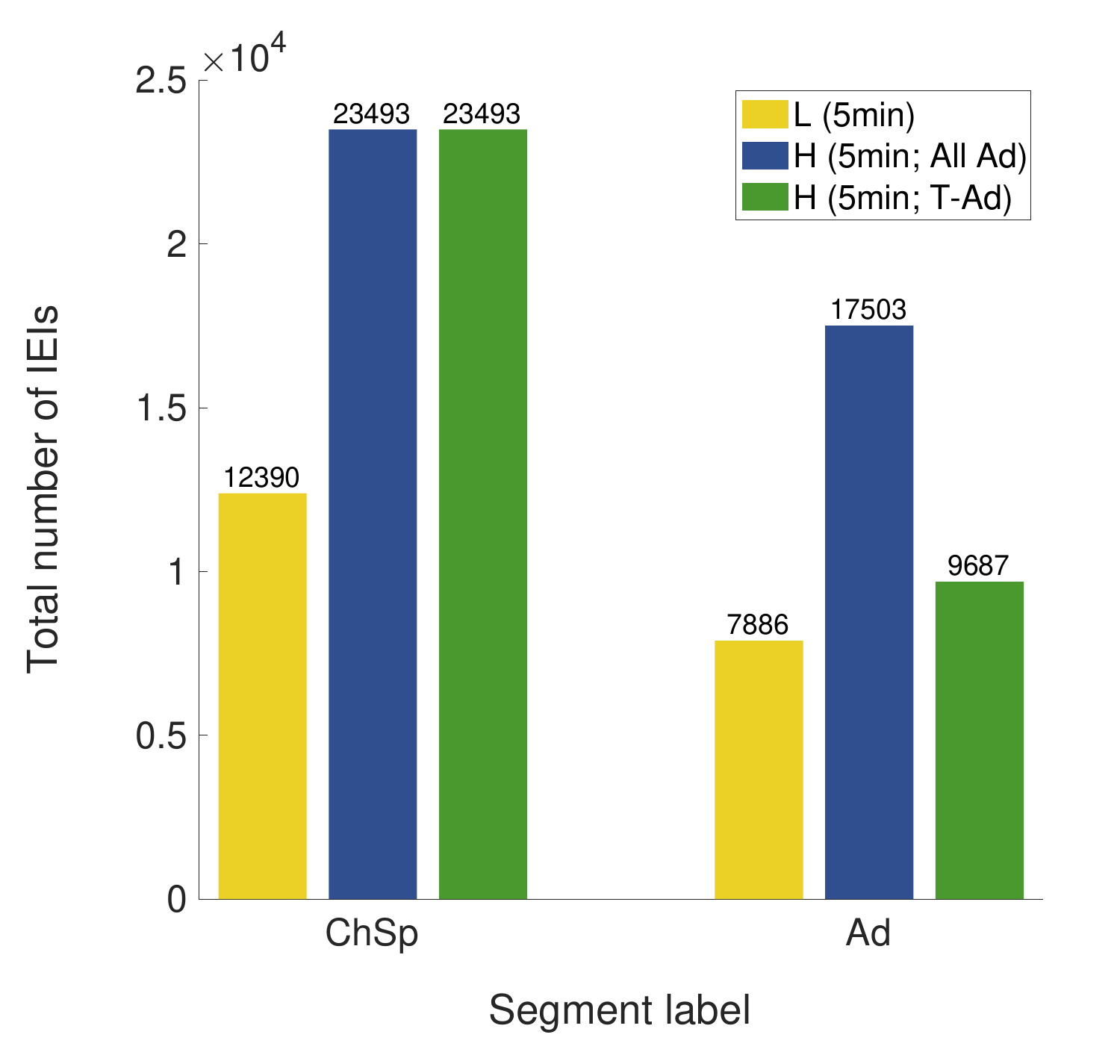}
\caption{\textbf{Total number of validation data IEIs by vocalization type.}
The total number of infant speech-related (ChSp) and adult (Ad) IEIs are shown as bar plots for the validation data: human listener-labeled 5-minute sections with all adult vocalizations (H (5min; All-Ad); blue); human listener-labeled 5-minute sections with only child-directed adult vocalizations included (H (5min; T-Ad); green); and the corresponding LENA subset of 5-minute sections (L (5min); yellow).
Note that the number of ChSp IEIs are the same for H (5min; All Ad) and H (5min; T-Ad) data since only the adult vocalizations differ between those datasets.}
\label{fig:NumIEIsValdata}
\end{figure}

%---Fig: NumIEIsResponseTypes_Valdata--------------------
\begin{figure}[H]
\centering
\includegraphics[width=\linewidth]{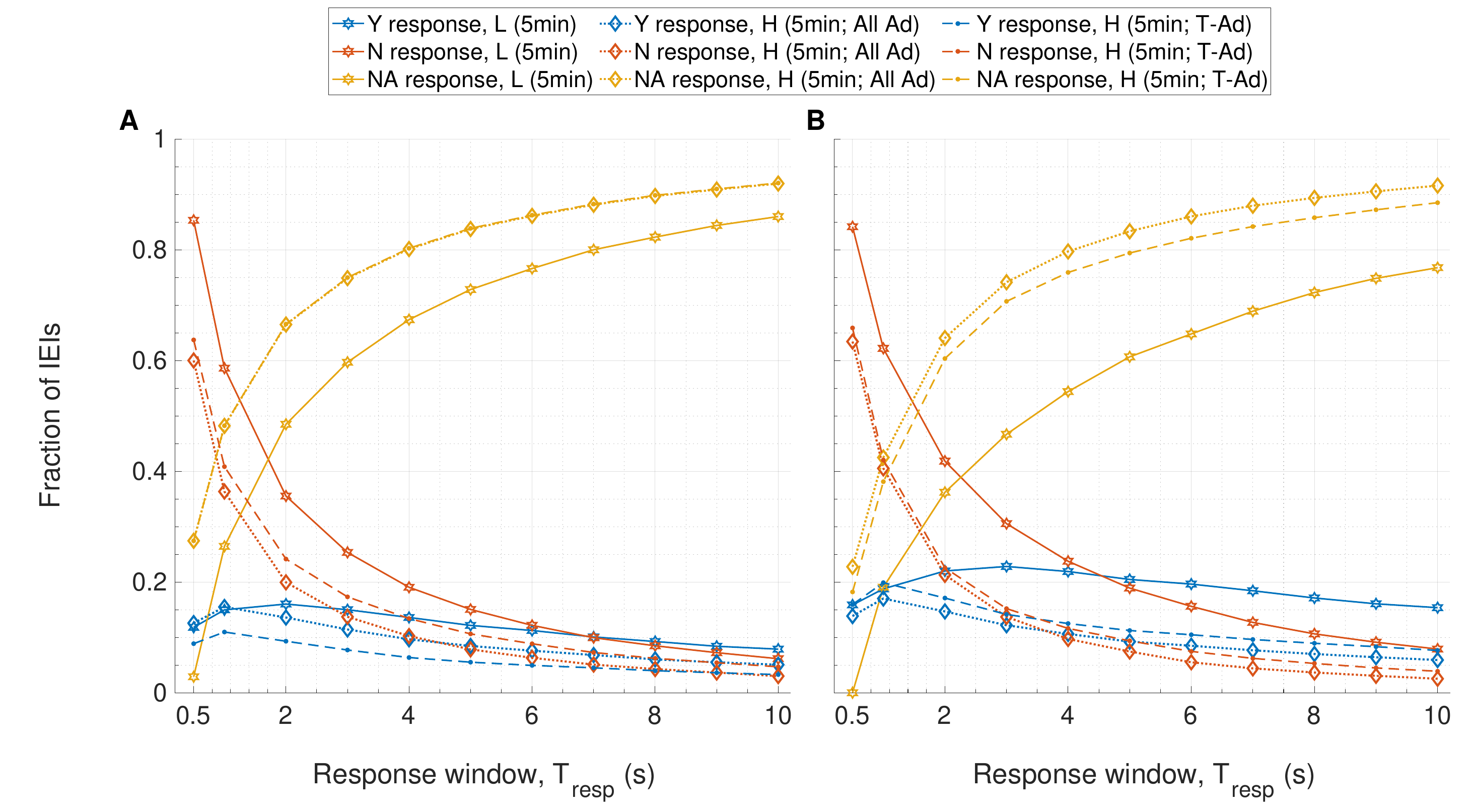}
\caption{\textbf{Number of IEIs associated with different response types for validation data. (A)} 
Fractions of infant speech-related (ChSp) IEIs associated with an adult (Ad) response (blue, Y response), not associated with an adult response (red, N response), and associated with NA response (yellow) are shown as a function of response window, $T_{\rm{resp}}$ (x-axis, seconds), for L (5min) (solid line, open hexagrams), H (5min; All Ad) (dotted line, open diamonds), and H (5min; T-Ad) (dot-dash line, solid circles) data.
All fractions are computed with respect to the total number of ChSp IEIs for that data type.
For example, the fraction of L (5min) ChSp IEIs associated with a Y response is computed with respect to the total number of ChSp IEIs for L (5min) data. 
\textbf{(B)} Fractions of Ad IEIs associated with a ChSp response (blue, Y response), not associated with a ChSp response (red, N response), and associated with NA response (yellow) are shown as a function of response window, $T_{\rm{resp}}$, for L (5min) (solid line, open hexagrams), H (5min; All Ad) (dotted line, open diamonds), and H (5min; T-Ad) (dot-dash line, solid circles) data.
All fractions are computed with respect to the total number of Ad IEIs for that data type.
A and B share the y-axis and legend.
%The Y axis is on a log scale to aid better visualisation.
}
\label{fig:NumIEIsResponseTypes_Valdata}
\end{figure}

\newpage
%---SECTION: ExtendedResults--------------------
\section{Extended results}\label{sec:ExtendedResults}

%---SUBSECTION: PrevIEIbeta--------------------
\subsection{Correlations between current and previous IEI}\label{subsec:PrevIEIbeta}

%---Tab: Lday_PrevStepBeta_ChSp--------------------
\begin{table}[H]
\centering
\begin{tabular}{cccc}
\hline
Infant age  & ChSp previous IEI $\beta$ & Num. IEIs & Marginal R\textsuperscript{2}\\
\hline
3 months & 0.30 (0.28, 0.31)& 44531 & 0.09 \\
6 months & 0.26 (0.25, 0.27)& 47166 & 0.07 \\
9 months & 0.25 (0.23, 0.26)& 48828 & 0.06 \\
18 months& 0.23 (0.21, 0.24)& 60092 & 0.05 \\
\hline
\end{tabular}
\caption{\textbf{Previous IEI $\beta$s for infant speech-related (ChSp) vocalizations  for LENA daylong data.}
Standardized regression coefficients ($\beta$s) from linear mixed effects analyses testing the effect of previous IEI on current IEI for ChSp vocalizations, with infant ID as a random effect are shown for LENA daylong data, by vocalization type and infant age (model structure: Current IEI $\sim$ Previous IEI + (1\textbar Infant ID)).
99.9\% confidence intervals are provided in parentheses. 
All effects are significant at the $p < 0.001$ level. 
Also shown are the number of IEIs that went into the relevant regression model as well as the marginal R\textsuperscript{2} value \cite{nakagawa2013general}.}
\label{tab:Lday_PrevStepBeta_ChSp}
\end{table}

%---Tab: Lday_PrevStepBeta_Ad--------------------
\begin{table}[H]
\centering
\begin{tabular}{cccc}
\hline
Infant age  & Ad previous IEI $\beta$ & Num. IEIs & Marginal R\textsuperscript{2}  \\
\hline
3 months & 0.33 (0.32, 0.33) & 129867 & 0.11 \\
6 months & 0.31 (0.30, 0.32) & 109028 & 0.09 \\
9 months & 0.31 (0.30, 0.32) & 104495 & 0.10 \\
18 months& 0.29 (0.28, 0.30) & 96451 & 0.08 \\
\hline
\end{tabular}
\caption{\textbf{Previous IEI $\beta$s for adult (Ad) vocalizations for LENA daylong data.}
Standardized regression coefficients ($\beta$s) from linear mixed effects analyses testing the effect of previous IEI on current IEI for Ad vocalizations, with infant ID as a random effect are shown for LENA daylong data, by vocalization type and infant age (model structure: Current IEI $\sim$ Previous IEI + (1\textbar Infant ID)).
99.9\% confidence intervals are provided in parentheses. 
All effects are significant at the $p < 0.001$ level. 
Also shown are the number of IEIs that went into the relevant regression model as well as the marginal R\textsuperscript{2} value.}
\label{tab:Lday_PrevStepBeta_Ad}
\end{table}

%---Tab: Valdata_PrevStepBeta_Chsp--------------------
\begin{table}[H]
\centering
\begin{tabular}{cccc}
\hline
Data type  & ChSp previous IEI $\beta$ & Num. IEIs & Marginal R\textsuperscript{2} \\
\hline
L (5min) & 0.10	(0.07, 0.13) & 12390 & 0.01 \\
H (5min; All Ad) &	0.19 (0.17, 0.21) & 23493 & 0.04 \\
H (5min; T-Ad) &	0.19 (0.17, 0.21)& 23493 & 0.04 \\
\hline
\end{tabular}
\caption{\textbf{Previous IEI $\beta$s for infant speech-related (ChSp) vocalizations for validation data.}
Standardized regression coefficients ($\beta$s) from linear mixed effects analyses testing the effect of previous IEI on current IEI for ChSp vocalizations, with infant ID, infant age, and infant ID-infant age interaction as random effects are shown for  validation datasets (model structure: Current IEI $\sim$ Previous IEI + (1\textbar Infant ID) + (1\textbar Infant Age) + (1\textbar Infant ID:Infant Age)).
99.9\% confidence intervals are provided in parentheses. 
All effects are significant at the $p < 0.001$ level. 
Also shown are the number of IEIs that went into the relevant regression model as well as the marginal R\textsuperscript{2} value.}
\label{tab:Valdata_PrevStepBeta_Chsp}
\end{table}

%---Tab: Valdata_PrevStepBeta_Ad--------------------
\begin{table}[H]
\centering
\begin{tabular}{cccc}
\hline
Data type &	Ad previous IEI	$\beta$ & Num. IEIs & Marginal R\textsuperscript{2} \\
\hline
L (5min) &	        0.17 (0.14,	0.21) &	7886	& 0.03 \\
H (5min; All Ad) &	0.17 (0.15, 0.20) &	17503	& 0.03 \\
H (5min; T-Ad) &	0.12 (0.08, 0.15) & 9687	& 0.01 \\
\hline
\end{tabular}
\caption{\textbf{Previous IEI $\beta$s for adult (Ad) vocalizations for validation data.}
Standardized regression coefficients ($\beta$s) from linear mixed effects analyses testing the effect of previous IEI on current IEI for Ad vocalizations, with infant ID, infant age, and infant ID-infant age interaction as random effects are shown for  validation datasets (model structure: Current IEI $\sim$ Previous IEI + (1\textbar Infant ID) + (1\textbar Infant Age) + (1\textbar Infant ID:Infant Age)).
99.9\% confidence intervals are provided in parentheses. 
All effects are significant at the $p < 0.001$ level. 
Also shown are the number of IEIs that went into the relevant regression model as well as the marginal R\textsuperscript{2} value.}
\label{tab:Valdata_PrevStepBeta_Ad}
\end{table}

%---Tab: PrevStepBeta_AgeEffs--------------------
\begin{table}[H]
\centering
\begin{tabular}[t]{llllcc}
\hline
Data type       & Voc. type & Age $b$    & Age\textsuperscript{2} $b$  & Num. recordings & Marginal R\textsuperscript{2} \\
\hline
L (day)         & ChSp     &\textbf{-0.013$^{***}$} & \textbf{4 $\times \: 10^{-4}$} & 200 & 0.12\\
                &          & \textbf{($p <$ 0.001)} & \textbf{($p$ = 0.01) }         & &\\
L (5min)        & ChSp     & 0.006                  & -3 $\times \: 10^{-4}$& 148 & 0.002\\
                &          &($p$ = 0.54)            & ($p$ = 0.52)          & &\\
H (5min; All Ad)& ChSp     & -0.014                 & 6 $\times \: 10^{-4}$ & 148 & 0.01\\
                &          &($p$ = 0.14)            & ($p$ = 0.17)          & &\\
H (5min; T-Ad)  & ChSp     & -0.014                 & 6 $\times \: 10^{-4}$ & 148 & 0.01\\
                &          & ($p$ = 0.14)           & ($p$ = 0.17)          & &\\
L (day)         & Ad       & -0.003                 & 5 $\times \: 10^{-5}$ & 200 & 0.05\\
                &          & ($p$ = 0.31)           &  ($p$ = 0.73)         & &\\
L (5min)        & Ad       & -0.009                 & 4 $\times \: 10^{-4}$ & 146 & 0.004\\
                &          &    ($p$ = 0.56)        &  ($p$ = 0.63)         & &\\
H (5min; All Ad)& Ad       & -0.005                 & 8 $\times \: 10^{-5}$ & 147 & 0.01\\
                &          & ($p$ = 0.65)           &($p$ = 0.87)           & &\\
H (5min; T-Ad)  & Ad       & 0.004                  & -4 $\times \: 10^{-4}$& 139 & 0.01\\
                &          & ($p$ = 0.85)           & ($p$ = 0.67) & &\\
\hline
\end{tabular}
\caption{\textbf{Effect of infant age on previous IEI $\beta$s.}
Unstandardized regression coefficients ($b$s) from linear mixed effects analyses testing the effect of infant age (in months) on previous IEI $\beta$s are shown for infant speech-related (ChSp) and adult (Ad) vocalizations for LENA daylong data (L (day)) and validation data: human listener-labeled 5-minute sections with all adult vocalizations included (H (5min; All-Ad)); human listener-labeled 5-minute sections with only child-directed adult vocalizations
included (H (5min; T-Ad)); and the corresponding LENA subset of 5-minute sections (L (5min)). 
Linear mixed effects analyses were run predicting recording-level previous IEI $\beta$s\textemdash correlations between current IEI and previous IEI at the recording level\textemdash with infant age and infant age\textsuperscript{2} as fixed effects and infant ID as a random effect (model structure: Previous IEI $\beta$ $\sim$ Infant Age + Infant Age\textsuperscript{2} + (1\textbar Infant ID)).
Infant age was not normalized for this analysis.
%$p$ values are in brackets
Significant $b$s at $p = 0.05$ are in bold while significant $b$s at $p = 0.001$ are indicated by \textsuperscript{***}.
The number of recordings (with each recording resulting in a recording-level previous IEI $\beta$) that contributed to the relevant regression model as well as the marginal R\textsuperscript{2} values are also shown.} 
\label{tab: PrevStepBeta_AgeEffs}
\end{table}

Detailed model outputs for all analyses testing the effect of the previous IEI on current IEI are available \href{https://osf.io/5xp7z/files/56ay8}{here}.\footnote{\url{https://osf.io/5xp7z/files/56ay8}}
Detailed model outputs for all analyses testing the effect of infant age on recording-level previous IEI $\beta$s are available \href{https://osf.io/npv3e}{here}.\footnote{\url{https://osf.io/npv3e}}

%---SUBSECTION: RespEffectResults--------------------
\subsection{Response effect on IEIs}\label{subsec:RespEffectResults}

Tables with response effect $\beta$s using a response window ($T_{\rm resp}$) value of 5 s as a representative case are shown below. 
Response $\beta$s for the effect of adult (Ad) responses on infant speech-related (ChSp) IEIs and for the effect of infant speech-related responses on adult IEIs are shown for LENA daylong data and validation data. 
Results for the full range of $T_{\rm{resp}}$ values analyzed are available in the \href{https://osf.io/5xp7z/}{OSF repository}\footnote{\url{https://osf.io/5xp7z/}} associated with this study.
Specifically, a spreadsheet summarizing model results for all vocalizers (ChSp and Ad), datasets (L (day), H (5min; All Ad), H (5min; T-Ad), and L (5min)), and response windows analyzed is available \href{https://osf.io/5xp7z/files/qzurx}{here}.\footnote{\url{https://osf.io/5xp7z/files/qzurx}} 
More detailed model outputs are available \href{https://osf.io/5xp7z/files/56ay8}{here}.\footnote{\url{https://osf.io/5xp7z/files/56ay8}}

%---Tab: Daylong_ChSpIEI_5s_RespBeta_Lday--------------------
\begin{table}[H]
\centering
\begin{tabular}{ccccc}
\hline
Infant age  & 	ChSp IEI Response $\beta$ & 99.9\% CI & Num. IEIs & R\textsuperscript{2} \\
\hline
3 months & -0.12 & (-0.16, -0.08) & 22458 & 0.004\\
6 months & -0.09 & (-0.13, -0.05) &	22384 & 0.002\\ 
9 months & -0.11 & (-0.15, -0.07) &	24490 & 0.003\\
18 months & -0.15 & (-0.19, -0.11)& 27482 & 0.006\\
\hline
\end{tabular}
\caption{\textbf{Effect of receiving an adult (Ad) response on infant speech-related (ChSp) IEI length for LENA daylong data for $T_{\rm resp} = 5$~s.}
Standardized regression coefficients ($\beta$s) from linear models testing the effect of receiving an Ad response on ChSp IEI residuals (after regressing current IEI on previous IEI with infant ID as a random effect) for $T_{\rm resp} = 5$ s are shown, by infant age (model structure: Residual IEI $\sim$ Response Receipt). 
99.9\% confidence intervals for the $\beta$s, the number of IEIs that went into the regression, and R\textsuperscript{2} values are also provided. 
All effects are significant at the $p < 0.001$ level. }
\label{tab:Daylong_ChSpIEI_5s_RespBeta_Lday}
\end{table}

%---Tab: Daylong_AdIEI_5s_RespBeta_Lday--------------------
\begin{table}[H]
\centering
\begin{tabular}{ccccc}
\hline
Infant Age  & 	Ad IEI Response $\beta$ & 99.9\% CI & Num. IEIs & R\textsuperscript{2} \\
\hline
3 months & -0.16 & (-0.21, -0.12) & 45037 & 0.004\\
6 months & -0.11 & (-0.15, -0.07) & 40722 & 0.002\\
9 months & -0.14 & (-0.18, -0.10) & 41120 & 0.003\\
18 months & -0.13 & (-0.17, -0.09)& 39828 & 0.003\\
\hline
\end{tabular}
\caption{\textbf{Effect of receiving an infant (ChSp) response on adult (Ad) IEI length for LENA daylong data for $T_{\rm resp} = 5$~s.}
Standardized regression coefficients ($\beta$s) from linear models testing the effect of receiving a ChSp response on Ad IEI residuals (after regressing current IEI on previous IEI with infant ID as a random effect) for $T_{\rm resp} = 5$ s are shown, by infant age (model structure: Residual IEI $\sim$ Response Receipt). 
99.9\% confidence intervals for the $\beta$s, the number of IEIs that went into the regression, and R\textsuperscript{2} values are also provided.
All effects are significant at the $p < 0.001$ level.}
\label{tab:Daylong_AdIEI_5s_RespBeta_Lday}
\end{table}

%---Tab: Valdata_ChSpIEI_5s_RespBeta--------------------
\begin{table}[H]
\centering
\begin{tabular}{ccccc}
\hline
Data type  & 	ChSp IEI Response $\beta$ & 99.9\% CI & Num. IEIs & R\textsuperscript{2} \\
\hline
L (5min) & \textbf{-0.07 ($p$ = 0.05)} & (-0.18, 0.05) & 3369 & 0.001\\
H (5min; All Ad) & -0.03 ($p$ = 0.30)& (-0.14, 0.07) & 3821 & 0.0003\\
H (5min; T-Ad) & \textbf{-0.07 ($p$ = 0.03)}& (-0.19, 0.04) & 3791 & 0.001\\
\hline
\end{tabular}
\caption{\textbf{Effect of receiving an adult (Ad) response on infant speech-related (ChSp) IEI length for validation datasets for $T_{\rm resp} = 5$~s.}
Standardized regression coefficients ($\beta$s) from linear models testing the effect of receiving an Ad response on ChSp IEI residuals (after regressing current IEI on previous IEI with infant ID, infant age, and infant ID-infant age interaction as random effects) for $T_{\rm resp} = 5$ s are shown for the validation datasets: human listener-labeled 5-minute sections with all adult vocalizations included (H (5min; All-Ad)); human listener-labeled 5-minute sections with only child-directed adult vocalizations included (H (5min; T-Ad)); and the corresponding LENA subset of 5-minute sections (L (5min)).
The model structure is: Residual IEI $\sim$ Response Receipt.
$p$ values are indicated in parentheses.
Significant $\beta$s at $p$ = 0.05 are in bold (the $p$ value for L (5min) data is 0.0478 and has been rounded to 0.05).
99.9\% confidence intervals for the $\beta$s, the number of IEIs that went into the regression, and R\textsuperscript{2} values are also provided.}
\label{tab:Valdata_ChSpIEI_5s_RespBeta}
\end{table}

%---Tab: Valdata_AdIEI_5s_RespBeta--------------------
\begin{table}[H]
\centering
\begin{tabular}{ccccc}
\hline
Data type  & 	Ad IEI Response $\beta$ & 99.9\% CI & Num. IEIs & R\textsuperscript{2} \\
\hline
L (5min) & \textbf{-0.10 ($p$ = 0.005)\textsuperscript{**}} & (-0.22, 0.02) & 3103 & 0.003 \\
H (5min; All Ad) & 0.06 ($p$ = 0.09) & (-0.06, 0.19) & 2918 & 0.001 \\
H (5min; T-Ad) & \textbf{-0.11 ($p =$ 0.01)} & (-0.26, 0.03) & 1995 & 0.003 \\
\hline
\end{tabular}
\caption{\textbf{Effect of receiving an infant (ChSp) response on adult (Ad) IEI length for validation datasets for $T_{\rm resp} = 5$~s.}
Standardized regression coefficients ($\beta$s) from linear models testing the effect of receiving a ChSp response on Ad IEI residuals (after regressing current IEI on previous IEI with infant ID, infant age, and infant ID-infant age interaction as random effects) for $T_{\rm resp} = 5$ s are shown for the validation datasets (model structure: Residual IEI $\sim$ Response Receipt). 
$p$ values are indicated in parentheses.
Significant $\beta$s at $p$ = 0.05 are in bold and significant $\beta$s at $p$ = 0.01 are indicated by \textsuperscript{**}. %, and significant $\beta$s at $p$ = 0.001 are indicated by \textsuperscript{***}.
99.9\% confidence intervals for the $\beta$s, the number of IEIs that went into the regression, and R\textsuperscript{2} values are also provided.}
\label{tab:Valdata_AdIEI_5s_RespBeta}
\end{table}

\newpage
\noindent
A spreadsheet summarizing the effect of infant age on recording-level response $\beta$s  for all vocalizers (ChSp and Ad), datasets (L (day), H (5min; All Ad), H (5min; T-Ad), and L (5min)), and response windows analyzed is available \href{https://osf.io/5xp7z/files/8sjvm}{here}.\footnote{\url{https://osf.io/5xp7z/files/8sjvm}} 
More detailed model outputs for these analyses are available \href{https://osf.io/npv3e}{here}.\footnote{\url{https://osf.io/npv3e}}

%---Fig: RespEffFigs_wAndWoCtrl_Lday_SI_99_9CI--------------------
\begin{figure}[H]
\centering
\includegraphics[width=0.95\linewidth]{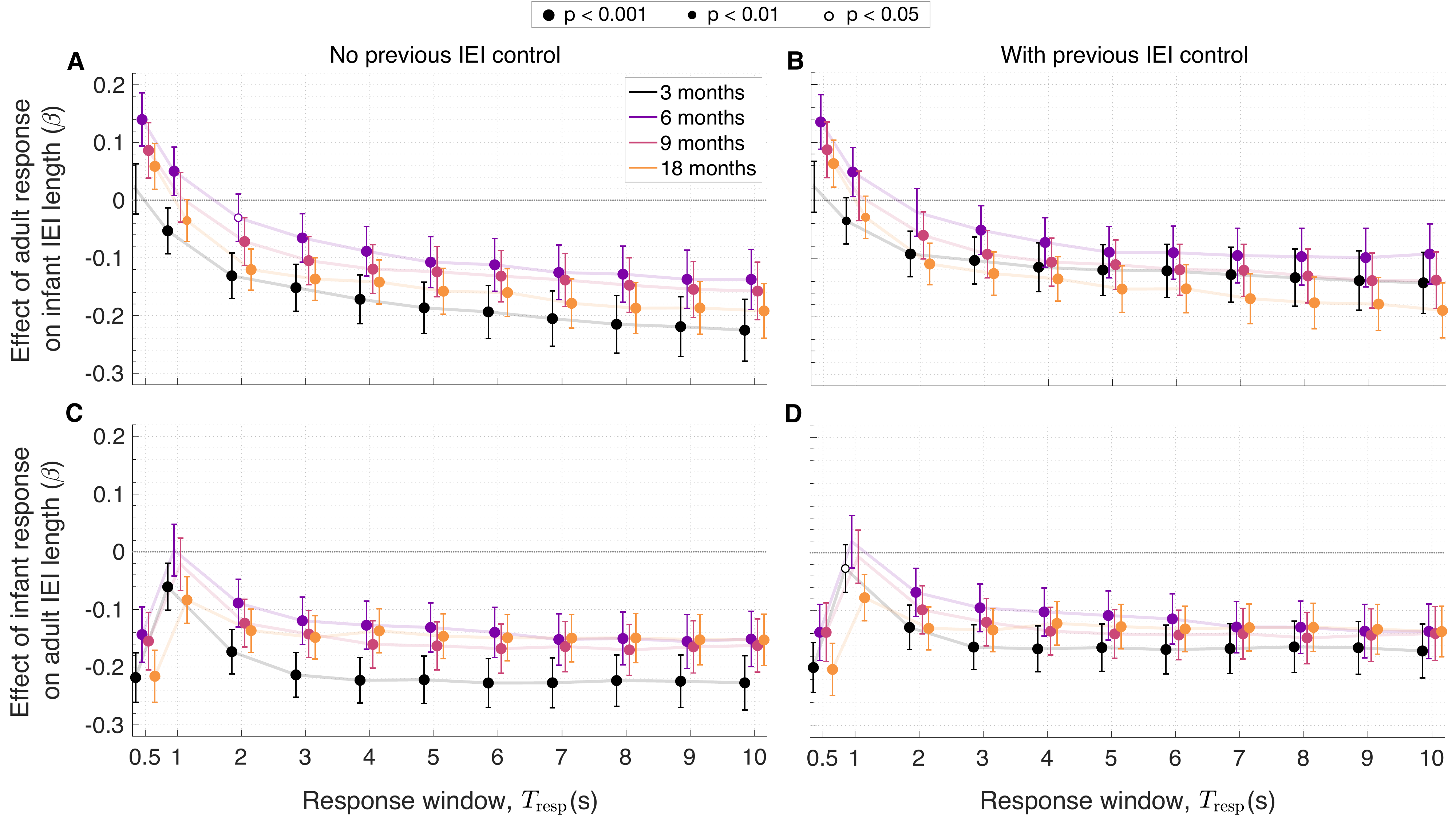}
\caption{\textbf{Response effects on IEIs with and without the previous IEI control: LENA daylong data.} 
\textbf{(A)} Standardized regression coefficients ($\beta$; y-axis) from linear mixed effects models testing the effect of receiving an adult response on infant IEIs \textit{without} controlling for the correlation between successive IEIs are shown as a function of response window duration, $T_{\rm{resp}}$ (x-axis, seconds).
$\beta$ values are staggered around the relevant $T_{\rm{resp}}$ value for easier visualization.
Statistical analyses were performed separately for each infant age group: 3, 6, 9, and 18 months (indicated by line and circle color; see legend). 
Infant ID was included as a random effect (model structure: Current IEI $\sim$ Response Receipt + (1\textbar Infant ID)). 
For all panels, bars show 99.9\% confidence intervals.
Significant $\beta$ values (at $p < 0.001$, $p < 0.01$, and $p < 0.05$) are indicated as differently-sized solid or open circles (see \textit{p}-value legend).
\textbf{(B)} $\beta$ values (y-axis scale shared with panel A) from linear models testing the effect of receiving an adult response on infant IEIs \textit{after} controlling for the correlation between successive IEIs (model structure: Residual IEI $\sim$ Response Receipt; see Section \ref{subsec:StatTests} for details) are shown as a function of response window duration, $T_{\rm{resp}}$.
\textbf{(C)} $\beta$ values from a similar model as in A are shown, testing the effect of receiving an infant response on adult IEIs \textit{without} controlling for the correlation between successive IEIs. 
\textbf{(D)} $\beta$ values (y-axis scale shared with panel C) from a similar model as in B are shown, testing the effect of receiving an infant response on adult IEIs \textit{after} controlling for the correlation between successive IEIs.
Note that all panels have the same y-axis scale and limits.}
%Note: this fig shows that the results do differ without the control. 
\label{fig:RespEffFigs_wAndWoCtrl_Lday_SI_99_9CI}
\end{figure}

%---Fig: RespEffFigs_wAndWoCtrl_Valdata_SI_99_9CI--------------------
\begin{figure}[H]
\centering
\includegraphics[width=0.95\linewidth]{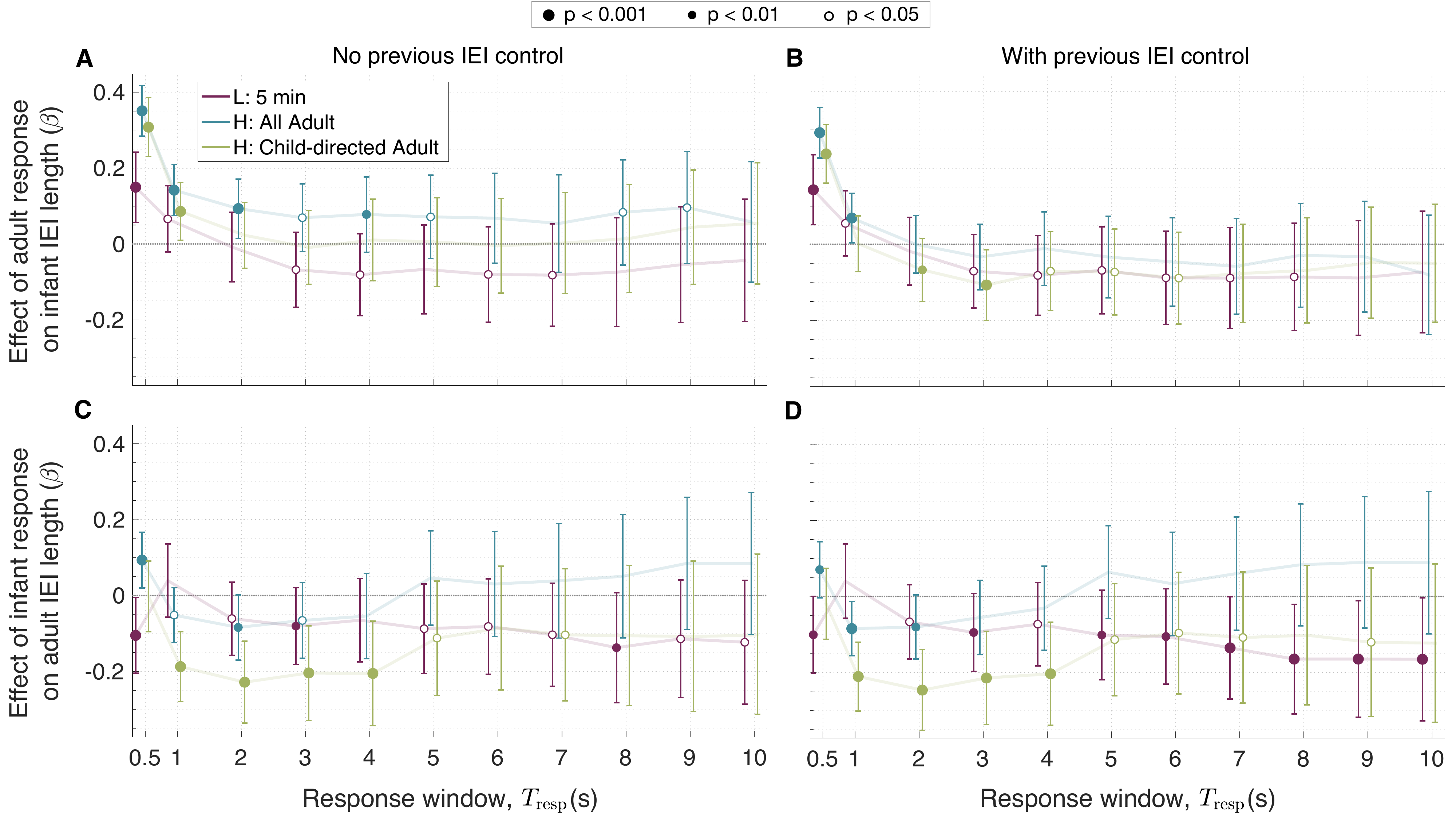}
\caption{\textbf{Response effects on IEIs with and without the previous IEI control: validation data.} 
\textbf{(A)} Standardized regression coefficients ($\beta$; y-axis) from linear mixed effects models testing the effect of receiving an adult response on infant IEIs \textit{without} controlling for the correlation between successive IEIs are shown as a function of response window duration, $T_{\rm{resp}}$ (x-axis, seconds) for validation datasets (see
legend): human listener-labeled 5-minute sections with all adult vocalizations included (H: All Adult; dark cyan), human listener-labeled 5-minute sections with only infant-directed
adult vocalizations included (H: Child-directed Adult; green), and corresponding 5-minute sections as labeled by LENA (L: 5 min; maroon). 
$\beta$ values are staggered around the relevant $T_{\rm{resp}}$ value for easier visualization.
Infant ID, infant age, and infant ID-infant age interaction were included as random effects (model structure: Current IEI $\sim$ Response Receipt + (1\textbar Infant ID) + (1\textbar Infant Age) + (1\textbar Infant ID:Infant Age)).
Infant age (in months) was normalized prior to these analyses.
For all panels, bars show 99.9\% confidence intervals.
Significant $\beta$ values (at $p < 0.001$, $p < 0.01$, and $p < 0.05$) are indicated as differently-sized solid or open circles (see \textit{p}-value legends).
\textbf{(B)} $\beta$ values (y-axis scale shared with panel A) from linear models testing the effect of receiving an adult response on infant IEIs \textit{after} controlling for the correlation between successive IEIs (model structure: Residual IEI $\sim$ Response Receipt; see Section \ref{subsec:ValDataMethods} for details) are shown as a function of response window duration, $T_{\rm{resp}}$.
\textbf{(C)} $\beta$ values from a similar model as in A are shown, testing the effect of receiving an infant response on adult IEIs \textit{without} controlling for the correlation between successive IEIs. 
\textbf{(D)} $\beta$ values (y-axis scale shared with panel C) from a similar model as in B are shown, testing the effect of receiving an infant response on adult IEIs \textit{after} controlling for the correlation between successive IEIs.
Note that all panels have the same y-axis scale and limits.}
\label{fig:RespEffFigs_wandwoCtrl_Valdata_99_9CI}
\end{figure}

The pattern of response $\beta$s (after the previous IEI control) quantifying the effect of adult response receipt on infant IEI length being positive (at $p < 0.001$; see Fig. \ref{fig:RespBetas} in the main text) for short response windows ($T_{\rm resp} < 2$ s) for both daylong (Fig. \ref{fig:RespEffFigs_wAndWoCtrl_Lday_SI_99_9CI}B) and validation datasets (Fig. \ref{fig:RespEffFigs_wandwoCtrl_Valdata_99_9CI}B) is likely due to the fact that adult vocalizations tend to be longer than most other sound types (see Section \ref{sec:SegmentNumAndDuration}). 
For these low response window values, the shortest IEIs in the data are included in response analyses (see Figs. \ref{fig:NumResponses_Ldaylong}, \ref{fig:NumIEIsResponseTypes_Valdata}). 
IEIs associated with a response necessarily have an intervening adult vocalization.
IEIs not associated with a response, on the other hand, can only have silence or sound types other than infant (speech and non-speech related) or adult\textemdash which may generally be shorter than adult vocalizations\textemdash intervening. 
At low response window thresholds, short IEIs for which the limiting factor on IEI length is the intervening sound type are more likely to be included in the analysis.
As a result, for small $T_{\rm resp}$ values, Resp IEIs may tend to be longer than NoResp IEIs purely as a function of the duration of the intervening adult response vocalization.\\

In contrast, while response $\beta$s (after the previous IEI control) quantifying the effect of infant responses on adult IEI length display non-monotonicity when response windows are small ($T_{\rm resp} \leq 2$ s) for both the daylong and 5-minute LENA-labeled datasets (see Fig. \ref{fig:RespBetas} in the main text as well as Figs. \ref{fig:RespEffFigs_wAndWoCtrl_Lday_SI_99_9CI}D, \ref{fig:RespEffFigs_wandwoCtrl_Valdata_99_9CI}D), these patterns are not present in the human listener labeled validation datasets.\footnote{Note that not all response $\beta$s are significant at $p < 0.001$ for $T_{\rm resp} \leq 2$ s.}
Thus, it is possible that this non-monotonic behavior is an artifact resulting from a combination of LENA’s minimum duration thresholds for different sound types and differences in typical vocalization durations. 
LENA child vocalizations have a 600 ms minimum and modal duration (Figs. \ref{fig:LdaySegsDurAndNumsSummaryStatswErrBars_ChnAd0IviMerged}, \ref{fig:DurDist_ChnspLday}), which is shorter than the 800--1000 ms minimum durations for all other sound type labels as reported in the LENA documentation \cite{gilkerson2020guide_SI} (our data indicate that most sound types\textemdash including adult vocalizations\textemdash have minimum durations lower than these published minima; see Figs. \ref{fig:LdaySegsDurAndNumsSummaryStatswErrBars_ChnAd0IviMerged} and \ref{fig:DurDist_AnLday}). 
Thus, for adult IEIs at $T_{\rm{resp}}=0.5$ s, it may be more likely that only one sound type intervenes, since IEIs as short as $T_{\rm{resp}}$ are included in the analyses.
For Resp IEIs, this sound type must, by definition, be the target child, which can be expected to be relatively short, resulting in short Resp IEIs.  
However, when $T_{\rm{resp}}=1$ s, Resp IEIs associated with a relatively short infant response may have a second intervening sound type\textemdash since IEIs must be greater than $T_{\rm{resp}}$ to be included in the analysis\textemdash and this could lead to Resp IEIs being longer.
For the human listener-labeled validation data, response effect patterns at low $T_{\rm resp}$ values are similar to results for infant IEIs, supporting the possibility that response effects at low $T_{\rm resp}$ values for the LENA datasets may be primarily driven by LENA's minimum duration thresholds for different sound types.

%---SUBSECTION: Prev1vsPrev2Iei--------------------
\subsection{Comparison of model results: controlling for one previous IEI vs. two previous IEIs}\label{subsec: Prev1vsPrev2Iei}

To test how including more history to account for temporal structure affected our results, we ran a model controlling for the previous 2 IEIs.
Here the 2-step analysis pipeline includes a linear mixed effects regression with the current IEI (\textit{i}\textsuperscript{th} IEI) as the dependent variable and the previous IEI (($i-1$)\textsuperscript{th} IEI) and the previous-to-previous IEI (($i-2$)\textsuperscript{th} IEI) as the independent variables. 
Infant ID was included as a random effect. 
For the LENA daylong data, these analyses were carried out at the age block level, while validation datasets were not separated into age blocks. 
As such, infant age and infant ID-infant age interaction were included as additional random effects for validation datasets. 
The model structure for the LENA daylong data was: Current IEI $\sim$ Previous IEI + Previous-to-previous IEI + (1\textbar Infant ID)).
The model structure for the validation data was: Current IEI $\sim$ Previous IEI + Previous-to-previous IEI + (1\textbar Infant ID) + (1\textbar Infant Age) + (1\textbar Infant ID:Infant Age)).
All fixed effects (as well as infant age in the random effect terms for the validation data) were normalized prior to analyses. 
Normalized residuals from the previous two IEIs mixed effects models were then used to test the effect of response receipt, using the same model structure (Residual IEI $\sim$ Response Receipt) as the response effect analyses outlined in Section \ref{subsec:StatTests}.
For the response effect analyses, we use a response window value ($T_{\rm resp}$) of 5 s as an illustrative example. \\

\begin{figure}[H]
    \centering
    \includegraphics[width=\linewidth]{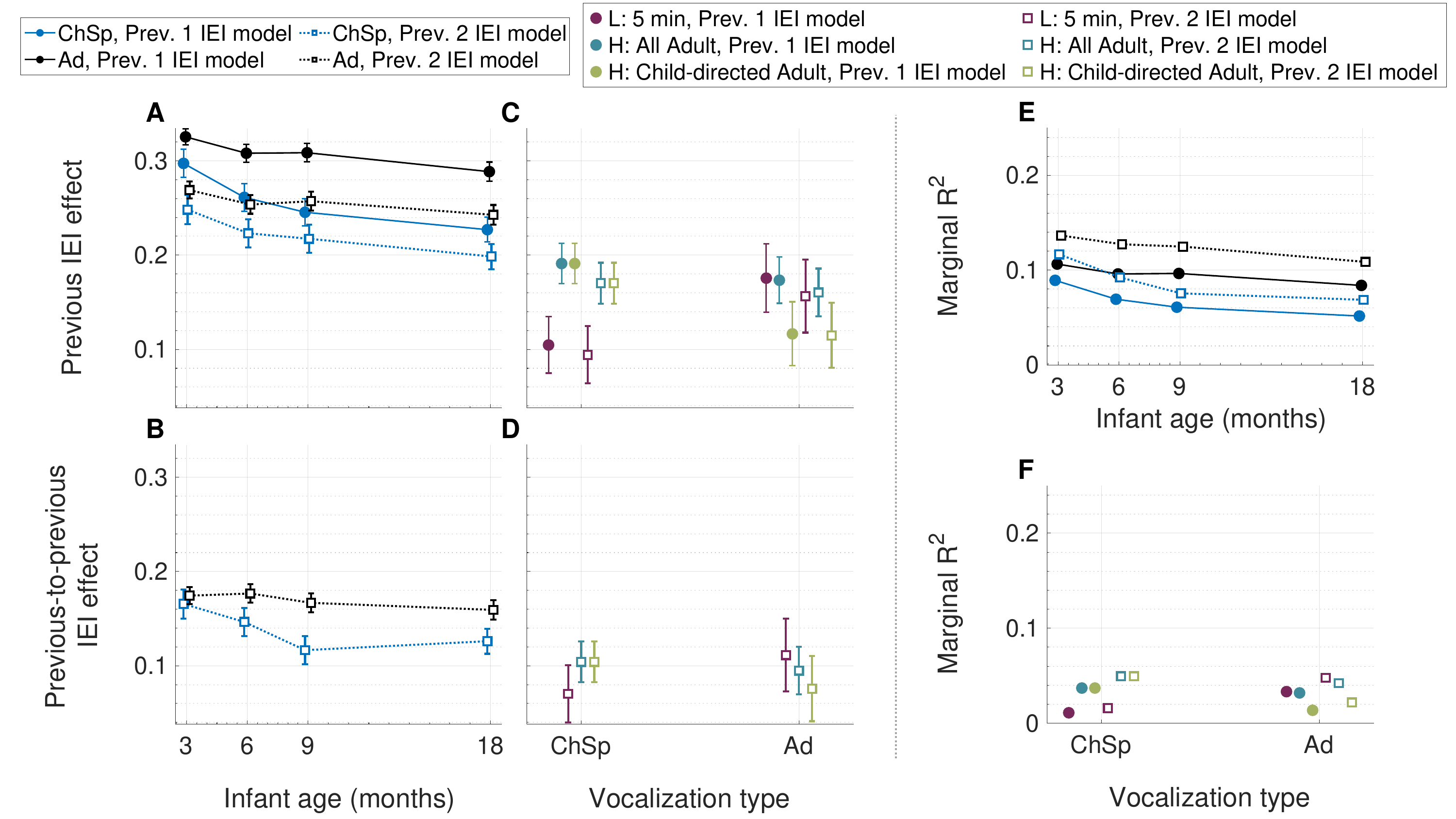}
    \caption{\textbf{Effect of previous IEI versus previous two IEIs on current IEI.
    (A)} Standardized regression coefficients ($\beta$; y-axis) quantifying the association between current IEI and previous IEI for infant speech-related (ChSp; blue; see legend on top left) and adult (Ad; black) vocalizers from linear mixed effects models with only the previous IEI as a fixed effect (Prev. 1 IEI model; solid circles) and models with previous IEI and previous-to-previous IEI as fixed effects (Prev. 2 IEI model; open squares)  are shown as a function of infant age in months (x-axis), for LENA daylong data. 
    99.9\% confidence intervals are shown as error bars.
    Analyses were performed separately for each age group (3, 6, 9, and 18 months). 
    Infant ID was treated as a random effect in all tests. 
    \textbf{(B)} shows $\beta$s (y-axis) quantifying the association between current IEI and previous-to-previous IEI for ChSp (blue) and Ad (black) vocalizers from the previous 2 IEI model as a function of infant age in months (x-axis) for the LENA daylong data.
    \textbf{(C)} $\beta$s (y-axis) quantifying the association between current IEI and previous IEI for ChSp and Ad vocalizers from the previous 1 IEI model (solid circles; see legend on top right) and the previous 2 IEI model (open squares) are shown for validation datasets: human listener-labeled 5-minute sections with all adult vocalizations included (H: All Adult; dark cyan), human listener-labeled 5-minute sections with only infant-directed adult vocalizations included (H: Child-directed Adult; green), and corresponding 5-minute sections as labeled by LENA (L: 5 min; maroon).  
    Validation data were not separated into age-blocks and instead, infant age (in months), infant ID, and infant ID-infant age interaction were treated as random effects. 
    Note that the $\beta$ values for ChSp vocalizations from H: Child-directed Adult and H: All Adult datasets are the same, since the child vocalizations included are the same for both datasets.
    \textbf{(D)} shows $\beta$s (y-axis) quantifying the association between current IEI and previous-to-previous IEI for ChSp and Ad vocalizers from the previous 2 IEI model for the validation datasets (see legend on top right).
    \textbf{(E)} shows the marginal $R^2$ values (y-axis) for ChSp (blue) and Ad (black) vocalizations for the previous 1 IEI model (solid circles) and the previous 2 IEI model (open squares) as a function of infant age (months; x-axis) for the LENA daylong data (see legend on top left).
    \textbf{(F)} shows the marginal $R^2$ values (y-axis) for ChSp and Ad vocalizations for the previous 1 IEI model (solid circles) and the previous 2 IEI model (open squares) for the validation data (see legend on top right).
    All $\beta$s are significant at $p <$ 0.001 for A--D, and $\beta$ (A, B) and marginal $R\textsuperscript{2}$ (E) values are staggered around the relevant infant age for easier visualization. 
    A, B, and E share a legend (top left) while C, D, and F share a legend (top right).}
    \label{fig:Prev1vs2IeiFig}
\end{figure}

\begin{figure}[H]
    \centering
    \includegraphics[width=\linewidth]{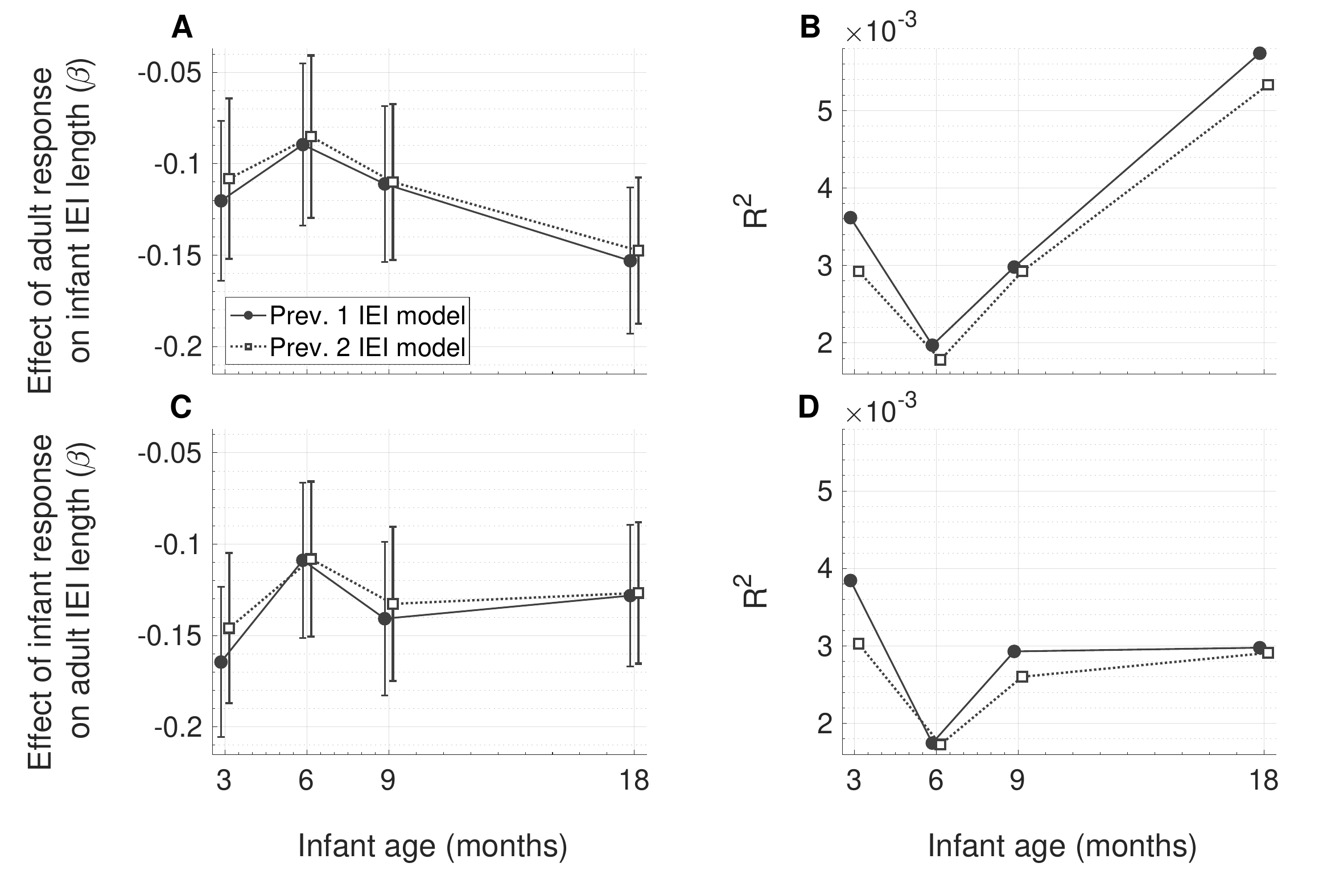}
    \caption{
    \textbf{Response effects for LENA daylong data ($T_{\rm resp}$ = 5 s) after controlling for previous IEI versus previous two IEIs.
    (A) } 
    Standardized regression coefficients ($\beta$; y-axis) from linear models testing the effect of receipt of an adult (Ad) response on the length of infant speech-related (ChSp) IEIs after controlling for the effect of previous IEI on current IEI (solid circles), and after controlling for the effect of the previous two IEIs on current IEI (open squares) are shown as a function of infant age (months; x-axis shared with C) for a response window value of $T_{\rm resp}$ = 5 s, for LENA daylong data.
    99.9\% confidence intervals are shown as error bars and all $\beta$ values are significant at $p < 0.001$.
    $\beta$ values are staggered around the relevant infant age for easier visualization.
    Analyses were performed separately for each infant age group (3, 6, 9, and 18 months). 
    \textbf{(B)} R\textsuperscript{2} values for the linear models whose $\beta$ values are shown in A are shown as a function of infant age (months; x-axis shared with D).
    \textbf{(C)} $\beta$s (y-axis) from linear models analogous to A, testing the effect of infant response on adult IEI length are shown for LENA daylong data as a function of infant age (x-axis).  
    All $\beta$ values are significant at $p < 0.001$.
    \textbf{(D)} R\textsuperscript{2} values for the linear models whose $\beta$ values are shown in C are shown as a function of infant age (months; x-axis). 
    A--D share a legend, while A and C, and B and D share y-axis scales.
    }    
    \label{fig:Prev1vs2IeiModelComp_RespEffLdayData}
\end{figure}

\begin{figure}[H]
    \centering
    \includegraphics[width=\linewidth]{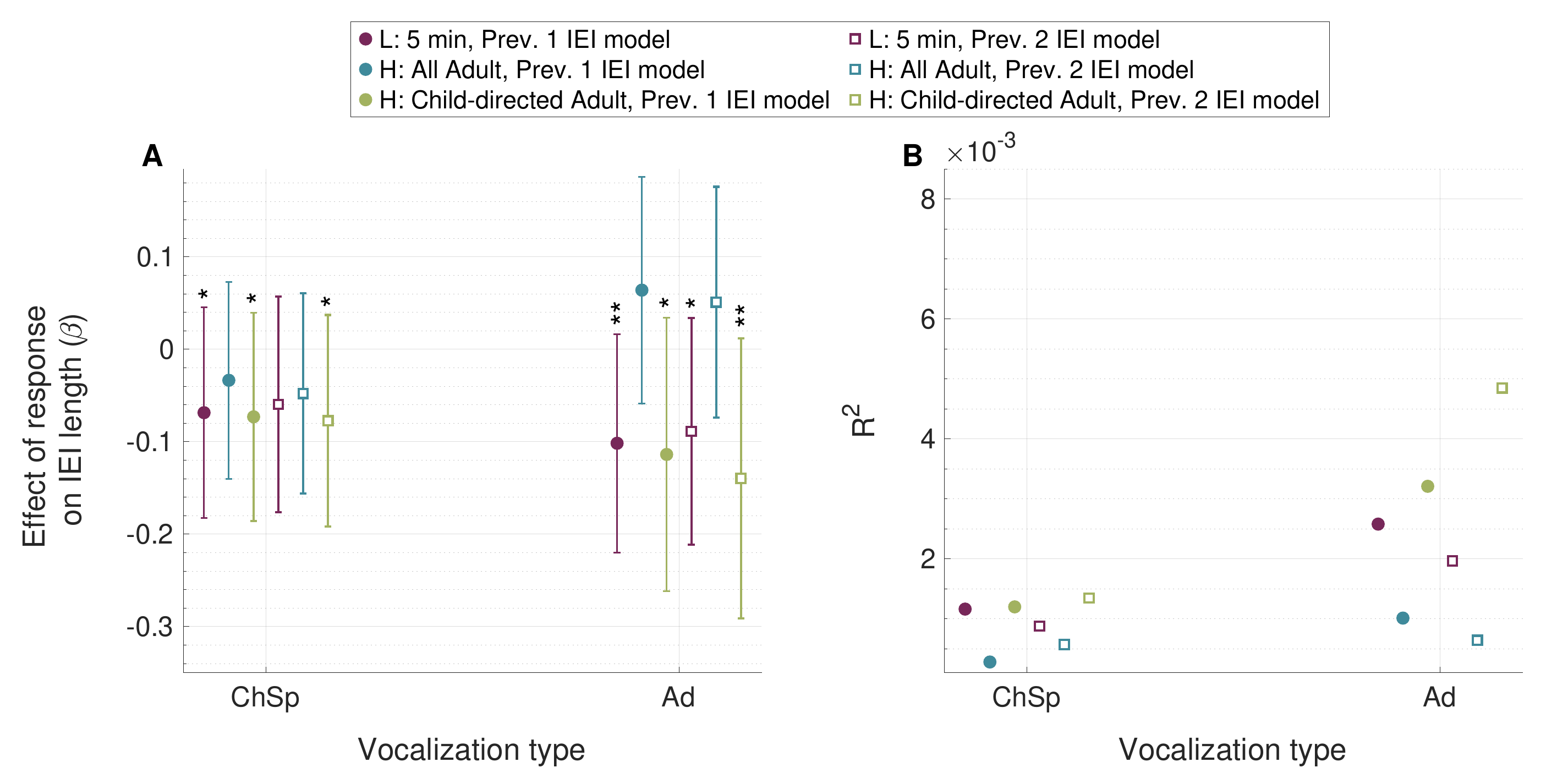}
    \caption{\textbf{Response effects for validation data ($T_{\rm resp}$ = 5 s) after controlling for previous IEI versus previous two IEIs. (A)}
    Standardized regression coefficients ($\beta$s; y-axis) from linear models testing the effect of response receipt on IEI length after controlling for the effect of previous IEI on current IEI (solid circles), and after controlling for the effect of the previous two IEIs on current IEI (open squares) are shown for a response window value of $T_{\rm resp}$ = 5 s, for the validation datasets (see legend): human listener-labeled 5-minute sections with all adult vocalizations included (H: All Adult; dark cyan), human listener-labeled 5-minute sections with only infant-directed adult vocalizations included (H: Child-directed Adult; green), and corresponding 5-minute sections as labeled by LENA (L: 5 min; maroon).
    $\beta$ values are grouped by vocalization type (x-axis), with $\beta$ values for the effect of adult (Ad) response receipt on infant speech-related (ChSp) IEI length on the left side, and $\beta$ values for the effect of ChSp response receipt on Ad IEI length on the right. 
    99.9\% confidence intervals are shown as error bars and significant $\beta$ values at $p < 0.05$ (*) and $p < 0.01$ (**) are indicated. 
    Validation data were not separated into age blocks.
    \textbf{(B)} R\textsuperscript{2} values for the linear models whose $\beta$ values are shown in A are shown. 
    A and B share a legend.}
\label{fig:Prev1vs2IeiModelComp_RespEffValData}
\end{figure}

We note that $\beta$ values for the effect of the previous IEI on current IEI are higher for models that control for the previous IEI versus the previous two IEIs for LENA daylong data and the validation data (Fig. \ref{fig:Prev1vs2IeiFig}A, C).
This is as expected, since the variance is split between the previous IEI and previous-to-previous IEI explanatory variables in the previous two IEI model.
R\textsuperscript{2} values are higher for the previous two IEI model versus the previous IEI model (Fig. \ref{fig:Prev1vs2IeiFig}E, F), which is also as expected. 
Response effect $\beta$s quantifying the effect of response receipt on IEI length, tested using residuals of the previous IEI models are comparable for both models, with no qualitative difference in findings for the LENA daylong data (Fig. \ref{fig:Prev1vs2IeiModelComp_RespEffLdayData}A, C) or the validation data (Fig. \ref{fig:Prev1vs2IeiModelComp_RespEffValData}A) for the effect of adult response on infant vocalizations and for that of infant response on adult vocalizations. 
R\textsuperscript{2} values for the response effect models for previous IEI versus previous two IEI residuals are comparable for the LENA daylong data (Fig. \ref{fig:Prev1vs2IeiModelComp_RespEffLdayData}B, D).
For the validation datasets, R\textsuperscript{2} values for the response effect models for previous IEI versus previous two IEI residuals diverge more but remain within the same order of magnitude (Fig. \ref{fig:Prev1vs2IeiModelComp_RespEffValData}B).

%---Fig: Prev1vs2IeiFig--------------------

%---Fig: Prev1vs2IeiModelComp_RespEffLdayData--------------------

%---Fig: Prev1vs2IeiModelComp_RespEffValData--------------------

\subsection{Alternative analytical approaches}\label{sec: AltAnalysis}

Some of the analytical approaches that may be employed as alternatives to the two-step framework we present include simultaneous regression models and autoregressive models.
We note that some of these alternatives are more readily amenable to the data in question than others. 
A simultaneous regression model which includes both previous IEI and response receipt as predictors of current IEI length provides an obvious contrast to the current approach where the effect of response receipt is assessed on IEI residuals from predicting the effect of the previous IEI on the current IEI.
These statistical approaches have somewhat different interpretations \cite{abdi2007part} and future research testing whether these approaches lead to different patterns of results\textemdash both for naturalistic datasets such as the one used in this study and simulated datasets where ground truth effects are known\textemdash would be insightful. 
More traditional autoregressive models are another potential alternative.
While these models can be useful in assessing the effects of temporal structure, we note that the nature of our data poses a challenge for most autoregression approaches. 
Specifically, naturalistic vocalization data are not temporally uniform but instead exhibit hierarchical clustering (see Section \ref{sec:burstiness}). 
The fact that each vocalization has a duration would also have to be taken into account.
Therefore, the relevant history of the system isn't easily expressed as a simple time series of the form $Y(t)$, as is required for straightforward implementations of autoregressive models.

% \clearpage
\end{document}